\documentclass[sigconf]{acmart}

\usepackage{balance}

\usepackage{multirow}
\usepackage{appendix}
\usepackage{subfig}
\usepackage{graphicx}
\usepackage{grffile}

\usepackage[font=bf]{caption}
\usepackage{color, url}
\usepackage{xspace} 
\usepackage{mathrsfs}
\usepackage{epstopdf}
\usepackage{balance}
\usepackage{bm}
\usepackage{rotating}
\usepackage{enumitem}
\usepackage{makecell}
\setlist{nolistsep}
\usepackage{algorithm}
\usepackage{algorithmic}
\usepackage{moresize}

\newcommand{\argmin}{\operatornamewithlimits{argmin}}
\newcommand{\myparatight}[1]{\smallskip\noindent{\bf {#1}:}~}

\newenvironment{packeditemize}{\begin{list}{$\bullet$}{\setlength{\itemsep}{1pt}\addtolength{\labelwidth}{0pt}\setlength{\leftmargin}{\labelwidth}\setlength{\listparindent}{\parindent}\setlength{\parsep}{0pt}\setlength{\topsep}{0pt}}}{\end{list}}

\newcommand{\ALAN}{}
\newcommand{\ALan}{}

\DeclareMathSizes{9}{8.5}{6}{5} 

\graphicspath{ {../figs/} }

\def\BibTeX{{\rm B\kern-.05em{\sc i\kern-.025em b}\kern-.08emT\kern-.1667em\lower.7ex\hbox{E}\kern-.125emX}}
    
\copyrightyear{2019} 
\acmYear{2019} 
\acmConference[CCS '19]{2019 ACM SIGSAC Conference on Computer and Communications Security}{November 11--15, 2019}{London, United Kingdom}
\acmBooktitle{2019 ACM SIGSAC Conference on Computer and Communications Security (CCS '19), November 11--15, 2019, London, United Kingdom}
\acmPrice{15.00}
\acmDOI{10.1145/3319535.3354206}
\acmISBN{978-1-4503-6747-9/19/11}
\begin{document}

\fancyhead{}

\title{Attacking Graph-based Classification via Manipulating the Graph Structure}

\author{Binghui Wang}
\affiliation{%
  \institution{ECE Department, Duke University}
}
\email{binghui.wang@duke.edu}

\author{Neil Zhenqiang Gong}
\affiliation{%
  \institution{ECE Department, Duke University}
}
\email{neil.gong@duke.edu}

\begin{abstract}
Graph-based classification methods are widely used for security analytics.
Roughly speaking, graph-based classification methods include \emph{collective classification} and \emph{graph neural network}. Attacking a graph-based classification method enables an attacker to evade detection in security analytics.
 However, existing adversarial machine learning studies mainly focused on machine learning for non-graph data. Only a few recent studies touched adversarial graph-based classification methods. However, they focused on graph neural network, leaving collective classification largely unexplored. 

We aim to bridge this gap in this work. 
We consider an attacker's goal is to evade detection via manipulating the graph structure. We formulate our attack as a graph-based optimization problem, solving which produces the edges that an attacker needs to manipulate to achieve its attack goal. However, it is computationally challenging to solve the optimization problem exactly. To address the challenge, we propose several approximation techniques to solve the optimization problem. 
We evaluate our attacks and compare them with a recent attack designed for graph neural networks using four graph datasets. 
Our results show that our attacks can effectively evade graph-based classification methods. Moreover,  
our attacks outperform the existing attack for evading collective classification methods and some graph neural network methods. 

\end{abstract}



\maketitle

\section{Introduction}
\label{intro}


Graph-based classification methods have been widely applied to various security problems such as malware detection~\cite{chau2011polonium,tamersoy2014guilt}, fraudulent user detection in social networks~\cite{Yu08,Yang12-spam,sybilrank,jia2017random,sybilbelief,wang2017sybilscar,Wang18RAID}, fake review detection~\cite{akoglu2013opinion,wang2019graph}, auction fraud detection~\cite{pandit2007netprobe},  APT infection detection~\cite{Oprea15}, and attribute inference~\cite{gong2014joint,GongAttributeInfer16,jia2017attriinfer}. 
Specifically, given 1) a graph (undirected or directed) and 2) a training dataset (some labeled \emph{positive} nodes and labeled \emph{negative} nodes), a graph-based classification method predicts the unlabeled nodes in the graph to be either positive or negative. 
The semantics of the graph, positive, and negative are different in different security 
problems. For instance, in fraudulent user detection, the graph could be a friendship graph where nodes are users and edges represent friendship between users, while positive means fraudulent user 
and negative means normal user. 

Roughly speaking, there are two types of  graph-based classification methods, i.e., \emph{collective classification}~\cite{Yu08,sybilrank,Yang12-spam,li2013finding,integro,jia2017random,pandit2007netprobe,chau2011polonium,sybilbelief,tamersoy2014guilt,akoglu2013opinion,gao2018sybilfuse,wang2017sybilscar,wang2017gang,wang2018structure,wang2019graph} and \emph{graph neural network}~\cite{scarselli2009graph,weston2012deep,kipf2017semi,battaglia2018relational}. Collective classification defines a \emph{prior reputation score} to each node based on the training dataset, assigns certain weights to edges, and propagates the prior reputation scores among the weighted graph to obtain a \emph{posterior reputation score} for each node. The posterior reputation scores are used to classify nodes. For instance, certain state-of-the-art collective classification methods assign the same weight to all edges and use \emph{Linearized Loopy Belief Propagation} (LinLBP) to propagate the reputation scores~\cite{jia2017attriinfer,wang2017gang}. Graph neural network generalizes neural networks to graph data. These methods learn feature vectors for nodes based on the graph structure and use them to classify nodes. For certain security problems, 
collective classification outperforms graph neural network~\cite{wang2019graph}. Moreover, collective classification methods were deployed in industry for malware detection~\cite{chau2011polonium,tamersoy2014guilt} and 
fraudulent user detection in social networks~\cite{sybilrank,integro}.

As graph-based classification methods are adopted to enhance security, an attacker is motivated to evade them. 
However, existing studies~\cite{nelson2010near,huang2011adversarial,biggio2013evasion,laskov2014practical,szegedy2013intriguing,goodfellow2014explaining,Papernot16,XuEvasionMalware,CarliniUsenixSecurity16,sharif2016accessorize,liu2016delving,biggio2012poisoning,xiao2015feature,poisoningattackRecSys16,YangRecSys17,jagielski2018manipulating,fang2018poisoning,fredrikson2015model,membershipInfer,membershipLocation,ganju2018property,tramer2016stealing,Ji18,wang2018stealing} on adversarial machine learning mainly focused on machine learning for non-graph data, with only a few studies~\cite{torkamani2013convex,liu2015exploiting,chen2017practical,dai2018adversarial,Sun18,dai2018adversarialnetwork,zugner2018adversarial,bojchevski2019adversarial,zugner2019adversarial,zugner2019certifiable} as exceptions. In particular, Chen et al.~\cite{chen2017practical} proposed an attack against graph-based \emph{clustering}, e.g., spectral clustering. Several recent work~\cite{dai2018adversarial,zugner2018adversarial,Sun18,dai2018adversarialnetwork,bojchevski2019adversarial,zugner2019adversarial} proposed attacks to graph neural network methods via manipulating the graph structure, i.e., inserting fake edges to the graph or deleting existing edges. However, collective classification under adversarial setting is largely unexplored. 

\noindent {\bf Our work:} 
In this work, we study attacks to collective classification via manipulating the graph structure. We generate the fake edges and the deleted edges based on the collective classification method LinLBP, because 1) LinLBP achieves state-of-the-art performance, and 2) LinLBP uses the same weight for all edges, making our formulated optimization problem easier to solve. However, we will empirically show that our attacks based on LinLBP can also transfer to other collective classification methods and graph neural network methods, as they all leverage the graph structure.

We consider an attacker has some positive nodes (e.g., malware, 
fraudulent users, fake reviews) and aims to spoof LinLBP to misclassify them as negative via manipulating the graph structure. In other words, the attacker aims to increase the \emph{False Negative Rate} (FNR)
of its positive nodes. We propose to characterize the attacker's background knowledge along three dimensions: \emph{Parameter}, \emph{Training dataset}, and \emph{Graph}, which characterize whether the attacker knows the parameters (i.e., prior reputation scores and edge weights) used by LinLBP,  the training dataset, and the complete graph, respectively. Inserting/Deleting different edges may incur different costs to the attacker. For instance, inserting an edge between the attacker's positive nodes is less expensive than inserting an edge between two negative nodes that are not under the attacker's control. 
 Therefore, we associate a (different) cost for inserting or deleting each edge. The attacker's goal is to achieve a high FNR (e.g., FNR=1) for its positive nodes with a minimal total cost of inserting/deleting edges.  

We formulate our attack as an optimization problem, where the objective function is the total cost of modifying the graph structure and the constraint is FNR=1 for the attacker's selected positive nodes. However, it is computationally challenging to solve such an optimization problem because 1) the constraint is highly nonlinear and 2) the optimization problem is a binary optimization problem (inserting or deleting an edge is a binary decision). We propose several techniques to approximately solve the optimization problem. For instance, we relax the binary variables in the optimization problem to be continuous variables and convert them to binary values after solving the optimization problem; we use posterior reputation scores as an alternative to FNR and translate the constraint FNR=1 into the objective function via Lagrangian multipliers; and we propose to alternately solve the optimization problem and compute the posterior reputation scores.

We first extensively evaluate our attacks using three real-world graphs with synthesized positive nodes and a real-world large-scale Twitter graph with real positive nodes. 
Our attacks are effective, e.g., our attacks can increase the FNR from 0 to be above 0.90 in many cases. Our attack is still effective even if the attacker does not have access  to the parameters of LinLBP (our attack can use substitute parameters), the training dataset (our attack can use a substitute training dataset), and the complete graph (a 20\% partial graph is sufficient). 
Second, our attacks can also transfer to other collective classification methods including \emph{random walk} based methods~\cite{jia2017random}, \emph{Loopy Belief Propagation} based methods~\cite{sybilbelief}, and recent \emph{Joint Weight Learning and Propagation} method~\cite{wang2019graph} that learns edge weights, as well as graph neural network methods including Graph Convolutional Network (GCN)~\cite{kipf2017semi}, LINE~\cite{tang2015line}, \ALAN{DeepWalk~\cite{perozzi2014deepwalk}, and node2vec~\cite{grover2016node2vec}}. 
Third, we compare our attack with a recent attack called Nettack~\cite{zugner2018adversarial} that generates fake edges and deleted edges based on GCN. 
We find that for GCN,  Nettack outperforms our attack; for other graph neural network methods,
our method outperforms Nettack; and for collective classification methods, our attack substantially outperforms Nettack.  

Our contributions can be summarized as follows:

\begin{itemize}
\item We perform the first systematic study on attacks to collective classification via manipulating the graph structure. 

\item We propose a threat model of attacking collective classification, formulate our attack as an optimization problem, and propose techniques to approximately solve the problem. 

\item We extensively evaluate our attacks and compare them with a recent attack using multiple datasets. 

\end{itemize}

\section{Related Work}
\label{related}

\subsection{Graph-based Classification Methods}

Roughly speaking, there are two types
of graph-based classification methods, i.e., \emph{collective classification} and \emph{graph neural network}. 

\subsubsection{Collective Classification} 
Collective classification has been applied for graph-based security 
analytics for over a decade. Specifically, given a training dataset, collective classification first defines a \emph{prior reputation score} for each node in the graph. Then, it assigns or learns weights for the edges in the graph and propagates the prior reputation scores among the graph to obtain a \emph{posterior reputation score} for each node. The posterior reputation scores are used to classify unlabeled nodes. Different collective classification methods use different ways to define the prior reputation scores,  assign/learn the edge weights, and propagate the prior reputation scores. In particular, state-of-the-art collective classification methods include 
\emph{Random Walk (RW)} based methods~\cite{zhu2003semi,gyongyi2004combating,Yu08,sybilrank,Yang12-spam,li2013finding,integro,jia2017random}, 
\emph{Loopy Belief Propagation (LBP)} based methods~\cite{pandit2007netprobe,chau2011polonium,sybilbelief,tamersoy2014guilt,akoglu2013opinion,gao2018sybilfuse}, 
\emph{Linearized Loopy Belief Propagation (LinLBP)} based methods~\cite{jia2017attriinfer,wang2017gang}, and \emph{Joint Weight Learning and Propagation (JWP)} method~\cite{wang2019graph}.


RW-based methods assign prior reputation scores 1, 0, and 0.5 to the labeled positive nodes, labeled negative nodes, and unlabeled nodes, respectively. Moreover, they assign the same weight (e.g.,~\cite{zhu2003semi}) or weights learnt using node attributes (e.g.,~\cite{integro}) to edges. Finally, they leverage random walk to propagate the reputation scores in the weighted graph. In particular, 
they iteratively distribute a node's current reputation score to its neighbors in proportion to the edge weights, and a node sums the reputation scores from its neighbors as the new reputation score.
LBP-based methods often assign the same weight for all edges. Moreover, they model the graph as a pairwise Markov Random Fields and leverage the standard LBP to perform inference, which obtains the posterior reputation scores. However, LBP has two well-known limitations: not guaranteed to converge for loopy graphs and not scalable (because it maintains messages on each edge). LinLBP linearizes LBP to address the two limitations. We will discuss details of LinLBP in Section~\ref{background}. LinLBP and LBP based methods were shown to outperform RW-based methods~\cite{jia2017attriinfer,wang2018structure}.

JWP~\cite{wang2019graph} jointly learns edge weights and propagates reputation scores. Given the current posterior reputation scores, JWP learns edge weights such that 1) the labeled positive nodes and labeled negative nodes have large and small posterior reputation scores in the next iteration, respectively, and 2) an edge has a large weight if the two corresponding nodes are predicted to have the same label using its current posterior reputation scores, otherwise the edge has a small weight. Given the learnt edge weights, JWP computes the posterior reputation scores in the next iteration, e.g., using LinLBP-based propagation. JWP outperforms LinLBP~\cite{wang2019graph}. However, attacking JWP directly is challenging because JWP learns edge weights. Specifically, when we insert fake edges to the graph or delete existing edges, the impact of these inserted/deleted edges on the classification of the attacker's positive nodes relies on the edge weights that need to be re-learnt by JWP.  Therefore, we will focus on attacking LinLBP and empirically show that our attacks also transfer to RW-based methods, LBP-based methods, and JWP.

\subsubsection{Graph Neural Network} 
These methods 
generalize neural networks to graph data. In particular, they learn feature vectors for nodes using neural networks and use them to classify nodes. 

Some graph neural network methods~\cite{scarselli2009graph,weston2012deep,kipf2017semi,battaglia2018relational}  learn the node feature vectors and classify nodes simultaneously.  
Roughly speaking, in these methods, neurons in the hidden layers of a neural network represent feature vectors of nodes, which are used to classify nodes in the last layer of the neural network. The architecture of the neural network is determined by the graph structure, e.g., a neuron in a layer that models a certain node is connected with the neurons in the previous layer that model the node's neighbors. In other words, the neural network iteratively computes a feature vector of a node via aggregating the feature vectors of the node's neighbors. The neural network parameters are learnt via minimizing a graph-based loss function on both labeled nodes in the training dataset and the remaining unlabeled nodes. In particular, the graph-based loss function is small if  nodes sharing the same label have similar feature vectors and nodes having different labels have distinct feature vectors.
For instance, Graph Convolutional Network (GCN)~\cite{kipf2017semi} 
leverages spectral graph convolutions~\cite{defferrard2016convolutional} in a neural network to learn feature vectors, and adopts a logistic regression classifier in the last layer.

Some graph neural network methods~\cite{perozzi2014deepwalk,tang2015line,cao2015grarep,grover2016node2vec} first learn feature vectors for nodes in an unsupervised learning fashion (also known as \emph{graph embedding}); then they use the feature vectors and the training dataset to learn a standard binary classifier (e.g., logistic regression); and finally they use the classifier to predict labels for the unlabeled nodes. Different methods learn the feature vectors using different  techniques. For instance, DeepWalk~\cite{perozzi2014deepwalk} learns feature vectors via generalizing the \emph{word to vector} technique developed for natural language processing to graph data. In particular, DeepWalk treats a node as a word in natural language, generates node sequences (like sentences in natural language) using truncated random walks, and leverages the skip-gram model~\cite{mikolov2013distributed} to learn a feature vector for each node. 
LINE~\cite{tang2015line} learns nodes' feature vectors by preserving both first-order and second-order proximity, where the first-order proximity captures the links in the graph (e.g., the first-order proximity between two nodes is 0 if they are not connected) and the second-order proximity between two nodes captures the similarity between their neighborhoods.

\subsection{Adversarial Graph-based Machine Learning}
Existing studies on adversarial machine learning mainly focused on machine learning for non-graph data. In particular, studies have demonstrated that machine learning is vulnerable to adversarial examples~\cite{nelson2010near,huang2011adversarial,biggio2013evasion,laskov2014practical,szegedy2013intriguing,goodfellow2014explaining,Papernot16,XuEvasionMalware,CarliniUsenixSecurity16,sharif2016accessorize,liu2016delving,jia2018attriguard}, poisoning attacks~\cite{biggio2012poisoning,xiao2015feature,poisoningattackRecSys16,YangRecSys17,jagielski2018manipulating,fang2018poisoning}, privacy attacks for users (e.g., model inversion attacks~\cite{fredrikson2014privacy,fredrikson2015model}, membership inference attacks~\cite{membershipInfer,membershipLocation}, property inference attacks~\cite{Ateniese15,ganju2018property}), as well as model parameter and hyperparameter stealing attacks~\cite{tramer2016stealing,wang2018stealing}.  



Graph-based machine learning under adversarial setting is much less explored. In particular, only a few studies
~\cite{torkamani2013convex,chen2017practical,dai2018adversarial,Sun18,dai2018adversarialnetwork,zugner2018adversarial,bojchevski2019adversarial,zugner2019adversarial,liu2015exploiting} 
focused on adversarial graph-based machine learning. 
For instance, Chen et al.~\cite{chen2017practical} proposed a practical attack against graph-based \emph{clustering}, e.g., spectral clustering. 
Torkamani and Lowd~\cite{torkamani2013convex} proposed an attack to associative Markov network-based classification methods. However, they considered that an attacker manipulates up to a fixed number of binary-valued node attributes instead of manipulating the graph structure. 
Several work~\cite{dai2018adversarial,zugner2018adversarial,Sun18,bojchevski2019adversarial,zugner2019adversarial} proposed attacks to graph neural network methods via inserting fake edges to the graph or deleting existing edges. 
 For instance, Z{\"u}gner et al.~\cite{zugner2018adversarial} proposed an attack called \emph{Nettack} against GCN~\cite{kipf2017semi}. 
 First, Nettack defines a graph structure preserving perturbation, which constrains that the node degree distributions of the graph before and after attack should be similar. 
 Then, Nettack learns a surrogate linear model of GCN.
 Finally, Nettack generates fake/deleted edges via maximizing a surrogate loss function with respect to the graph structure, where the optimization problem is subject to the node degree distribution constraint. Several recent works~\cite{zugner2019certifiable,bojchevski2019adversarial,zugner2019adversarial} further studied attacks and defenses for graph neural networks, which are concurrent to our work. 
 
 Our attack is designed for collective classification and is complementary to existing attacks that target graph neural networks. 

\section{Background and Problem Setup}


\subsection{Linearized Loopy Belief Propagation}
\label{background}

Suppose we are given an undirected graph $G=(V,E)$\footnote{Our attacks can also be generalized to directed graphs.}, where $u \in V$ is a node and $(u,v) \in E$ is an edge.
$|V|$ and $|E|$ are the number of nodes and edges, respectively. 
Moreover, we have a training dataset $L$, which consists of a set of labeled positive nodes $L_P$ and a set of labeled negative nodes $L_N$. 
LinLBP~\cite{jia2017attriinfer} assigns the prior reputation score $q_u$ for a node $u$ as follows:
{\small
\begin{align}
\label{prior}
q_u = 
\begin{cases}
\theta &\text{ if } u \in L_P \\
-\theta &\text{ if } u \in L_N \\
0 &\text{ otherwise, }
\end{cases}
\end{align}
}%
where $0<\theta \leq 1$ is a parameter of LinLBP. 
 LinLBP assigns the same weight $w$ in the interval (0, 0.5] for all edges. A larger weight means that two linked nodes are more likely to have the same label. We denote by $\mathbf{A}$ the adjacency matrix of the graph, and by $\mathbf{W}$ a $|V|\times |V|$ matrix,  
 every entry of which is the weight $w$. Then, the posterior reputation scores in LinLBP are a solution of the following system:
 {\small
\begin{align}
\label{sybilscar}
\mathbf{p}  = {\mathbf{q}} +  {\mathbf{A} \odot \mathbf{W}} {\mathbf{p}},
\end{align}
}%
where $\mathbf{q}$ and $\mathbf{p}$ are the column vector of prior reputation scores and posterior reputation scores of all nodes, respectively.
$\odot$ means element-wise product of two matrices. We note that $\mathbf{A} \odot \mathbf{W}$ essentially is the weight matrix of the graph. However, as we will see in the next section, splitting the weight matrix into the element-wise produce of $\mathbf{A}$ and $\mathbf{W}$ makes it easier to present our techniques to solve the optimization problem that models our attack. 
The posterior reputation scores are  iteratively computed as follows:
{\small
\begin{align}
\label{sybilscar-iter}
\mathbf{p}^{(t)}  = {\mathbf{q}} +  {\mathbf{A} \odot \mathbf{W}} {\mathbf{p}^{(t-1)}},
\end{align}
}%
where $\mathbf{p}^{(t)}$ is the column vector of posterior reputation scores in the $t$th iteration. When the posterior reputation scores converge, a node $u$ is predicted to be negative if 
its posterior reputation score is negative, i.e., $p_u < 0$. We note that LinLBP 
has two parameters, i.e., the prior reputation score parameter $\theta$ and the edge weight 
$w$. 

\subsection{Threat Model}
\label{threat}

\myparatight{Attacker's background knowledge} A LinLBP system is characterized by three major components: parameters of LinLBP, a training dataset $L$, and a graph $G=(V,E)$. Therefore, we characterize an attacker's background knowledge along three dimensions:
\begin{packeditemize}
\item {\bf Parameter}. This dimension characterizes whether the attacker knows the parameters (i.e., $\theta$ and $w$) of LinLBP. 

\item {\bf Training}. This dimension characterizes whether the attacker knows the training dataset. 

\item {\bf Graph}. This dimension characterizes whether the attacker knows the complete graph. 
\end{packeditemize}

For convenience, we denote an attacker's background knowledge as a triple (Parameter, Training, Graph), where Parameter can be \emph{Yes} or \emph{No}, Training can be \emph{Yes} or \emph{No}, and Graph can be \emph{Complete} or \emph{Partial}.  
For instance, a triple (Yes, Yes, Complete) means that the attacker (e.g., an insider) knows the parameters, the training dataset, and the complete graph. An attacker knows its positive nodes and edges between them. An attacker could also know a subgraph of the negative nodes. 
For instance, in social networks, an attacker can develop a web crawler to collect a partial social graph of the negative nodes (i.e., normal users). We will empirically show that our attack is still effective even if the attacker uses substitute parameters, a substitute training dataset, and a partial graph.  

\myparatight{Attacker's capability} 
We consider an attacker can change the graph structure, i.e., the attacker can insert fake edges to the graph or delete existing edges.  
For instance, in social networks, an attacker can create new edges among its created 
fraudulent users
(positive nodes) or remove existing edges between its created 
fraudulent users.
In addition, an attacker may also compromise normal users (negative nodes) and create edges between the compromised users and its 
fraudulent users.
However, adding or removing different edges may incur different costs for the attacker. 
For instance, 
in the 
fraudulent user
detection problem, adding an edge between two 
fraudulent users
requires a smaller cost than adding an edge between a 
fraudulent user
and a compromised normal user, as an attacker usually needs an extra effort to compromise a normal user. 
Thus, we associate a cost with inserting/deleting each edge, where the costs may depend on different security 
applications. Moreover, in certain applications, the maximum number of edges that can be inserted/deleted for a node is bounded. For instance, in online social networks, the number of friends (i.e., edges) a user can have is often bounded. Thus, we introduce a parameter $K$, which is the maximum number of edges that can be inserted/deleted for a node.

\myparatight{Attacker's goal} 
Suppose an attacker has some selected positive nodes (e.g., malware, 
fraudulent user,
fake review), which we call \emph{target nodes}. We consider the attacker's goal is to make the target nodes \emph{evade} the detection of LinLBP.  Specifically, the attacker aims to manipulate the graph structure with a minimal total cost such that LinLBP misclassifies the target nodes to be negative. In other words, the attacker's goal is to achieve a high \emph{False Negative Rate} (FNR) for its target nodes via manipulating the graph structure with a minimal total cost. We note that the target nodes may be a subset of the positive nodes under the attacker's control.

\subsection{Problem Definition}
Given 
our threat model, we 
formally define our problem as follows:
\vspace{-3mm} 
\begin{definition}[Attack to LinLBP]
Given some target nodes, attacker's background knowledge, cost of inserting/deleting each edge, and the maximum number of inserted/deleted edges for each node,  an attack to LinLBP is to manipulate the graph structure with a minimal total cost such that LinLBP achieves a as high FNR for the target nodes as possible. 
\end{definition}

\section{Our Attacks}
We first discuss our attacks under the threat model where the attacker has full knowledge, i.e., the attacker knows the parameters of LinLBP, the training dataset, and the complete graph. Then, we adjust our attacks to the scenarios where the attacker 
only has partial knowledge 
about the parameters, the training dataset, and/or the graph.

\ALan{\subsection{Adversary with Full Knowledge}}

\myparatight{Overview} We first formulate our attack as an optimization problem. Specifically, we associate a binary variable with each pair of nodes, where a binary variable has a value of 1 if and only if our attack changes the connection state between the corresponding two nodes. The objective function of the optimization problem is the total cost of manipulating the graph structure and the constraints are 1) FNR=1 for the attacker's target nodes and 2) the maximum number of inserted/deleted edges per node is bounded by $K$. However, it is computationally challenging to solve the optimization problem 
because 
it is a binary optimization problem that has $O(|V|^2)$ binary variables and the constraint FNR=1 is highly nonlinear.

To address the challenges, we propose techniques to approximately solve the optimization problem. Specifically, we relax the binary variables to continuous variables whose values are in the interval [0, 1] and convert them to binary values after solving the optimization problem; and we reduce the optimiztion to the binary variables that are related to the attacker's target nodes. Moreover, we translate the constraint FNR=1 to be a constraint on the posterior reputation scores of the attacker's target nodes and add the constraint to the objective function using Lagrangian multipliers. However, the converted optimization problem still faces a computational challenge because the posterior reputation scores depend on the graph structure (i.e., the continuous variables in the optimization problem) in a complex way, i.e., each variable could influence the posterior reputation score of each node. To address the challenge, we propose to alternately optimize the continuous variables in the optimization problem and compute the posterior reputation scores. Our key idea is that posterior reputation scores are computed iteratively. Instead of using the final posterior reputation scores, we use the intermediate posterior reputation scores to optimize the continuous variables in the optimization problem.


\noindent {\bf Formulating our attack as an optimization problem:} We associate a binary variable $B_{uv}$ with each pair of nodes $u$ and $v$. $B_{uv}=1$ if our attack changes the connection state between $u$ and $v$, otherwise $B_{uv}=0$. Specifically, if $u$ and $v$ are already connected in the original graph, then $B_{uv}=1$ if our attack deletes the edge between $u$ and $v$; if $u$ and $v$ are not connected in the original graph, then $B_{uv}=1$ if our attack inserts an edge between $u$ and $v$. As we discussed in our threat model in Section~\ref{threat}, inserting/deleting different edges may incur different costs for the attacker. Therefore, we associate a variable $C_{uv}$ with each pair of nodes $u$ and $v$ to represent the cost of modifying the connection state between $u$ and $v$. In particular, $B_{uv}=1$ incurs a cost $C_{uv}$ for the attacker. Therefore, the total cost of an attack is 
$\sum_{u,v\in V, u<v}B_{uv}C_{uv}$, 
where $V$ is the set of nodes in the graph and the constraint $u<v$ essentially means that we count once for each pair of nodes. 
We note that, for the same node pair, we could also associate 
different costs for inserting an edge and deleting the existing edge between them.
Our method is also applicable for such fine-grained cost. However, for simplicity, we use
the same cost for inserting/deleting an edge.

The attacker has a set of selected target nodes $\mathcal{T}$ and aims to achieve a high FNR (i.e., FNR=1 in our formulation) for the target nodes with a minimal total cost. 
Formally, our attack aims to find the variables $B_{uv}$ via solving the following optimization problem:
{\footnotesize
\begin{align}
& \min_{\mathbf{B}} \sum_{u,v \in V, u<v} B_{uv} C_{uv}, \label{rawobj} \\
& \textrm{s.t.} \quad FNR = 1, \label{const1}\\ 
& \quad \quad B_{uv} \in \{0, 1\}, \text{ for } u,v\in V, \label{const2} \\ 
& \quad \quad \sum_{v} B_{uv} \leq K \label{const3}, \text{ for } u \in V,  
\end{align}
}%
where the symmetric matrix $\mathbf{B}$ includes all the binary variables for each pair of nodes, the objective function is the total cost, the first constraint means all target nodes are misclassified as negative, the second constraint means the variables in the optimization problem are binary, and the third constraint means the maximum number of inserted/deleted edges for each node is bounded by $K$. For convenience, we call the matrix  $\mathbf{B}$ \emph{adversarial matrix}.

\myparatight{Challenges for solving the optimization problem} Solving the optimization problem in Equation~\ref{rawobj} exactly faces several challenges. First, the variables $B_{uv}$ are binary. Second, the optimization problem has $|V|(|V|-1)/2$ binary variables. Third, the constraint FNR=1 is highly nonlinear.

\myparatight{Our techniques for approximately solving the optimization problem} We propose several approximation techniques to address the challenges. To address the first challenge, we relax $B_{uv}$ to a continuous variable whose value is in the interval [0, 1] and convert it to a binary value after solving the optimization problem. 

In practice, it is often expensive to modify the connection states between negative nodes 
not under the attacker's control. Thus, to address the second challenge, we reduce the optimization space to the edges between the attacker's target nodes and the remaining nodes 
and the edges between the target nodes. Specifically, we solve the optimization problem over the 
variables related to the attacker's target nodes, i.e., the variables $B_{uv}$ where ${\small u\in \mathcal{T}, v\in V-\mathcal{T}}$ and 
$B_{uv}$ where ${\small u,v\in \mathcal{T}, u<v}$, where the first set of variables characterize the connection states between the target nodes and the remaining nodes while the second set of variables characterize the connection states between the target nodes. 

Recall that a node is predicted to be negative if its posterior reputation score is negative. Therefore, to address the third challenge, we replace the constraint FNR=1 as the constraint $p_u < 0$ for each $u \in \mathcal{T}$. Moreover, we convert the constraint on posterior reputation scores to the objective function via Lagrangian multipliers. Summarizing our approximation techniques, we aim to solve the following optimization problem: 
{\footnotesize
\begin{align}
& \min_{\mathbf{B}} \mathcal{F}(\mathbf{B}) = \sum_{u \in \mathcal{T}, v \in V-\mathcal{T}} B_{uv} C_{uv} +  \sum_{u, v \in \mathcal{T}, u < v} B_{uv} C_{uv} + \lambda \sum_{u \in \mathcal{T}} p_u \label{aa} \\
&  \textrm{s.t.} \quad B_{uv} \in [0, 1], \text{ for } u\in  \mathcal{T}, v\in V  \\
& \quad \quad \sum_{v} \bar{B}_{uv} \leq K, \text{ for } u\in  V \label{const_final}
\end{align}
}%
where $\lambda > 0$ is a Lagrangian multiplier, $\bar{B}_{uv}$ is the binarized value of the continuous variable $B_{uv}$, and the posterior reputation scores are a solution of the following system: 
{\small
\begin{align}
\label{sybilscar-new}
\mathbf{p}  = {\mathbf{q}} +  {|\mathbf{A} - \bar{\mathbf{B}}| \odot \mathbf{W}} {\mathbf{p}},
\end{align}
}%
where $\mathbf{A}$ is the adjacency matrix of the original graph, $\bar{\mathbf{B}}$ is the binarized adversarial matrix, and $\mathbf{W}$ is a matrix with every entry as the edge weight $w$.  ${\scriptsize |\mathbf{A} - \bar{\mathbf{B}}|}$ essentially is the adjacency matrix of the graph after our attack. 

A popular method to solve the above optimization problem is to use gradient descent. However, it is still computationally challenging because the posterior reputation scores depend on the variables $\bar{\mathbf{B}}$ in a complex way, i.e., every variable $B_{uv}$ influences each posterior reputation score according to the system in Equation~\ref{sybilscar-new}. To address the challenge, we propose to alternately optimize the variables in the optimization problem and compute the posterior reputation scores. Our key insight is that posterior reputation scores are iteratively computed. Instead of using the final posterior reputation scores, we use the intermediate posterior reputation scores to solve the optimization problem. Then, given the intermediate variables $B_{uv}$, we update the posterior reputation scores. Specifically, we repeat the following two steps.

{\bf Step I: updating posterior reputation scores}. We update the posterior reputation scores in the $t$th iteration using the adversarial matrix ${\scriptsize \bar{\mathbf{B}}^{(t-1)}}$ in the $(t-1)$th iteration as follows:
{\small
\begin{align}
\mathbf{p}^{(t)}  = {\mathbf{q}} +  {|\mathbf{A} - \bar{\mathbf{B}}^{(t-1)}| \odot \mathbf{W}} {\mathbf{p}^{(t-1)}}. 
\end{align}
}%

{\bf Step II: updating adversarial matrix}. In this step, we update the adversarial matrix in the $t$th iteration while fixing the posterior reputation scores in the $t$th iteration. For convenience, we transform the optimization problem in Equation~\ref{aa} to the following optimization problem:
{\footnotesize
\begin{align}
\label{aa1}
& \min_{\mathbf{B}^{(t)}} \mathcal{F}(\mathbf{B}^{(t)}) = \sum_{u \in \mathcal{T}} \mathbf{b}_{u}^{(t)} \mathbf{c}_{u}^T + \lambda \sum_{u \in \mathcal{T}} p_u^ {(t+1)} \\
\label{aa2}
&  \textrm{s.t.} \quad \mathbf{b}_{u}^{(t)} \in [0, 1], \text{ for } u\in  \mathcal{T}  \\
& \quad \quad \sum_{v} \bar{B}_{uv}^{(t)} \leq K, \text{ for } u\in   V \label{aa3}
\end{align}
}%
where ${\scriptsize \mathbf{b}_{u}^{(t)}}$ is the $u$th row of the matrix ${\small \mathbf{B}^{(t)}}$ (i.e., the modified connection states between a node $u$ and the remaining nodes), $\mathbf{c}_{u}$ is the adjusted $u$th row of the cost matrix $\mathbf{C}$, $T$ is transpose of a vector, and ${\scriptsize \mathbf{p}^{(t+1)} =  {\mathbf{q}} +  {|\mathbf{A} - \bar{\mathbf{B}}^{(t)}| \odot \mathbf{W}} {\mathbf{p}^{(t)}}}$. Specifically, ${c}_{uv}=C_{uv}$ if $u\in  \mathcal{T}$ and $v \notin  \mathcal{T}$, while ${\small {c}_{uv}=C_{uv}/2}$ if $u, v\in  \mathcal{T}$ (because each pair of nodes between the target nodes in $\mathcal{T}$ is counted twice in the objective function in Equation~\ref{aa1}).

We use a \emph{projected gradient descent} to solve the optimization problem in Equation~\ref{aa1}. Specifically, we iteratively apply the following steps for each target node $u$:
{\footnotesize
\begin{align}
& {\mathbf{s}}_{u}^{(i+1)} = \tilde{\mathbf{b}}_{u}^{(i)} - \eta \frac{\partial \mathcal{F}(\tilde{\mathbf{B}}^{(i)})}{\partial \tilde{\mathbf{b}}_{u}^{(i)}} \label{prox_grad_des_vec}, \\
& \tilde{\mathbf{b}}_{u}^{(i+1)}  = \textrm{proj}\big( {\mathbf{s}}_{u}^{(i+1)} \big) \label{prox},
\end{align}
}%
where the first equation means we update the variables $\tilde{\mathbf{b}}_{u}$ using gradient descent and the second equation means that we project the variables to satisfy the two constraints in Equation~\ref{aa2} and Equation~\ref{aa3}.  Note that the variables $\tilde{\mathbf{b}}_{u}$ are initialized using the adversarial matrix in the $(t-1)$th iteration, i.e., ${\scriptsize \tilde{\mathbf{b}}_{u}^{(0)}={\mathbf{\bar{b}}}_{u}^{(t-1)}}$. Specifically, the gradient can be computed as follows:
{\footnotesize
\begin{align}
& \frac{\partial \mathcal{F}(\mathbf{\tilde{B}}^{(i)})}{\partial \mathbf{\tilde{b}}_{u}^{(i)}} = \mathbf{c}_{u} + \lambda \frac{\partial p_u^{(t+1)}}{\partial \mathbf{\tilde{b}}_{u}^{(i)}} =  \mathbf{c}_{u} + \lambda  \textrm{sign}(\mathbf{\tilde{b}}_{u}^{(i)} - \mathbf{a}_{u}) \odot \mathbf{w}_{u} \mathbf{P}^{(t)},
\end{align}
}%
where $\mathbf{a}_{u}$ is the $u$th row of the adjacency matrix, the $\textrm{sign}$ operator applies to every entry of the vector ${\scriptsize \mathbf{\tilde{b}}_{u}^{(i)} - \mathbf{a}_{u}}$, and ${\scriptsize \mathbf{P}^{(t)}}$ is a diagonal matrix with the diagonal elements ${\scriptsize P_{u,u}^{(t)} = p_u^{(t)}}$. Note that when computing the gradient, we approximate ${\scriptsize \bar{\mathbf{B}}}$ as ${\scriptsize \tilde{\mathbf{B}}}$ in the computation of  ${\scriptsize \mathbf{p}^{(t+1)}}$. Moreover, the operator \emph{proj} is defined as follows:
{\footnotesize
\begin{align}
\label{prox_oper}
\textrm{proj}(\mathbf{s}_u^{(i+1)}) = \argmin_{0 \leq \mathbf{\tilde{b}}_{u}^{(i+1)} \leq 1,\  \mathbf{\tilde{b}}_{u}^{(i+1)} \mathbf{1}^T \leq K} \| \mathbf{\tilde{b}}_{u}^{(i+1)} - \mathbf{s}_u^{(i+1)} \|_2^2,
\end{align}
}%
which means that the \emph{proj} operator aims to find a vector ${\scriptsize \mathbf{\tilde{b}}_{u}^{(i+1)}}$ that is the closest to the vector  ${\scriptsize \mathbf{s}_u^{(i+1)}}$ in terms of Euclidean distance and that satisfies the constraint ${\scriptsize 0 \leq \mathbf{\tilde{b}}_{u}^{(i+1)} \leq 1}$ and a relaxed constraint ${\scriptsize \mathbf{\tilde{b}}_{u}^{(i+1)} \mathbf{1}^T \leq K}$ on the maximum number of inserted/deleted edges for node $u$. The optimization problem in Equation~\ref{prox_oper} can be solved exactly by the \emph{break point search method}~\cite{yuan2017exact}. 
We repeat the Equation~\ref{prox_grad_des_vec} and~\ref{prox} multiple iterations to solve  $\mathbf{\tilde{b}}_{u}$. However, more iterations make our method less efficient. In our experiments, we repeat 4 iterations to solve $\mathbf{\tilde{b}}_{u}$ as we find that 4 iterations achieve a good trade-off between accuracy and efficiency. 

After solving $\mathbf{\tilde{b}}_{u}$ with 4 iterations, we convert them to be binary values. Specifically, we first convert each ${\tilde{b}}_{uv}$ to 1 if ${\tilde{b}}_{uv} > 0.5$ and 0 otherwise. Then, 
we select the largest $K$ entries in $\mathbf{\tilde{b}}_{u}$ and convert them to be 1, while we convert the remaining entries in the vector to be 0. Finally, we assign the converted binary vector as the corresponding row $\mathbf{\bar{b}}_{u}^{(t)}$ in the adversarial matrix  in the $t$th iteration. Note that such converted adversarial matrix may not be symmetric for the targeted nodes. Specifically, we may have $\bar{B}_{uv}\neq \bar{B}_{vu}$ for a pair of target nodes $u$ and $v$. Therefore, we perform another postprocessing for the edges between target nodes. Specifically, we set $\bar{B}_{uv}=\bar{B}_{vu} \leftarrow \bar{B}_{uv}\cdot \bar{B}_{vu}$, i.e., we modify the connection state between $u$ and $v$ only if both $\bar{B}_{uv}=1$ and  $\bar{B}_{vu}=1$.

\myparatight{Computational complexity}
We first analyze the time complexity in one iteration. 
In \textbf{Step I}, updating posterior reputation scores traverses all edges in the graph and the time complexity is $O(|E|)$.
In \textbf{Step II}, for each target node $u \in \mathcal{T}$, updating $\mathbf{s}_u^{(i+1)}$ in Equation~\ref{prox_grad_des_vec} traverses all nodes and requires a time complexity $O(|V|)$; and the proj operator requires a time complexity $O(|V| \log |V|)$~\cite{yuan2017exact}. Therefore, computing the adversarial matrix $\mathbf{\bar{B}}^{(t)}$ in the $t$th iteration requires a time complexity of 
$O(m |\mathcal{T}| |V| \log|V|)$, where $m$ is the number of iterations used to compute $\mathbf{\tilde{b}}_{u}$. Suppose we alternate between \textbf{Step I} and \textbf{Step II} for $n$ iterations, then the total time complexity is $O(n (|E| + m |\mathcal{T}| |V| \log|V| ))$. 

\ALan{\subsection{Adversary with Partial Knowledge}}

{\myparatight{Parameter=No, Training=Yes, Graph=Complete}}
When the attacker does not know the parameters of LinLBP, i.e., the prior reputation score parameter $\theta$ and/or the edge weight  $w$, the attacker can randomly select parameters from their corresponding domains ($0<\theta \leq 1$ and $0<w\leq 0.5$) as the substitute parameters. 
Then, the attacker uses our attack with the substitute parameters to generate the inserted fake edges and deleted existing edges.  

{\myparatight{Parameter=Yes, Training=No, Graph=Complete}}
The attacker can sample a substitute training dataset from the original graph when the attacker does not have the training dataset used by the LinLBP. Specifically, the attacker knows its positive nodes. Therefore, the attacker can sample some nodes from its positive nodes as labeled positive nodes in the substitute training dataset. Moreover, the attacker can sample some nodes from the nodes that are not its positive nodes as labeled negative nodes. Then, the attacker applies our attack using the substitute training dataset to generate inserted fake edges and deleted existing edges. 


{\myparatight{Parameter=Yes, Training=Yes, Graph=Partial}}
The attacker at least knows its positive nodes and edges between them. Suppose the attacker also knows a subgraph of the negative nodes. For instance, in online social networks, the attacker can develop a web crawler to collect at least a partial social graph of the negative nodes (i.e., normal users). 
Then, the attacker applies our attack on the partial graph, which includes the subgraph of the positive nodes and the subgraph of the negative nodes, to generate inserted fake edges and deleted existing edges.   

{\myparatight{Parameter=No, Training=No, Graph=Partial}}
In this scenario, the attacker has the least knowledge of a LinLBP system. The attacker uses the substitute parameters, samples a substitute training dataset, and leverages the partial graph to perform our attack.

\vspace{+2mm}
\section{Evaluation}
\label{eval_syn}




\begin{table}[!t]
\centering
\small
\caption{Dataset statistics.}
\label{dataset_stat}
\begin{tabular}{|c|c|c|c|} 
 \hline
 {\small \textbf{Dataset}} & {\small \#Nodes} & {\small \#Edges}  &  {\small Ave. degree} \\ \hline
 {\small \textbf{Facebook}} & {\small 4,039} & {\small 88,234} & {\small 44} \\ \hline
 {\small \textbf{Enron}}& {\small 33,696} & {\small 180,811} & {\small 11} \\ \hline
 {\small \textbf{Epinions}}& {\small 75,877} & {\small 811,478} & {\small 21} \\ \hline
 {\small \textbf{Twitter}}& {\small 21,297,772} & {\small 265,025,545} & {\small 25} \\ \hline
\end{tabular}
\end{table}

\subsection{Experimental Setup}
\label{setup}

\myparatight{Dataset description} 
\ALan{
We use three real-world graphs with synthesized positive nodes and a large-scale Twitter graph with real positive nodes to evaluate our attacks (See Table~\ref{dataset_stat}). We adopt three graphs with synthetic positive nodes in order to study our attacks for graphs with different properties, e.g., size, average degree, etc..
}


{\bf \emph{1) Graphs with synthesized positive nodes.}} We use three real-world graphs that represent different security applications. 
We obtained the largest connected component of each graph from SNAP (http://snap.stanford.edu/ data/index.html).
The three graphs are Facebook (4,039 nodes and 88,234 edges), Enron (33,696 nodes and 180,811 edges), and Epinions (75,877 nodes and 811,478 edges), respectively. 
In the Facebook graph, a node represents a user; two users are connected if they are friends to each other; a node is negative if it is a normal user; and a node is positive if it is a malicious user.  In the Enron graph, a node  represents an email address; an edge between two nodes indicates that at least one email was exchanged between the two corresponding email addresses; a node is negative if it is a normal email address; and a node is positive if it is a spamming email address. Epinions was a general review site, which enables users to specify which other users they trust. In the Epinions graph, a node represents a reviewer; an edge between two nodes indicates that they trust each other; negative means a normal reviewer; and positive means a fake reviewer. 

The nodes in a graph are negative nodes and thus we synthesize positive nodes. We follow previous studies (e.g., \cite{alvisiSybil13,sybilbelief,wang2017sybilscar}) on graph-based security analytics to synthesize positive nodes. Specifically, to avoid the impact of structural differences between negative nodes and positive nodes, we replicate the negative nodes and their edges as positive nodes and edges in each of the three graphs. Moreover, we assume an attacker has already inserted some edges between positive nodes and negative nodes, \ALan{which we call \emph{attack edges},} to make positive nodes connected with negative nodes. 


{\bf \emph{2) Twitter graph with real positive nodes.}}  We obtained an undirected Twitter dataset with real postive nodes (fraudulent users) from Wang et al.~\cite{wang2017sybilscar}. Specifically, the Twitter network has 21M users, 265M edges, where 18M edges are attack edges, and an average degree of 25. An undirected edge $(u,v)$ means that user $u$ and user $v$ follow each other.
A user suspended by Twitter is treated as a 
fraudulent user
(positive node), while an active user is treated as benign (negative node). 
In total, 145,183 nodes are 
fraudulent users, 
2,566,944 nodes are benign, and the remaining nodes are unlabeled.

\ALan{\noindent {\bf Training dataset and testing dataset:}
For each graph with synthesized positive nodes, we randomly select 100  positive nodes and 100 negative nodes to form a training dataset. 
For the Twitter graph, we sample 3,000 positive nodes and 3,000 negative nodes uniformly at random as the training dataset, due to its large size. 
The remaining labeled nodes form a testing dataset. We use a small training dataset because graph-based classification assumes so. 
}

\noindent {\bf Attacker's target nodes:}
An attacker aims to maintain some positive nodes (called \emph{target nodes}) that evade detection. The target nodes could be a subset of the positive nodes, i.e., the attacker uses some other positive nodes to support the target nodes. The attacker could use different ways to select the target nodes. We consider the following three ways:  

{\bf \emph{1) RAND:}} In this method, the attacker samples some positive nodes uniformly at random as the target nodes. 

{\bf \emph{2) CC:}}  The attacker selects a \emph{connected component} of positive nodes as the target nodes. Specifically, the attacker first randomly picks a positive node as target node. Then, the attacker uses breadth first search to find other target nodes. 

{\bf \emph{3) CLOSE:}} The attacker samples some positive nodes that are close to negative nodes. In particular, we adopt a variant of \emph{closeness centrality} in network science to measure the closeness between a positive node and the negative nodes. Specifically, for a positive node, we compute the inverse of the sum of the distances between the positive node and all negative nodes as the closeness  centrality of the positive node. Then, the attacker selects the positive nodes that have large closeness  centrality as the target nodes. 

The target nodes selected by CC are more structurally similar. 
If one of them evades detection, then others are also likely to evade. 
\ALAN{The target nodes selected by CLOSE could also evade detection, as they are close to negative nodes.}
Thus, we expect that our attacks are more effective when the target nodes are selected by CC \ALAN{and CLOSE than by RAND}.

\begin{table*}[t]
\centering
\small
\caption{Results of our attacks and the baseline attacks in different scenarios. The column ``\#Edges'' shows the number of edges modified by our attack; the column ``\#Add/\#Del'' shows the respective number of inserted edges and deleted edges; and the column ``Cost'' shows the total cost of our attack. 
FNR is the fraction of target (positive) nodes misclassified as negative nodes. 
Our attacks can substantially increase the FNRs and  significantly outperform the baseline attacks.}
\addtolength{\tabcolsep}{-3.5pt}
\begin{tabular}{|c|c|c|c|c|c|c|c|c|c|c|c|c|c|c|c|}
\hline 
\multicolumn{16}{|c|}{\textbf{RAND}}  \\ \hline
\multirow{2}{*}{{\bf Dataset}} & {\textbf{No attack}} & {\textbf{\makecell{Random attack}}} & {\textbf{\makecell{Del-Add  attack}}} & \multicolumn{4}{c|}{\textbf{Equal}} & \multicolumn{4}{c|}{\textbf{Uniform}} & \multicolumn{4}{c|}{\textbf{Categorical}} \\ \cline{2-16} 
   &  FNR &   FNR  &  FNR &   FNR &  \#Edges  &  \#Add/\#Del   &  Cost     &    FNR  &   \#Edges  &  \#Add/\#Del &   Cost     &  FNR    & \#Edges  & \#Add/\#Del &  Cost  \\ \hline
  {\bf Facebook} &  0 & 0.04 & 0.39 &   0.78  &   1960    & 1855/105 &   1960    &     0.66 &   1988  & 1853/135 &  2082    &   0.60   & 1891 & 1884/7    &  19342  \\ \hline
   {\bf Enron} & 0 & 0.06 & 0.67 &   1.00  &   2000  & 2000/0  &  2000     &   0.96   &   1998  & 1993/5 &    2193   &   0.89 &    1884  & 1879/5 &  19257    \\ \hline
    {\bf Epinions} & 0 & 0.03 & 0.73 &   0.98  &   2000   & 1931/69 &  2000     &    0.94  &   2000   & 1995/5 &   2078    &    0.92  &   1971 & 1971/0 &   30010  \\ \hline
 {\bf Twitter} & 0 & 0.02 &  0.40 &   0.85   &   1966   & 1566/400 &  1966     &    0.85  &   1942   & 1530/412 &   2326    &    0.82  &   1855 & 1666/189 &   26550  \\ \hline \hline
 
 \multicolumn{16}{|c|}{\textbf{CC}}  \\ \hline
\multirow{2}{*}{{\bf Dataset}} & {\textbf{No attack}} & {\textbf{\makecell{Random attack}}} & {\textbf{\makecell{Del-Add attack}}} & \multicolumn{4}{c|}{\textbf{Equal}} & \multicolumn{4}{c|}{\textbf{Uniform}} & \multicolumn{4}{c|}{\textbf{Categorical}} \\ \cline{2-16} 
  &  FNR &   FNR  & FNR &  FNR    &  \#Edges  &  \#Add/\#Del   &   Cost       &   FNR    &  \#Edges  &  \#Add/\#Del   &   Cost       &   FNR    &  \#Edges  & \#Add/\#Del   &   Cost \\ \hline
  {\bf Facebook} &   0 & 0.02 & 0.43 &   0.94   &  1980  &   1747/233 &  1980    &  0.94     &    1996   &  1858/138 &  2184   & 0.68	 &  1934  & 1540/394 & 	25349 \\ \hline
   {\bf Enron} &   0 & 0.03 & 0.76 &   1.00   &   2000  & 2000/0 &  2000     & 0.99     &   2000   & 2000/0 &    2820   &   0.92  &  1764 & 1497/267 & 18237 \\ \hline
    {\bf Epinions} & 0 & 0.02 & 0.63 &    0.99 &  2000  &  1886/114 &  2000     &   0.94  &    2000  & 1993/7 &   2128    &  0.88 &  2000 & 1749/251 &  17741 \\ \hline
 {\bf Twitter} & 0 & 0.02  & 0.43 &   0.88   &   1997   & 1976/21 &  1997     &    0.86  &   1990   & 1906/84 &   3102    &    0.85  &   1969 & 1858/111 &   29800  \\ \hline \hline
 
    \multicolumn{16}{|c|}{\textbf{CLOSE}}  \\ \hline
\multirow{2}{*}{{\bf Dataset}} & {\textbf{No attack}} & {\textbf{\makecell{Random attack}}} & {\textbf{\makecell{Del-Add attack}}} & \multicolumn{4}{c|}{\textbf{Equal}} & \multicolumn{4}{c|}{\textbf{Uniform}} & \multicolumn{4}{c|}{\textbf{Categorical}} \\ \cline{2-16} 
  &  FNR &   FNR  &  FNR &   FNR  &  \#Edges  &  \#Add/\#Del   &   Cost       &   FNR    &  \#Edges  &  \#Add/\#Del   &   Cost       &   FNR    &  \#Edges  & \#Add/\#Del   &   Cost \\ \hline
  {\bf Facebook} &   0 & 0.02 & 0.31 &   0.96 &  2000  &   1455/545 &  2000    &  0.93   &    1980   &  1385/595 &  2076   & 0.69	&  1983  & 1759/224 & 	27124 \\ \hline
   {\bf Enron} &   0 & 0.02 & 0.47 &  1.00 &   2000  & 1828/172 &  2000     & 1.00 &   2000   & 1942/58 &    3781   &   0.90 & 1705 & 1705/0 & 21705 \\ \hline
    {\bf Epinions} & 0 & 0.02 & 0.56 &    1.00 &  2000  &  1960/40 &  2000     &   0.96 &    2000  & 1992/8 &   3854    &  0.86 & 1997 & 1651/346 &  21252 \\ \hline
{\bf Twitter} & 0 & 0.01 & 0.52 &  0.87  &   1956   & 1774/182 &  1956     &    0.85  &   1942   & 1748/194 &   4204    &    0.83  &   1872 & 1728/144 &   27440  \\ \hline
\end{tabular}
\label{overall}
\end{table*}

\noindent {\bf Simulating costs:}
We associate a cost of modifying the connection state for each pair of nodes. While the costs are application-dependent, we simulate multiple scenarios in our experiments. 


{\bf \emph{1) Equal cost:}} In this scenario, modifying the connection state for any pair of nodes has the same cost for the attacker. In particular, we assume the cost to be 1. We note that a recent attack called Nettack~\cite{zugner2018adversarial} for GCN essentially assumes the equal cost.


{\bf \emph{2) Uniform cost:}} We assume the costs for different node pairs are uniformly distributed among a certain interval. In particular, for each node pair, we sample a number from the interval as the cost of modifying the connection state between the node pair. In our experiments, we consider [1, 10] as the interval. 



{\bf \emph{3) Categorical cost:}} In this scenario, we categorize node pairs into different groups and assign the same cost for each group. Specifically, it is easy for an attacker to insert edges between the positive nodes or remove existing edges between them. Therefore, we associate a small cost (i.e., 1 in our experiments) with modifying the connection state between two positive nodes. In practice, the attacker could have compromised some negative nodes, e.g., the attacker compromised some normal users in social networks via stealing their credentials through social engineering attacks. It is relatively easy for the attacker to modify the connection states between the compromised negative nodes and the positive nodes. However, it incurs costs for the attacker to compromise the negative nodes. Therefore, we associate a modest cost (i.e., 10) with modifying the connection state between a compromised negative node and a positive node. Finally, it is relatively hard for the attacker to modify the connection state between positive nodes and the remaining negative nodes 
 that are not compromised, because the attacker has to bribe or compromise them to establish mutual connections between them and the positive nodes. Therefore, we associate a large cost (i.e., 100) with modifying the connection state between a positive node and an uncompromised negative node. 
Note that our attack does not change the connection states between negative nodes, so the costs for them do not affect our attack. We randomly sample 100 negative nodes as the compromised ones. 

\ALan{{We stress that the specific cost values (e.g., 1, 10, 100) may not have semantic meanings. We just use them to simulate different scenarios and indicate different costs of manipulating different types of edges.}}




\ALan{
\noindent {\bf Baseline attacks:}
We compare with two baseline attacks: Random attack and Del-Add attack. We will also compare with a recent attack~\cite{zugner2018adversarial} designed for GCN in Section~\ref{compare_attack}.

{\bf \emph{Random attack:}} 
We  randomly modify the connection states between the target nodes and other nodes.
Specifically, for each target node $u$, we randomly select $K$ nodes and we modify the connection state between $u$ and each selected node. If $u$ and a selected node are connected in the original graph, then we delete the edge between them, otherwise we insert an edge between them.  

{\bf \emph{Del-Add attack:}}
Suppose a target node is connected with $d$ positive nodes. If $ d > K$, then we randomly delete $K$ edges between the target node and its connected positive nodes, otherwise we first delete $d$ edges between the target node and its connected positive nodes and then we add edges between the target node and $(K-d)$ randomly selected negative nodes. The intuition of the Del-Add attack is that target nodes, which are sparsely connected with other positive nodes and densely connected with negative nodes,  are likely to be misclassified as negative nodes. 
}

\noindent {\bf Parameter setting:}
We set the Lagrangian multiplier $\lambda$ as $\lambda$=10,000 for Categorical cost and $\lambda$=1,000 for the other two types of costs, considering their different magnitude of cost values. We set the learning rate $\eta=0.1$ in our attack. We also explore the impact of $\lambda$ and $\eta$ and show the results in Figure~\ref{impact-lambda-eta-Equal}. Without otherwise mentioned, \ALan{we add 10K attack edges (AE) between negative nodes and positive nodes uniformly at random in each graph with synthesized positive nodes.}
Moreover, we assume CC as the method of selecting target nodes, 100 target nodes, $K$=20 (i.e., the number of modified edges is bounded by 20 per node), and Equal cost (as previous attacks~\cite{dai2018adversarial,zugner2018adversarial,Sun18} used this cost). For LinLBP, we set $\theta=0.5$ and $w=0.01$. 
We implement our attack against LinLBP~\cite{wang2017sybilscar} in C++.
We obtain the publicly available C++ source code for LinLBP from the authors~\cite{wang2017sybilscar}. 
We perform all our experiments on a Linux machine with 512GB memory and 32 cores.

\ALAN{Our attack alternately performs \textbf{Step I} and \textbf{Step II}. 
In our experiments, we check the FNR of the target nodes in each iteration and stop the algorithm when it does not change in two consecutive iterations or we reached the predefined maximum number of iterations, e.g., 10 in our experiments. We found the FNR converges within 10 iterations in our experiments. 
}

\subsection{Evaluation with Full Knowledge}
We first evaluate our attacks in the threat model where the attacker knows the parameters of LinLBP, the training dataset, and the complete graph. 
There are seven parameters that could affect our attack performance. The seven parameters are: 
cost type (Equal cost, Uniform cost, and Categorical cost), the method used to select target nodes (RAND, CC, and CLOSE), \ALan{the number of attack edges}, the number of target nodes, maximal number of modified edges $K$ per target node, the hyperparameter $\lambda$, and  the hyperparameter $\eta$. 
When studying the impact of a specific parameter, we fix the remaining parameters to their default values. 
Next, we first show an overall result that demonstrates the effectiveness of our attacks. 
Then, we study the impact of each parameter. 


\begin{figure*}[!tbp]
\vspace{-6mm}
\center
\subfloat[Facebook]{\includegraphics[width=0.24\textwidth]{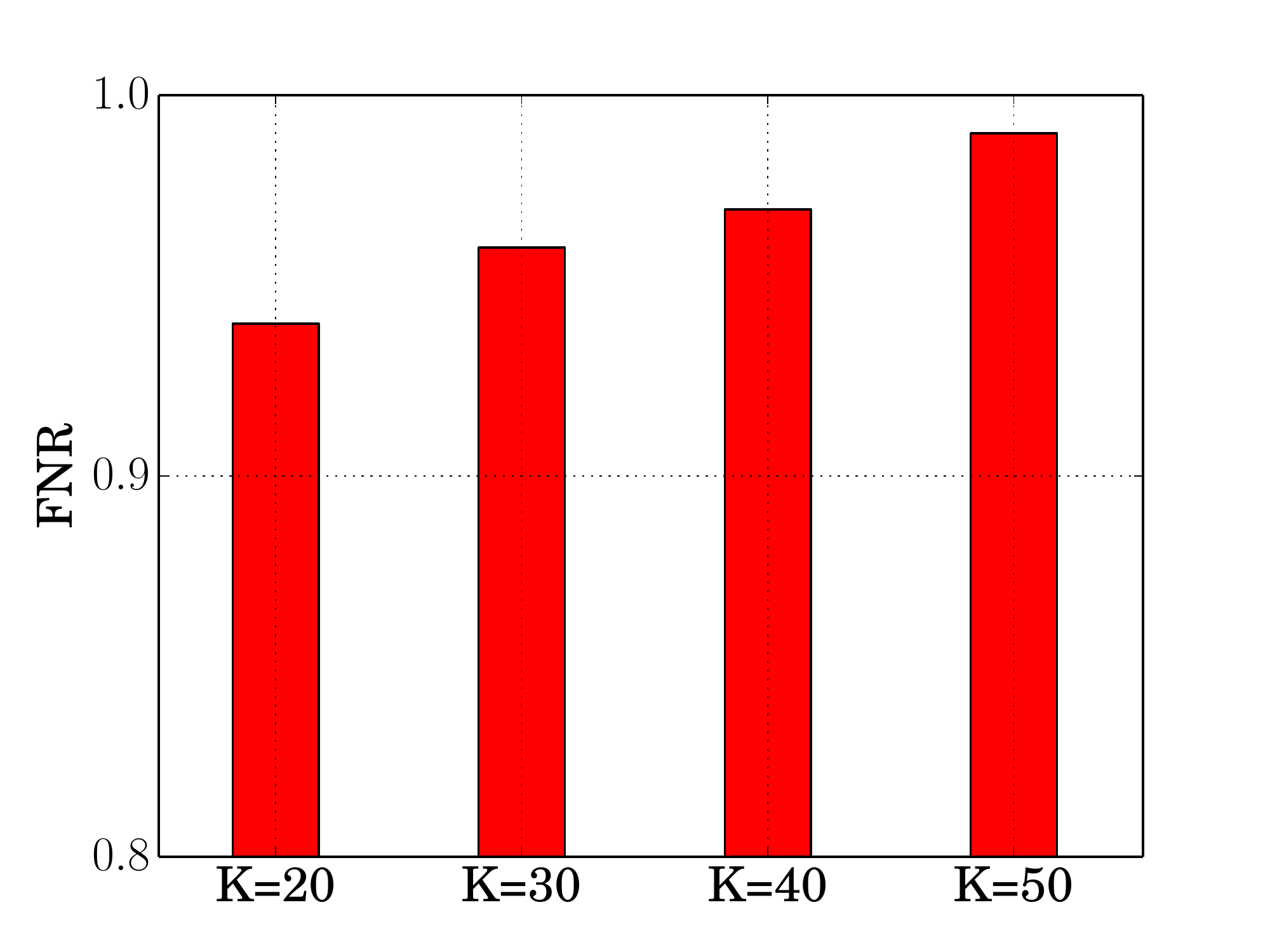}\label{}} 
\subfloat[Enron]{\includegraphics[width=0.24\textwidth]{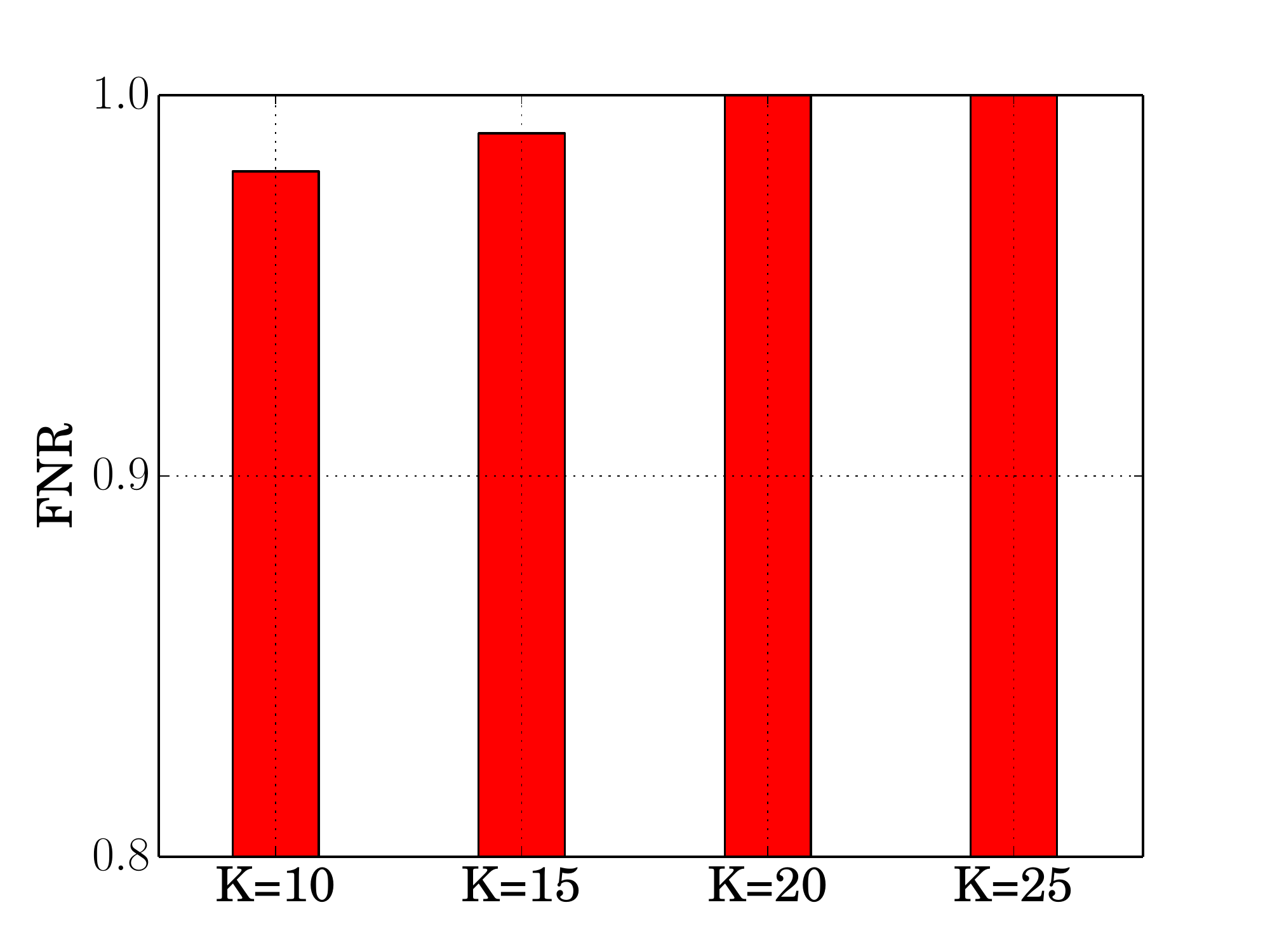}\label{}} 
\subfloat[Epinions]{\includegraphics[width=0.24\textwidth]{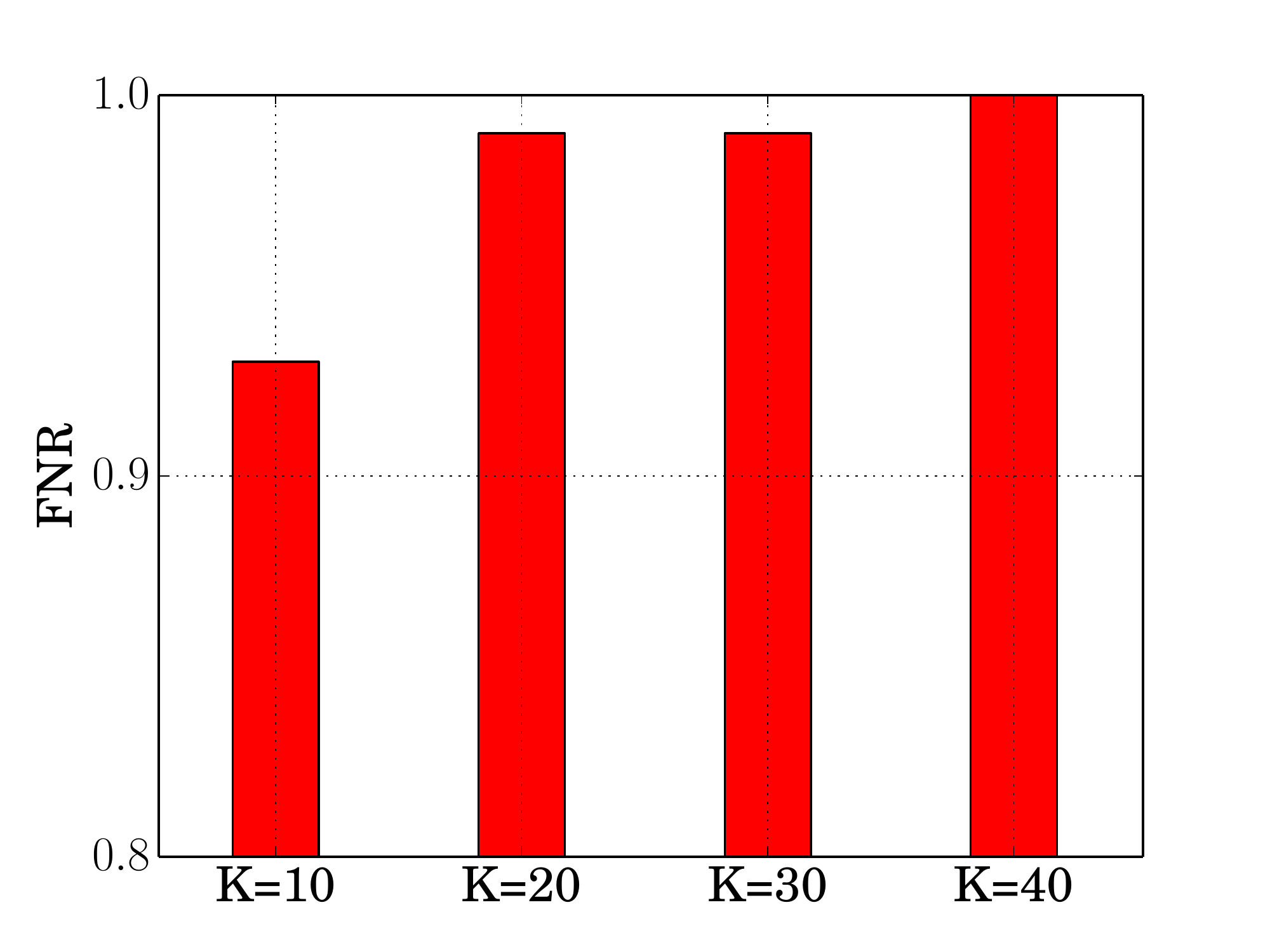}\label{}} 
\subfloat[Twitter]{\includegraphics[width=0.24\textwidth]{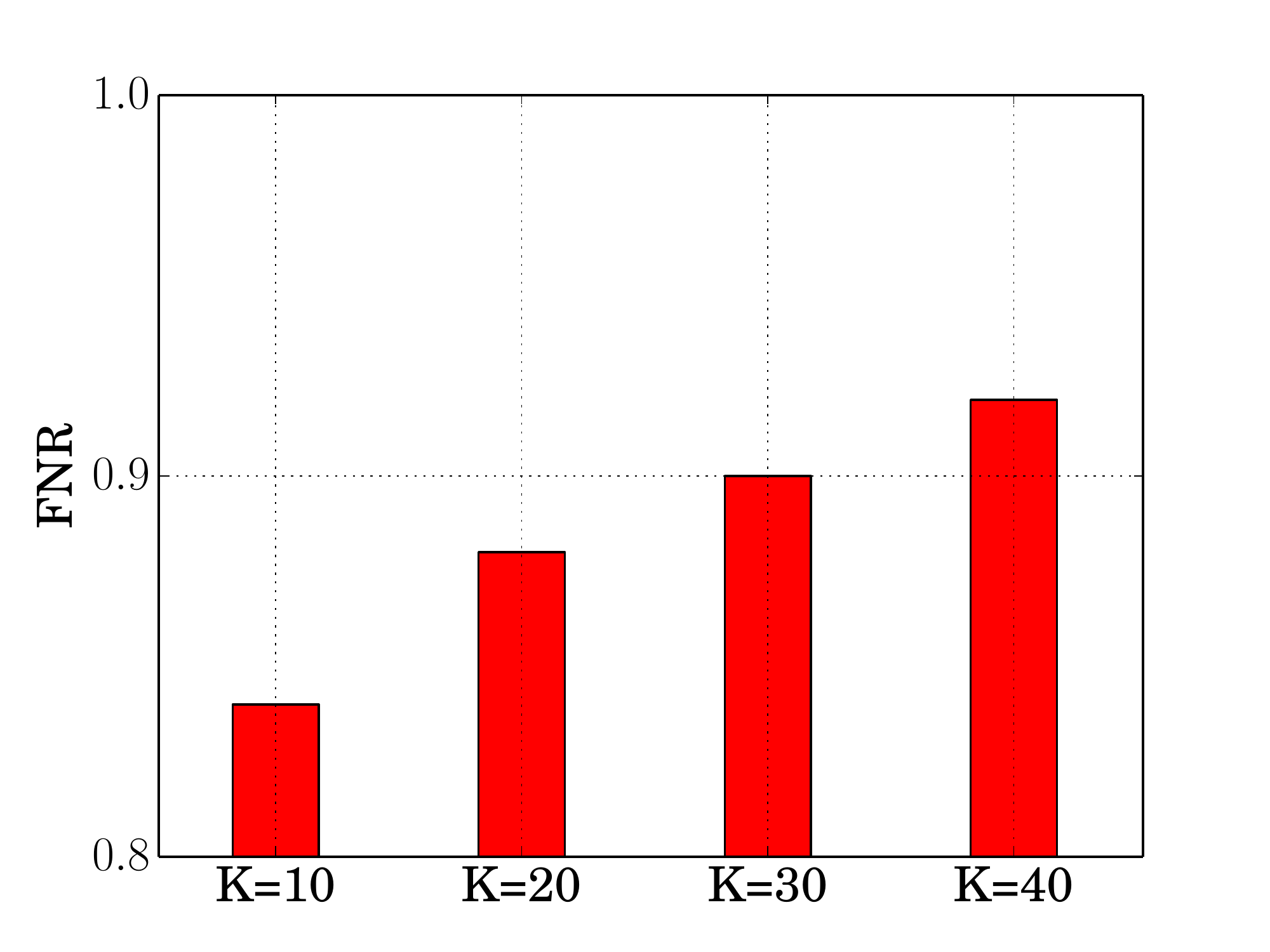}\label{}} 
\caption{Impact of $K$ on our attacks on the four graphs. We observe that our attacks are more effective as $K$ increases.}
\label{impact-K-Equal}
\end{figure*}

\vspace{-3mm}
\myparatight{Our attacks are effective and significantly outperform the baseline attacks} Table~\ref{overall} shows the results of our attacks and the baseline attacks 
on the 
four 
graphs for each method of selecting the target nodes and each type of cost. 
 First, our attacks are effective. Specifically, in most scenarios, our attacks increase the FNR from 0 to be above 0.85. 
\ALan{Moreover, our attacks are significantly more effective than the Random attack and Del-Add attack. In particular, our attacks have 0.2 to 0.6 higher FNRs than the best attack performance of the baseline attacks on the four graphs.}
 Second, our attacks use much more inserted edges than deleted edges. A possible reason is that the graphs are sparse, i.e., only a small fraction of node pairs are connected in the original graphs. Therefore, the space of inserting edges is much larger than the space of deleting edges, leading our attacks to generate much more inserted edges. 
 Overall, our attacks are 
 the least effective on Facebook, i.e., our attack achieves 
 a lowest FNR for Facebook given the same method of selecting target nodes and the same cost type.  The reason is that Facebook has the largest average degree (see Table~\ref{dataset_stat}).
 With the same maximum number (i.e., $K=20$) of modified edges per node, Facebook can better tolerate our attacks.

\myparatight{Impact of different types of costs} The results in Table~\ref{overall} allow us to compare different types of costs. We observe that our attacks are the most effective for Equal cost and the least effective for Categorical cost. The reason is that Equal cost gives our attack the most flexibility at generating the inserted edges and deleted edges. However, for the Categorical cost scenario, it is much more expensive to modify edges between target nodes and compromised negative nodes; and it is even more expensive to modify edges between target nodes and uncompromised negative nodes. These cost constraints essentially restrict the search space of our attacks, making our attacks less effective for Categorical cost. However, we want to stress that our attacks are still effective for Categorical cost, e.g., on the Enron, Epinions, and Twitter graphs, our attacks still increase the FNR to be around 0.85. 
\ALAN{Furthermore, Table~\ref{res_overlap} shows the number of overlapped modified edges among the 3 types of costs. We observe that the number of overlapped modified edges is much smaller than the number of modified edges shown in Table~\ref{overall}, indicating that our attack indeed explores different search spaces for different types of costs. 
} 

\ALAN{\myparatight{Impact of different target node selection methods} The results in Table~\ref{overall} also compare RAND, CC, and CLOSE for selecting target nodes. We observe that our attacks are more effective when using CC and CLOSE than using RAND to select the target nodes, and CC and CLOSE have similar attack performance. Specifically, given the same graph and cost type, our attack achieves a higher FNR for CC and CLOSE than for RAND. For instance, with the Facebook graph and Equal cost, our attacks achieve FNRs of 0.78, 0.94, and 0.96 for RAND, CC, and CLOSE,  respectively. The reason is that the target nodes selected by CC are structurally close and similar. Therefore, it is more likely for our attacks to  make them evade detection ``collectively''. Moreover, the target nodes selected by CLOSE are close to negative nodes and thus they are relatively easy to bypass detection.  However, the target nodes selected by RAND are more structurally dissimilar; different target nodes require different efforts (inserting different edges or deleting different edges) to evade detection, and thus it is harder for our attacks to make the target nodes evade.   
}

\begin{table}[!t]
\centering
\ssmall
\caption{Number of overlapped modified edges on different types of costs and different target node selection methods. }
\begin{tabular}{|c|c|c|c|}
\hline
 \textbf{RAND} &  \textbf{Equal vs. Uniform}  & \textbf{Equal vs. Categorical}   & \textbf{Uniform vs. Categorical}   \\ \hline
 \textbf{Facebook}  & 28 & 0 & 11  \\ \hline
 \textbf{Enron}  & 94 & 0 & 12 \\ \hline
 \textbf{Epinions}  & 67 & 0 & 2 \\ \hline 
 \textbf{Twitter}  & 361 & 17 & 259 \\ \hline \hline

  \textbf{CC} &  \textbf{Equal vs. Uniform}  & \textbf{Equal vs. Categorical}   & \textbf{Uniform vs. Categorical}   \\ \hline
 \textbf{Facebook}  & 156 & 72 & 46  \\ \hline
 \textbf{Enron}  & 176 & 0 & 9 \\ \hline
 \textbf{Epinions}  & 191 & 67 & 10 \\ \hline
 \textbf{Twitter}  & 192 & 40 & 42 \\ \hline \hline

  \textbf{CLOSE} &  \textbf{Equal vs. Uniform}  & \textbf{Equal vs. Categorical}   & \textbf{Uniform vs. Categorical}   \\ \hline
 \textbf{Facebook}  & 190 & 5 & 13  \\ \hline
 \textbf{Enron}  & 416 & 24 & 21 \\ \hline
 \textbf{Epinions}  & 428 & 260 & 147 \\ \hline
 \textbf{Twitter}  & 471 & 190 & 470 \\ \hline
 \end{tabular}
\label{res_overlap}
\end{table}

\begin{figure}[!t]
\vspace{-6mm}
\center
\subfloat{\includegraphics[width=0.32\textwidth]{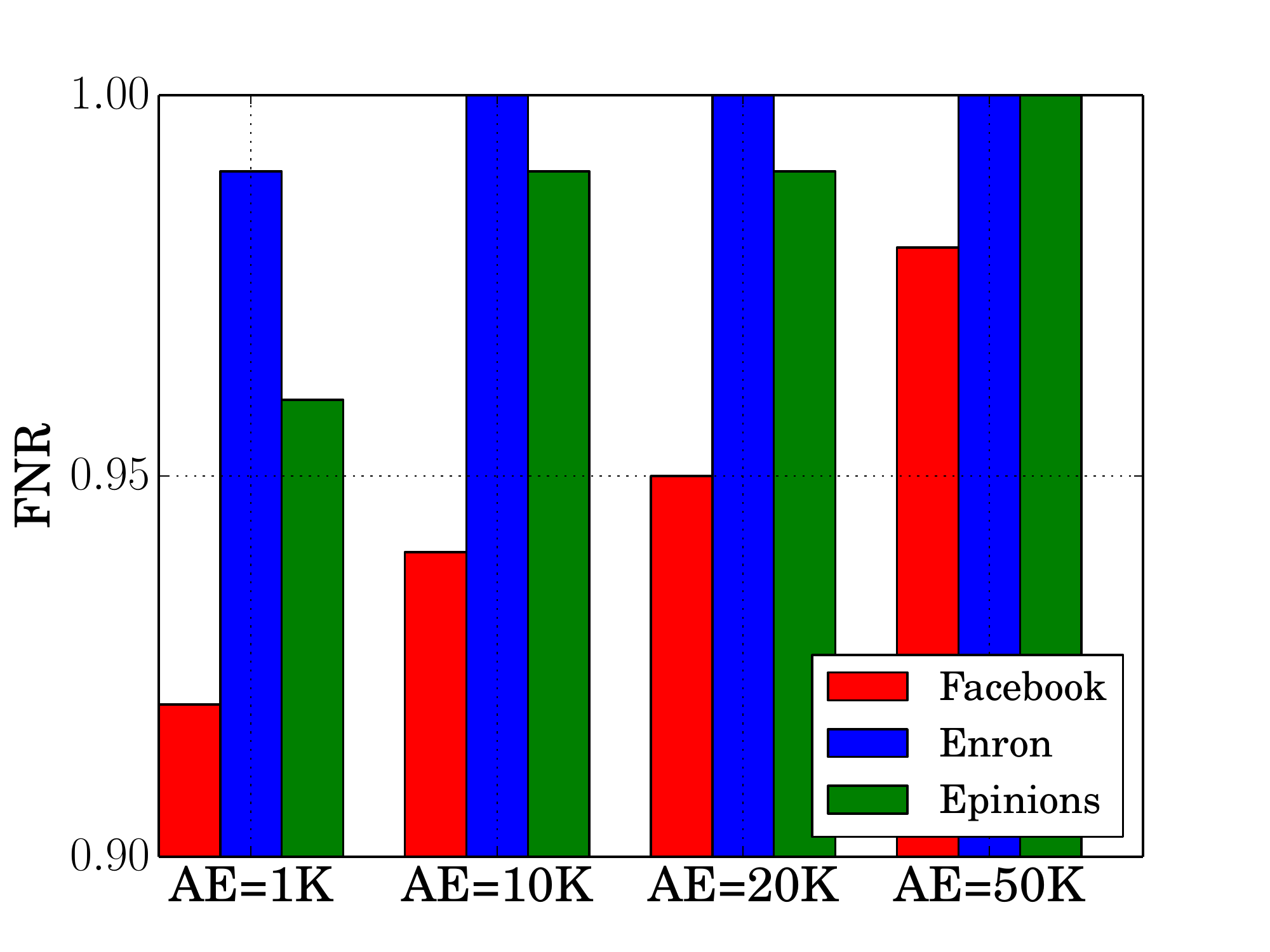}}
\caption{Impact of the number of attack edges ($AE$) on the three graphs with synthesized positive nodes for Equal cost. We observe that as $AE$ increases, FNR also increases. }
\label{impact-AE-Equal}
\vspace{-6mm}
\end{figure}



\myparatight{Impact of $K$}
Figure~\ref{impact-K-Equal} shows the FNRs vs. $K$ on the 
four graphs
with Equal cost and CC as the method of selecting target nodes. 
The FNRs vs. $K$ with Uniform cost and Categorical cost are shown in Appendix (See Figure~\ref{impact-K-CC-Other}). 
Considering that the four graphs have different average degrees, we set different ranges of $K$ for the four graphs.  
We observe that as $K$ increases, FNR also increases on all the four graphs. This is because a larger $K$ allows an attacker to modify more edges per node, and thus our attacks can increase the FNRs more. Note that the total costs increase linearly as $K$, and thus we omit the results on the total cost for simplicity.

\ALan{
\myparatight{Impact of the number of attack edges (AE)} 
Figure~\ref{impact-AE-Equal} shows the FNRs vs. the number of attack edges on the three  graphs with synthesized positive nodes for Equal cost (note that the Twitter graph has a fixed number of attack edges). 
The FNRs vs. $AE$ for Uniform cost and Categorical cost are shown in Appendix (See Figure~\ref{impact-AE-Uni-Cat}).
We observe that as $AE$ increases, our attacks achieve higher FNRs. This is because the accuracy of collective classification decreases as $AE$ increases, even without our attacks. In other words, when $AE$ is large, the FNR of the positive nodes is already large without our attacks. 
}

\begin{figure}[tbp]
\vspace{-6mm}
\center
\subfloat[]{\includegraphics[width=0.24\textwidth]{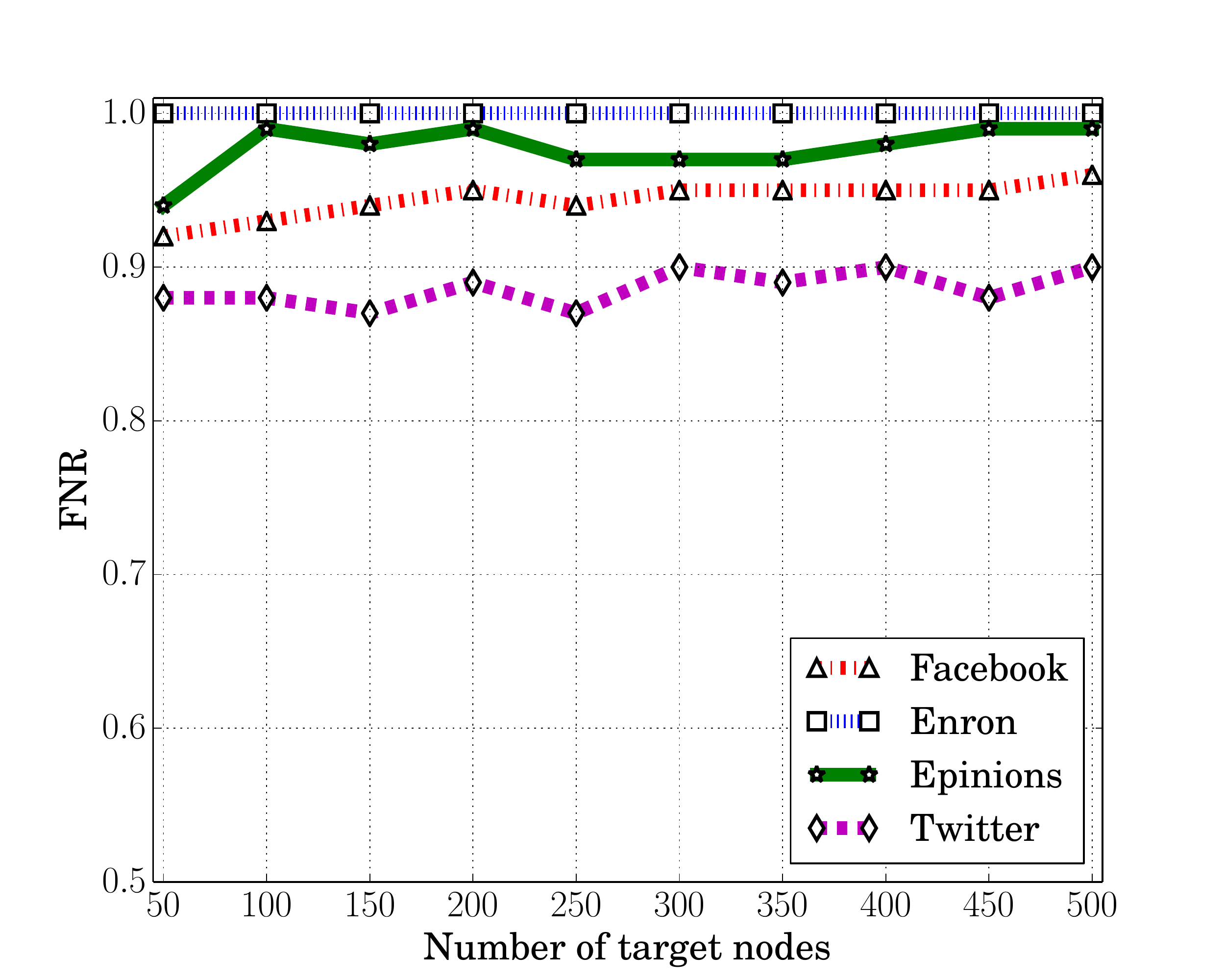}\label{impact-attacker-nodes}}
\subfloat[]{\includegraphics[width=0.24\textwidth]{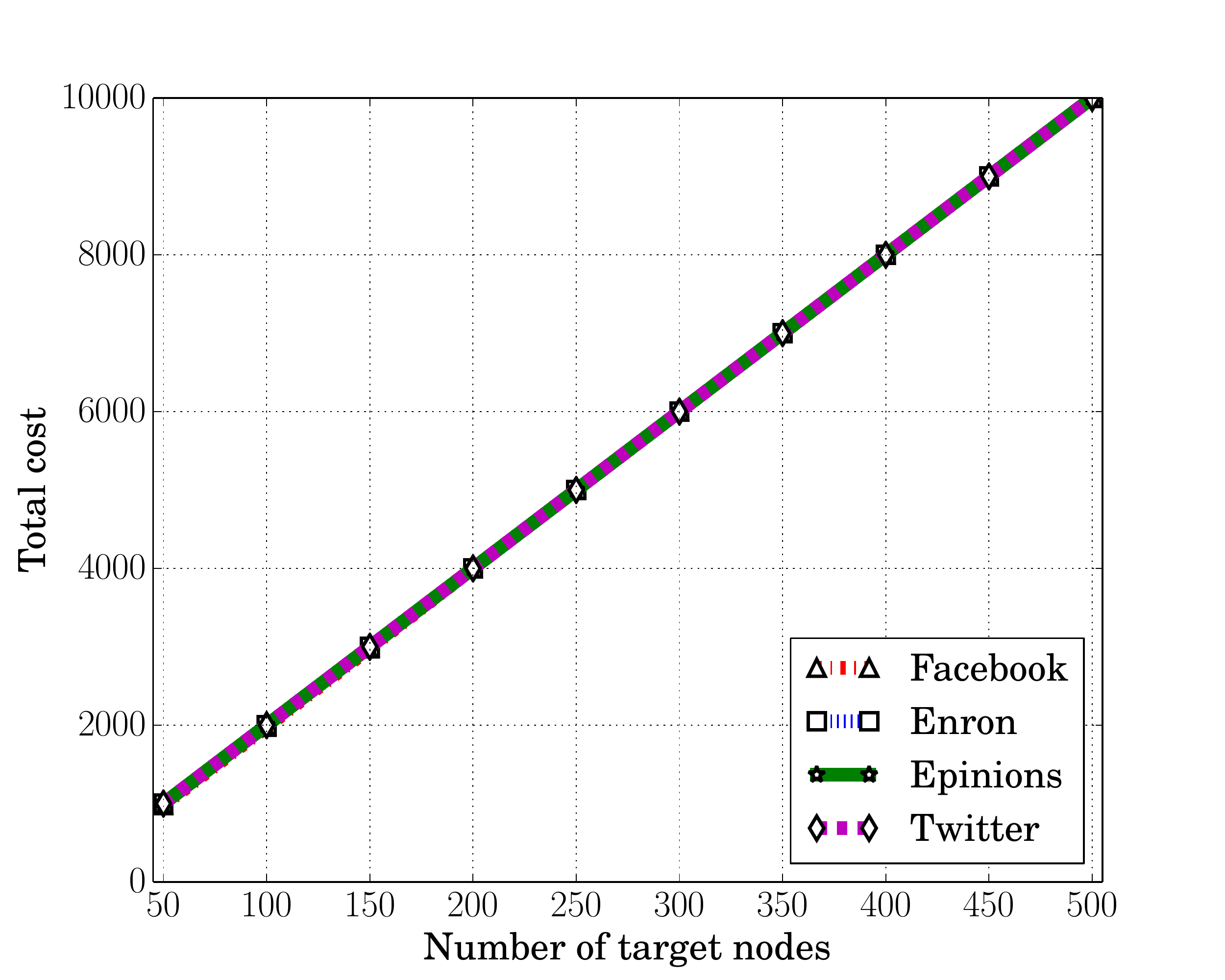}\label{impact-attacker-nodes-cost}}
\caption{(a) Impact of the number of target nodes with Equal cost. 
We observe that FNRs are stable. 
(b) Costs with respect to the number of target nodes. We observe that the costs are linear to the number of target nodes. 
}
\vspace{-4mm}
\end{figure}


\myparatight{Impact of the number of target nodes} 
Figure~\ref{impact-attacker-nodes} shows the FNRs vs. the number of target nodes on the 
four graphs, while Figure~\ref{impact-attacker-nodes-cost} shows the total cost vs. the number of target nodes, where Equal cost and CC are used. Due to the limited space, we show the FNRs vs. the number of target nodes with Uniform cost and Categorical cost in Appendix (See Figure~\ref{impact-attacker-nodes-other-cc}). On one hand, we observe that FNRs are stable across different number of target nodes. This is because, for each target node, our attack iteratively computes the modified edges. On the other hand, our attacks require a larger total cost when the attacker uses more target nodes.  Our results indicate that, our attacks can make more positive nodes evade detection when the attacker can tolerate a larger cost.

\begin{figure}[tbp]
\vspace{-6mm}
\center
\subfloat[$\lambda$]{\includegraphics[width=0.24\textwidth]{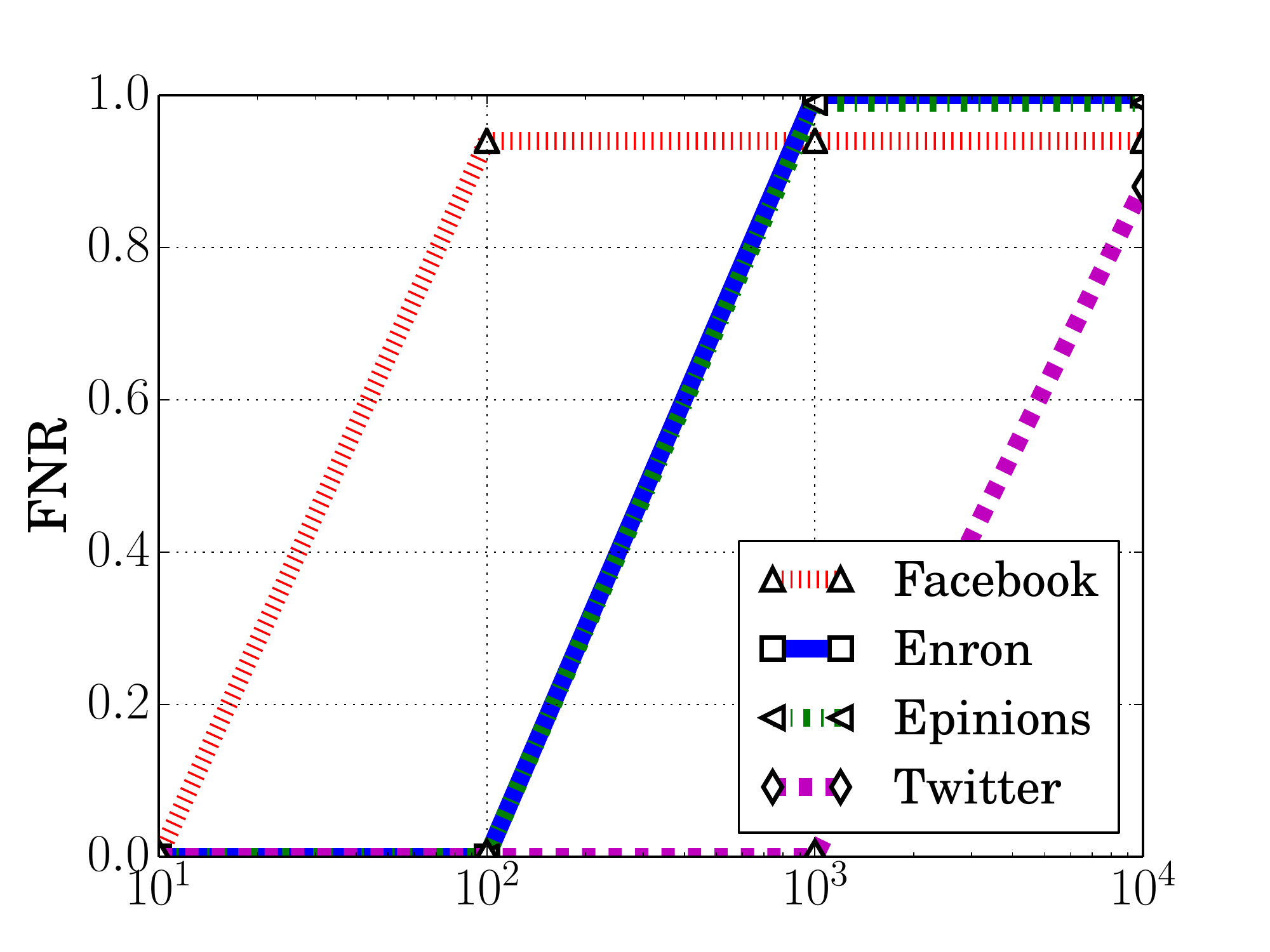}}
\subfloat[$\eta$]{\includegraphics[width=0.24\textwidth]{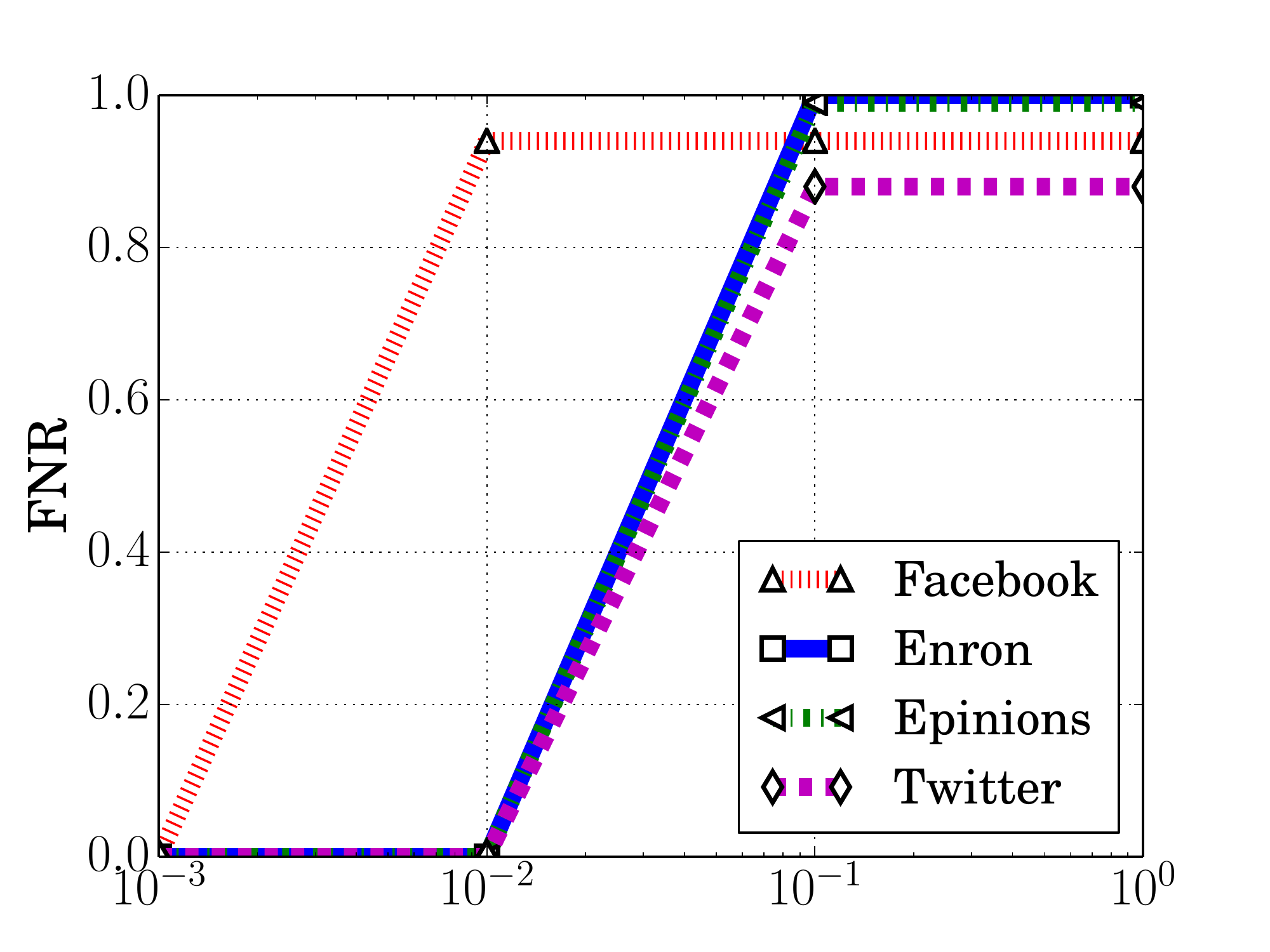}}
\caption{Impact of $\lambda$ and $\eta$ on our attacks.
We observe that our attacks achieve stable FNRs when $\lambda$ and $\eta$ are larger than certain thresholds.
}
\label{impact-lambda-eta-Equal}
\vspace{-4mm}
\end{figure}

\myparatight{Impact of hyperparameters}
Figure~\ref{impact-lambda-eta-Equal} shows the impact of $\lambda$ and $\eta$ on our attacks in the four graphs with Equal cost, where the curves for Enron and Epinions overlap. 
The impact of $\lambda$ and $\eta$ on our attacks with Uniform cost and Categorical cost are reported in Appendix (See Figure~\ref{impact-lambda-eta-Uni} and Figure~\ref{impact-lambda-eta-Cat}). 
We observe phase transition phenomena for both $\lambda$ and $\eta$. Specifically, when $\lambda$ and $\eta$ are larger than certain thresholds (e.g., 100 for $\lambda$ and 0.01 for $\eta$ on Facebook), our attacks become effective. However, our attacks are less effective or ineffective when $\lambda$ and $\eta$ are smaller than the thresholds. Moreover, once $\lambda$ and $\eta$ are larger than the thresholds, our attacks are insensitive to them, e.g., the FNRs are stable. Facebook has smaller thresholds, and we speculate the reason is that Facebook is much denser than the other three graphs.   



\subsection{Evaluation with Partial Knowledge}
We consider multiple cases where the attacker does not have access to the parameters of LinLBP, the training dataset, and/or the complete graph. We only show results on the Epinions graph as we observe similar patterns on the other three graphs, and the results for the other three graphs are shown in Appendix. Our results show that our attacks can still effectively increase the FNRs for the target nodes even if the attacker does not have access to the parameters of LinLBP, the training dataset, and/or the complete graph. 

\begin{figure}[!tbp]
\center
\subfloat[$\theta$]{\includegraphics[width=0.24\textwidth]{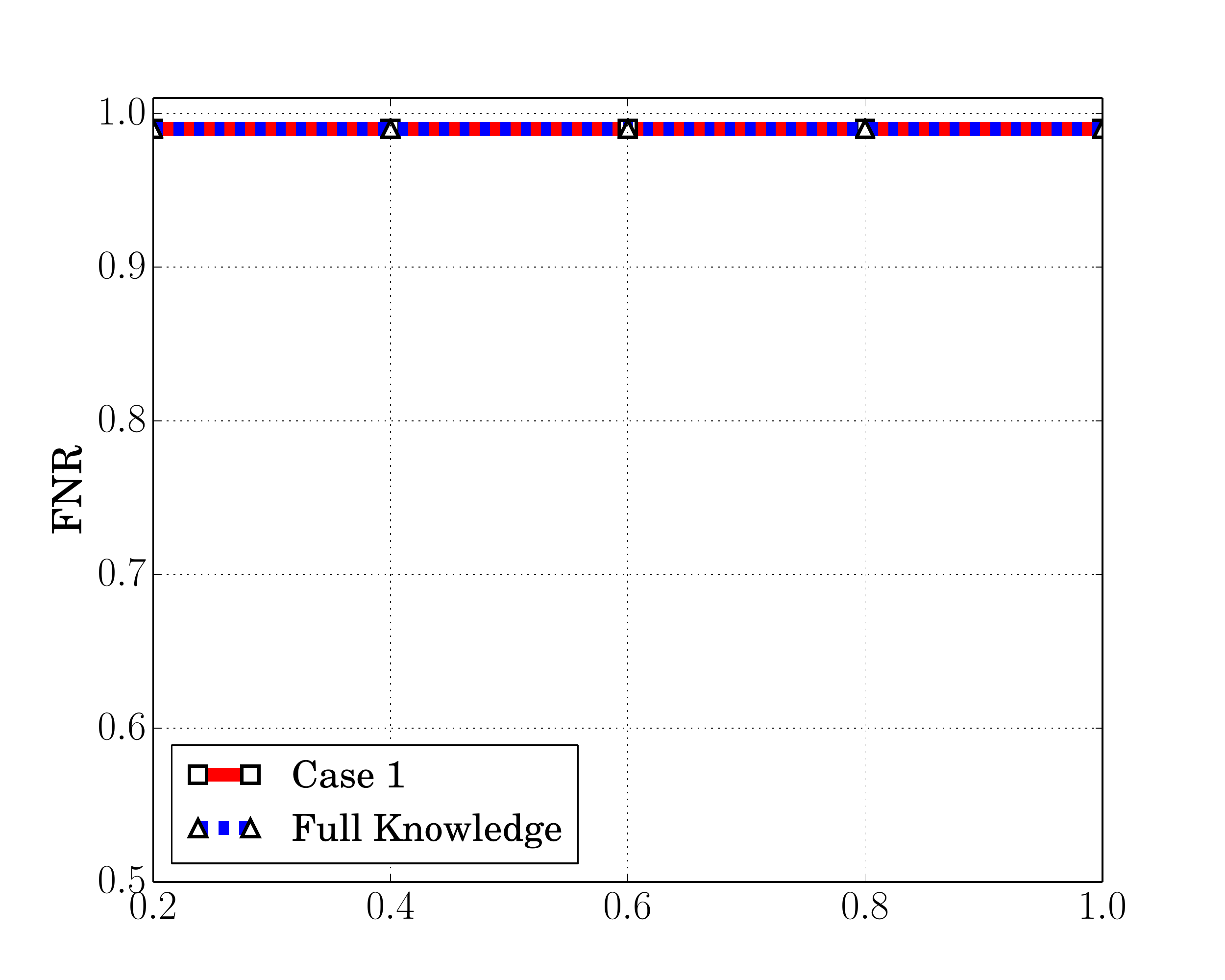}\label{impact-substitute-prior}}
\subfloat[$w$]{\includegraphics[width=0.24\textwidth]{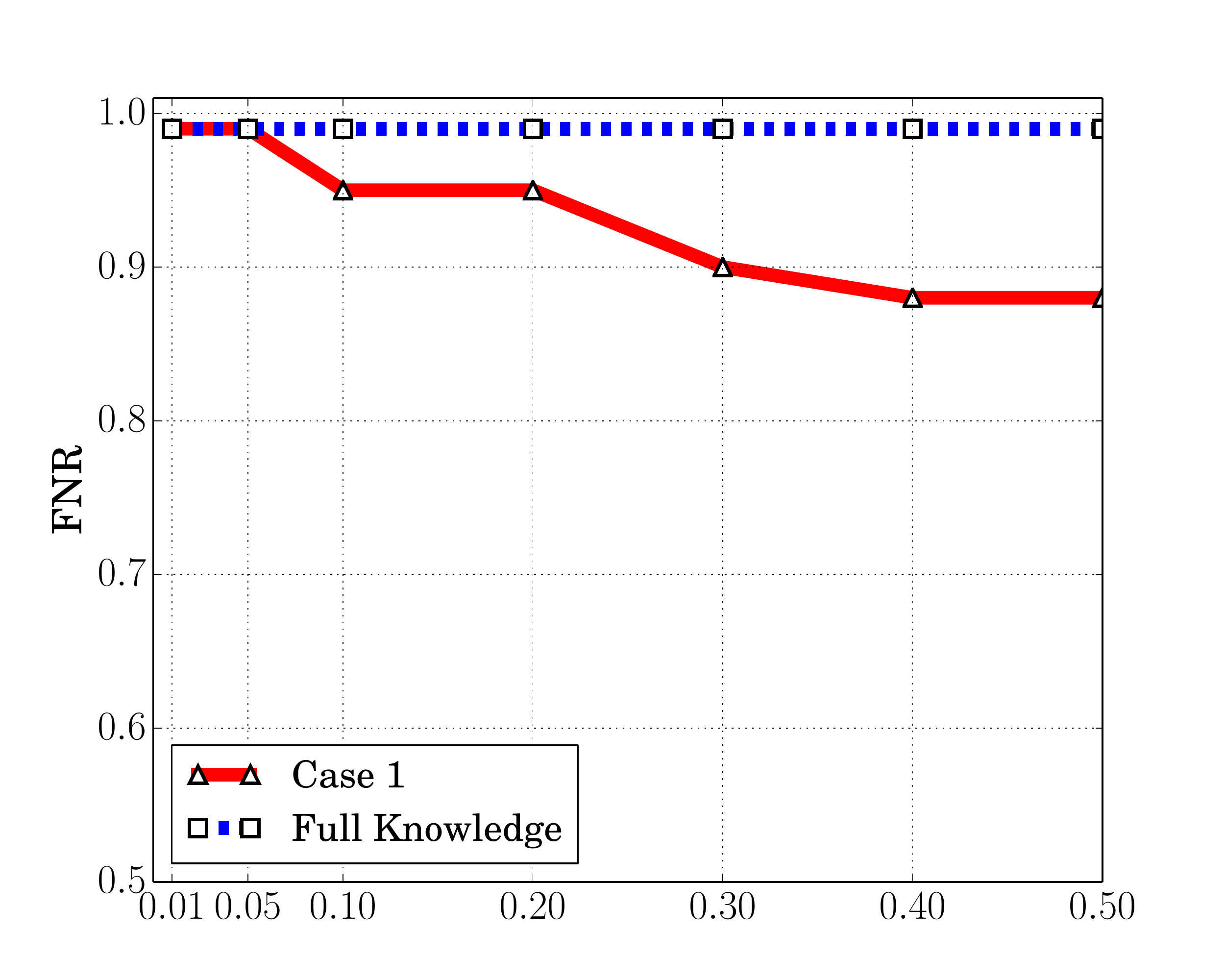}\label{impact-substitute-weight}}
\caption{FNRs of our attacks with different substitute parameters of LinLBP on Epinions. 
}
\label{impact-substitute-parameter}
\vspace{-6mm}
\end{figure}

\noindent \textbf{Parameter=No, Training=Yes, Graph=Complete (Case 1):}
In this case, the attacker does not know the parameters (i.e., $\theta$ and $w$) of LinLBP.
The attacker randomly samples numbers from their corresponding domains (i.e., $0 < \theta \leq 1$,  $0 < w \leq 0.5$) as the substitute parameters and apply our attacks with them. Figure~\ref{impact-substitute-parameter} shows the FNRs of our attacks with different substitute parameters on Epinions, where $w=0.01$ when exploring different $\theta$ in Figure~\ref{impact-substitute-prior} and $\theta=0.5$ when exploring different $w$ in Figure~\ref{impact-substitute-weight}. 
The FNRs of our attacks with different substitute parameters on Facebook, Enron, and Twitter are shown in Figure~\ref{impact-substitute-parameter-FB}, Figure~\ref{impact-substitute-parameter-Enron}, and Figuer~\ref{impact-substitute-parameter-Twitter} in Appendix, respectively. 
We observe that our attacks are insensitive to $\theta$. Moreover, our attacks are more effective when the substitute weight parameter is closer to the true weight. Specifically, the true weight is 0.01. Our attacks achieve FNRs that are close to 1 when the substitute weight is around 0.01, 
and FNRs decrease as the substitute weight increases. However, even if the substitute weight is far from the true weight, our attacks are still effective. For instance, the FNR is still 0.88 when our attack uses 0.5 as the substitute weight. 


\begin{figure}[tbp]
\center
\subfloat[]{\includegraphics[width=0.24\textwidth]{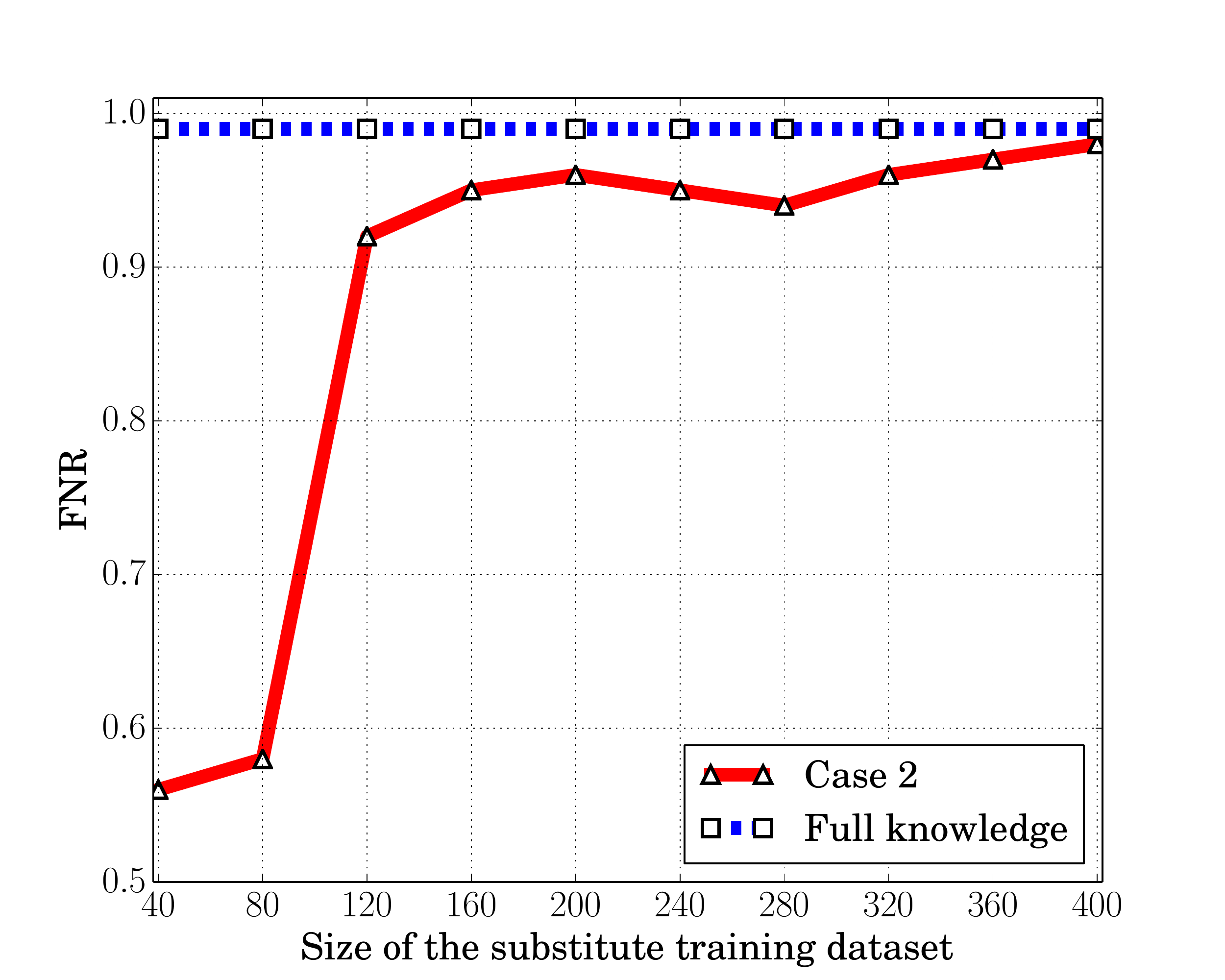} \label{impact-substitute-nodes}}
\subfloat[]{\includegraphics[width=0.24\textwidth]{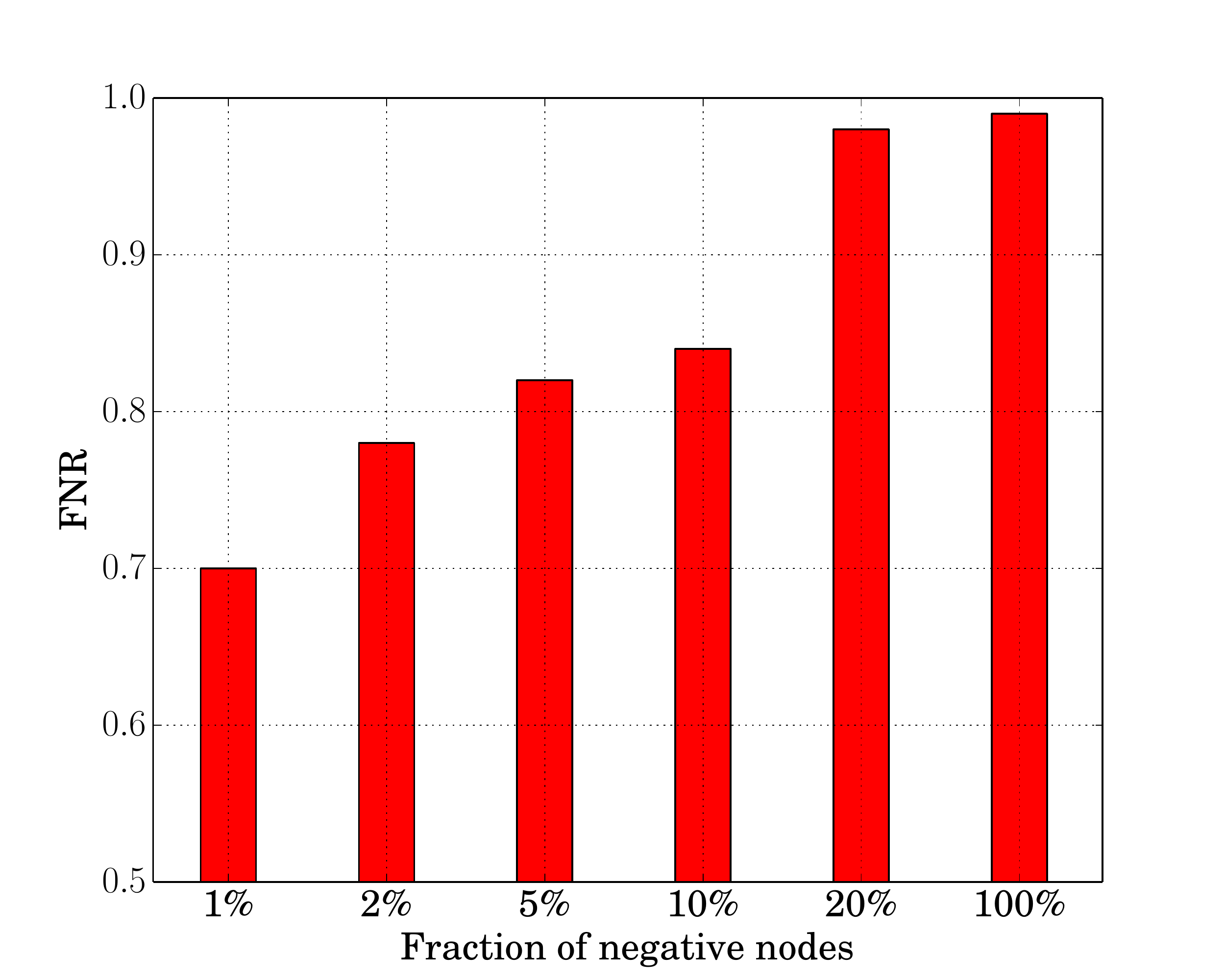} \label{impact-partial-Equal}}
\caption{(a) FNRs vs. size of the substitute training dataset on Epinions; (b) FNRs when the attacker knows $\tau$\% of the negative nodes and edges between them on Epinions.}
\end{figure}

\begin{figure*}[tbp]
\vspace{-6mm}
\center
\subfloat[]{\includegraphics[width=0.3\textwidth]{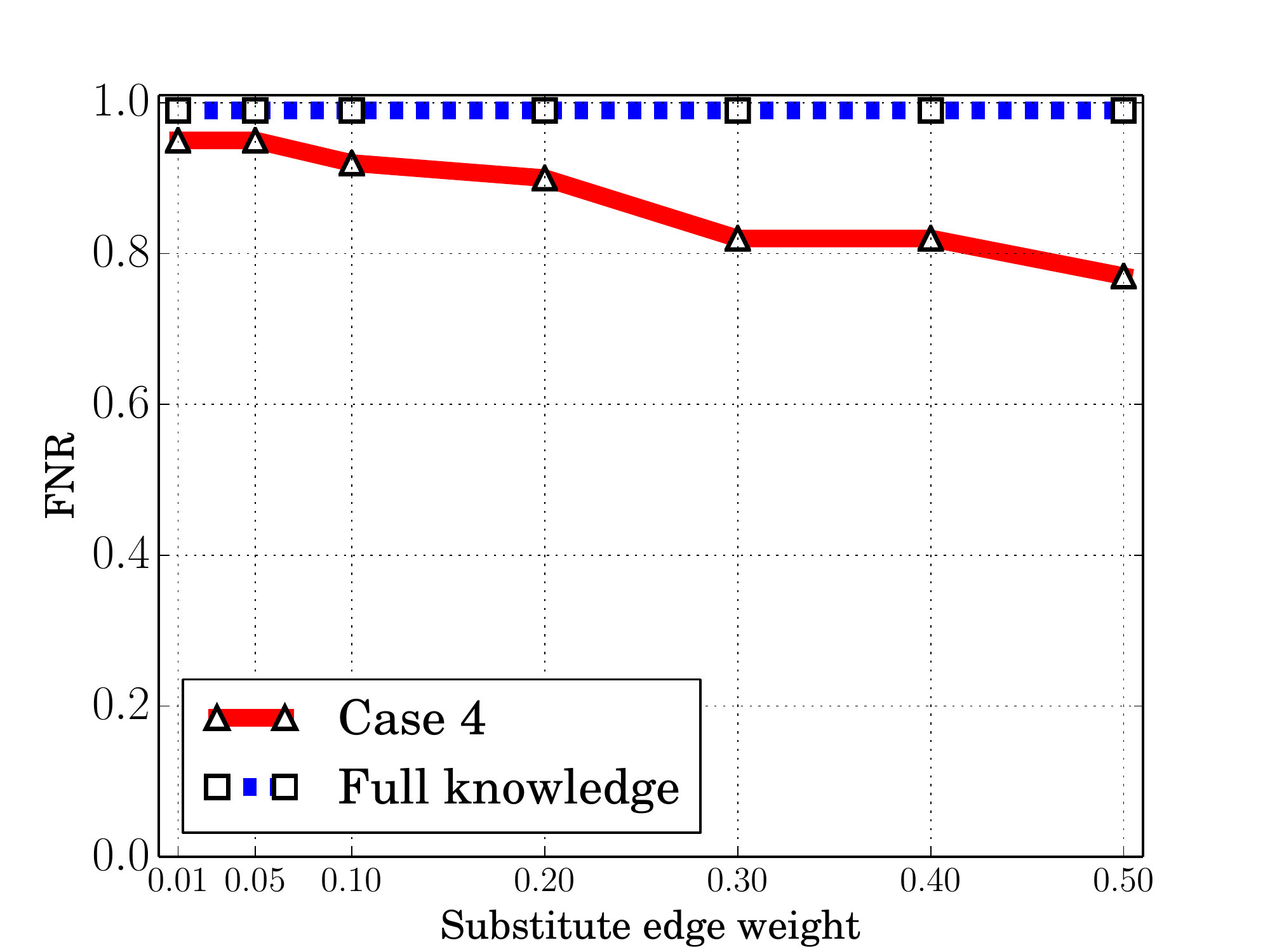} \label{least-Epinions-weight}} 
\subfloat[]{\includegraphics[width=0.3\textwidth]{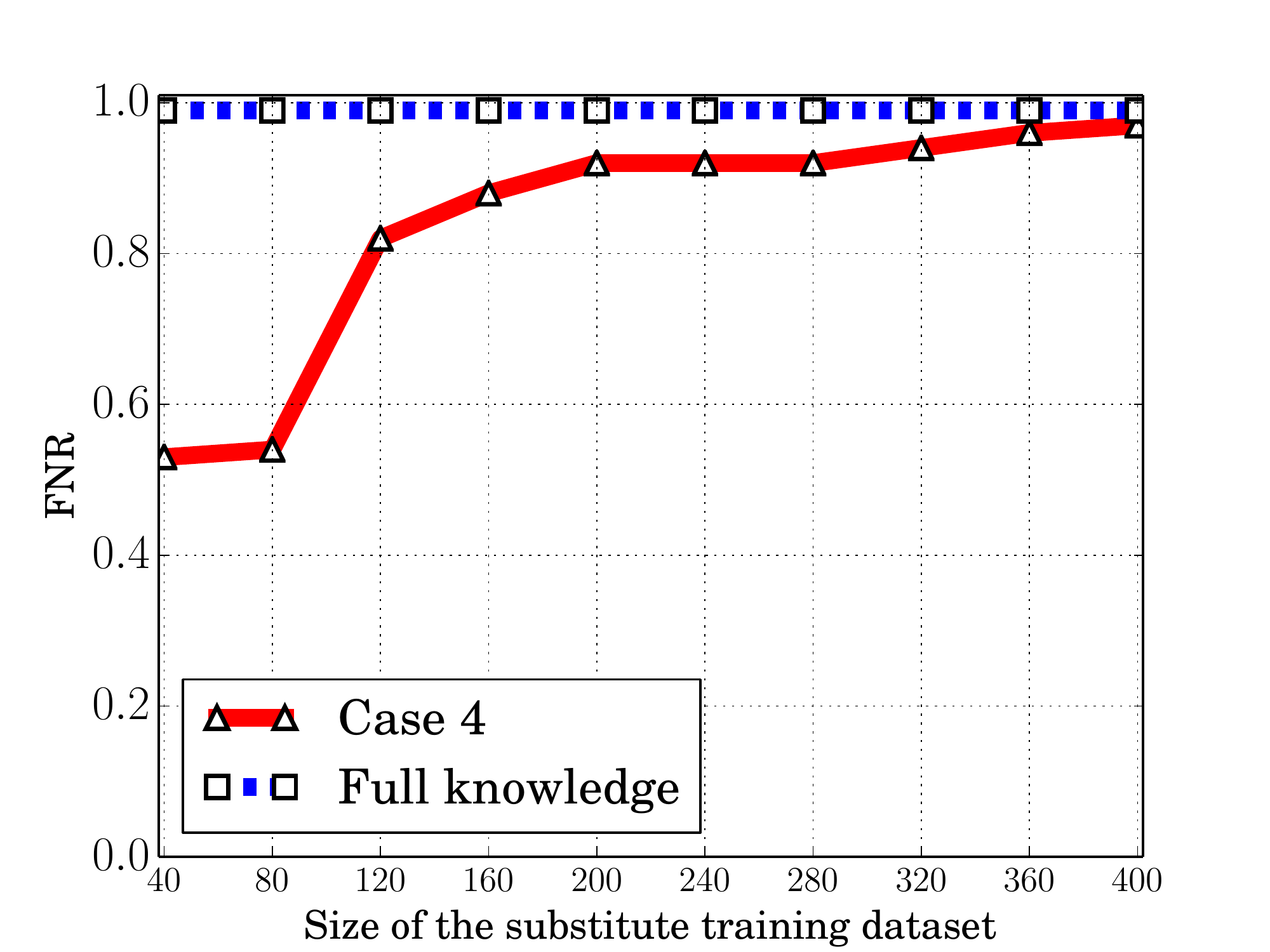} \label{least-Epinions-training}}
\subfloat[]{\includegraphics[width=0.3\textwidth]{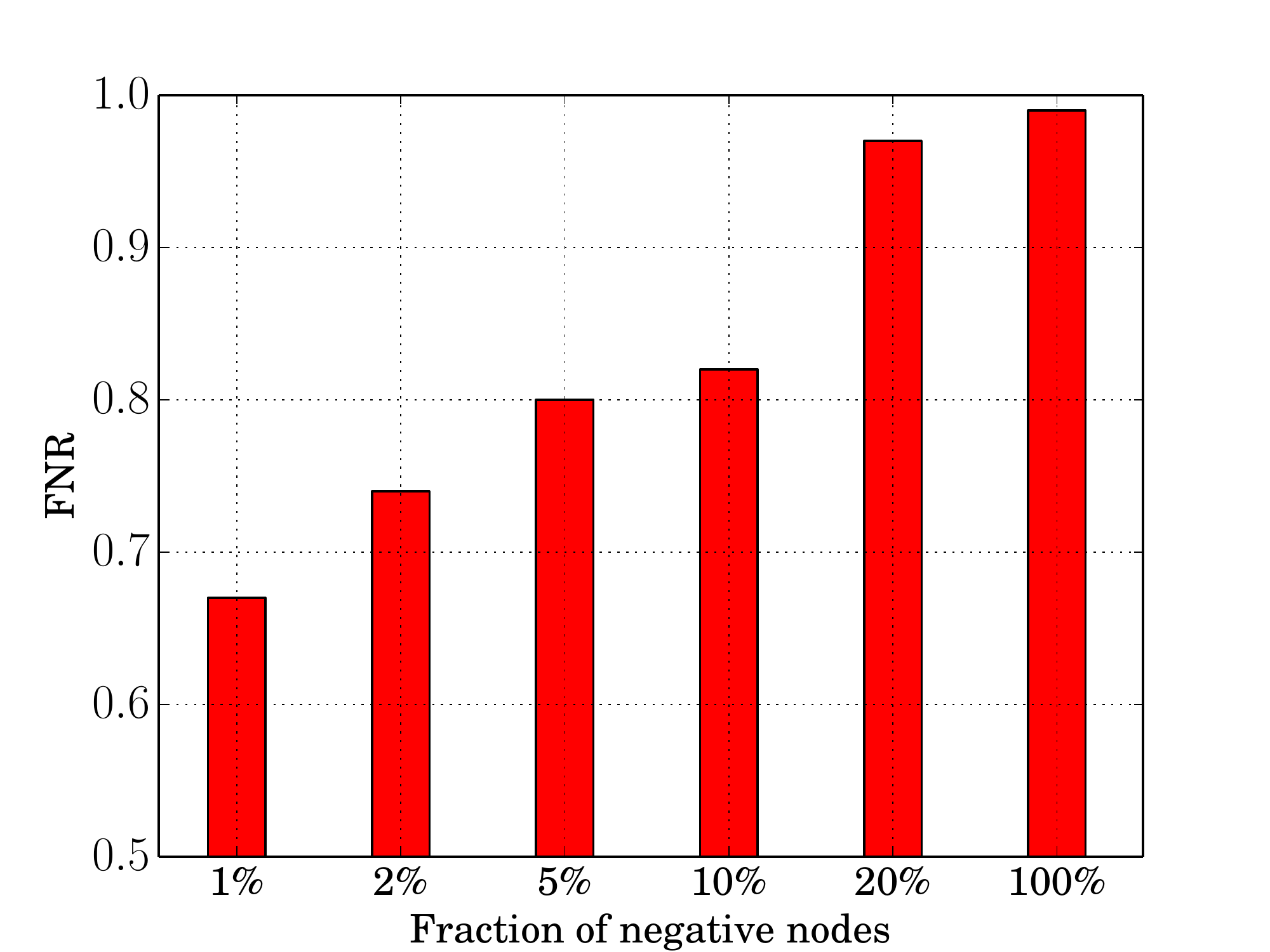} \label{least-Epinions-partial}}
\caption{FNRs of our attacks vs. (a) substitute edge weight; (b) size of the substitute training dataset; (c) fraction of negative nodes on Epinions when the attacker uses substitute parameters, substitute training dataset, and a partial graph.}
\label{impact-least}
\end{figure*}

\noindent \textbf{Parameter=Yes, Training=No, Graph=Complete (Case 2):}
In this case, the attacker does not know the true training dataset used by LinLBP. However, the attacker knows the positive nodes as they were created by itself and the negative nodes as well. Therefore, the attacker can sample some positive nodes as labeled positive nodes and sample equal number of negative nodes as labeled negative nodes, which are used as a substitute training dataset.  Then, the attacker applies our attack using the substitute training dataset. Figure~\ref{impact-substitute-nodes} shows the FNRs as a function of the size of the substitute training dataset on the Epinions graph. 
The FNRs vs. size of the substitute training dataset on Facebook, Enron, and Twitter are displayed in Figure~\ref{substitute-FB}, ~\ref{substitute-Enron}, and~\ref{substitute-Twitter} in Appendix, respectively.
We observe that as the size of the substitute training dataset increases, the FNRs of our attacks first increase dramatically and then increase slowly until reaching to the FNRs in the full knowledge scenario. 

%


\noindent \textbf{Parameter=Yes, Training=Yes, Graph=Partial (Case 3):}
In this scenario, the attacker knows the subgraph of positive nodes and their edges as well as a connected subgraph of some negative nodes and edges between them.
Specifically, we consider the attacker knows $\tau \%$ of the negative nodes. In particular, we randomly select a negative node and span a connected component that includes $\tau \%$ of the negative nodes via breadth first search. Then, the attacker's partial graph consists of the connected component of negative nodes, positive nodes, and the edges between them. 
The attacker applies our attacks using this partial graph to generate the inserted/deleted edges. Note that LinLBP still detects positive nodes using the original complete graph with the edges modified by our attacks. 
Figure~\ref{impact-partial-Equal} shows the FNRs of our attacks as the attacker knows a larger subgraph of negative nodes (i.e., $\tau$\% is larger) on Epinions. 
The FNRs vs. $\tau \%$ of negative nodes on Facebook, Enron, and Twitter are shown in Figure~\ref{partial-FB}, ~\ref{partial-Enron}, and~\ref{partial-Twitter} in Appendix, respectively.
We observe that as $\tau$ increases, our attacks are more effective. Moreover, when the attacker only knows $20\%$ of negative nodes, our attacks can obtain FNRs close to those when the attacker knows the complete graph.  


\myparatight{Parameter=No, Training=No, Graph=Partial (Case 4)}
In this case, the attacker knows the \emph{least knowledge} of the three dimensions of a LinLBP system. 
\ALAN{To study our attacks in this case, we fix two dimensions and observe FNR vs. the third dimension. By default, we assume the attacker knows 20\% of negative nodes and edges between them, samples a substitute training dataset with 400 nodes on {Facebook, Enron, and Epinions and 3,000 nodes on Twitter}, and has a substitute edge weight 0.05. 
Figure~\ref{impact-least} shows FNRs of our attacks on Epinions vs. each dimension. Figure~\ref{least-FB}, Figure~\ref{least-Enron}, and Figure~\ref{least-Twitter} in Appendix show FNRs vs. each dimension on Facebook, Enron, and Twitter, respectively. 
We observe similar patterns as in Case 1, Case 2, and Case 3 (see Figure~\ref{impact-substitute-parameter}, Figure~\ref{impact-substitute-nodes}, and Figure~\ref{impact-partial-Equal}), respectively. The reason is that our attacks do not require the true parameters, the true training dataset, nor the complete graph. 
}


\begin{table}[!tbp]
\centering
\ssmall
\caption{Transferability of our attacks to other graph-based classification methods on Facebook. 
}
\begin{tabular}{|c|c|c|c|c|c|}
\hline
\multicolumn{6}{|c|}{\textbf{RAND}}             \\ \hline
\multicolumn{2}{|c|}{\textbf{Method}} &  \textbf{No attack}  &  \textbf{Equal} & \textbf{Uniform}   & \textbf{Categorical}  \\ \hline
\multirow{4}{*}{{\bf \makecell{Collective \\ Classification}}}  & {\bf LinLBP}  & 0 & 0.78 & 0.66 & 0.6 \\ \cline{2-6} 
                   &  {\bf JWP} & 0 & 0.76 & 0.63 & 0.59 \\ \cline{2-6} 
                   &  {\bf LBP} & 0.07 & 0.7 & 0.65 & 0.6 \\ \cline{2-6} 
                   & {\bf RW}  & 0.13 & 0.68 & 0.61 & 0.55 \\ \hline
\multirow{4}{*}{{\bf \makecell{Graph  \\ Neural Network}}}   &  {\bf LINE} & 0.07 & 0.61 & 0.61 & 0.45 \\ \cline{2-6} 
				& {\bf DeepWalk}  & 0.25 & 0.55 & 0.52 & 0.52 \\  \cline{2-6} 
				& {\bf node2vec}  & 0.2 & 0.5 & 0.45 & 0.45 \\ \cline{2-6}		
				& {\bf GCN}  & 0.12 & 0.52 & 0.52 & 0.41 \\ \hline
\multicolumn{6}{|c|}{\textbf{CC}}             \\ \hline
\multicolumn{2}{|c|}{\textbf{Method}} &  \textbf{No attack}  &  \textbf{Equal} & \textbf{Uniform}   & \textbf{Categorical}  \\ \hline
\multirow{4}{*}{{\bf \makecell{Collective \\ Classification}}}  & {\bf LinLBP}  & 0 & 0.94 & 0.94 & 0.68 \\ \cline{2-6} 
                   &  {\bf JWP} & 0 & 0.93 & 0.93 & 0.63 \\ \cline{2-6} 
                   &  {\bf LBP} & 0.01 & 0.92 & 0.92 & 0.64 \\ \cline{2-6} 
                   & {\bf RW}  & 0.03 & 0.92 & 0.92 & 0.63 \\ \hline
\multirow{4}{*}{{\bf \makecell{Graph  \\ Neural Network}}}  &  {\bf LINE} & 0.01 & 0.85 & 0.75 & 0.62 \\ \cline{2-6} 
				& {\bf DeepWalk}  & 0.04 & 0.71 & 0.65 & 0.45 \\ \cline{2-6} 
				& {\bf node2vec}  & 0.04 & 0.69 & 0.61 & 0.45 \\ \cline{2-6}
                   & {\bf GCN}  & 0.05 & 0.54 & 0.53 & 0.45 \\ \hline
\multicolumn{6}{|c|}{\textbf{CLOSE}}             \\ \hline
\multicolumn{2}{|c|}{\textbf{Method}} &  \textbf{No attack}  &  \textbf{Equal} & \textbf{Uniform}   & \textbf{Categorical}  \\ \hline
\multirow{4}{*}{{\bf \makecell{Collective \\ Classification}}}  & {\bf LinLBP}  & 0 & 0.96 & 0.93 & 0.69 \\ \cline{2-6} 
                   &  {\bf JWP} & 0 & 0.94 & 0.93 & 0.65 \\ \cline{2-6} 
                   &  {\bf LBP} & 0.01 & 0.94 & 0.92 & 0.64 \\ \cline{2-6} 
                   & {\bf RW}  & 0.04 & 0.93 & 0.92 & 0.63 \\ \hline
\multirow{4}{*}{{\bf \makecell{Graph  \\ Neural Network}}}  &  {\bf LINE} & 0.01 & 0.86 & 0.77 & 0.59 \\ \cline{2-6} 
				& {\bf DeepWalk}  & 0.05 & 0.72 & 0.69 & 0.48 \\ \cline{2-6} 
				& {\bf node2vec}  & 0.05 & 0.77 & 0.69 & 0.50 \\ \cline{2-6}
                & {\bf GCN}  & 0.05 & 0.54 & 0.53 & 0.42 \\ \hline
\end{tabular}
\label{transfer_FB}
\vspace{-4mm}
\end{table}

\subsection{Transferring to Other Graph-based Classification Methods}
Our attacks generate inserted/deleted edges based on LinLBP method. A natural question is: are our inserted/deleted edges also effective for other graph-based classification methods? To answer this question, we use our attacks to generate the modified edges based on LinLBP, modify the graph structure accordingly, and apply another graph-based classification method to detect the target nodes. 
Recall that graph-based classification methods can be roughly categorized as \emph{collective classification} and \emph{graph neural network}. We consider three representative collective classification methods, i.e., JWP~\cite{wang2019graph}, LBP~\cite{sybilbelief}, and RW~\cite{wang2018structure}, and four representative graph neural network methods, i.e., LINE~\cite{tang2015line}, DeepWalk~\cite{perozzi2014deepwalk}, node2vec~\cite{grover2016node2vec}, and GCN~\cite{kipf2017semi}.
We obtained 
source code of all these methods from their corresponding authors.

%
Table~\ref{transfer_FB} shows the effectiveness of our attacks at attacking the considered graph-based classification methods on Facebook. The results on Enron, Epinions, and Twitter are shown in Table~\ref{transfer_Enron}, Table~\ref{transfer_Epinions}, and Table~\ref{transfer_Twitter} in Appendix, respectively. We make several observations.
First, our attacks are the most effective when using LinLBP to detect the target nodes. Specifically, when using the same method to select the target nodes and same cost type, our attacks achieve the largest FNR for the target nodes when the detector uses LinLBP. This is reasonable because our attacks generate the modified edges based on LinLBP. Second, our attacks can transfer to other graph-based classification methods, though the transferability varies among different graph-based classification methods. Third, our attacks can better transfer to other collective classification methods than other graph neural network methods in most cases. For instance, with Equal cost and CC, our attacks achieve $\geq 0.92$ FNRs for collective classification methods, while our attacks achieve 0.85 and 0.54 FNRs for LINE and GCN, respectively. The reason is collective classification methods and graph neural network methods use different mechanisms to leverage the graph structure. Fourth, our attacks have the least transferability to GCN, i.e., our attacks achieve the smallest FNRs for GCN. We speculate the reason is that GCN uses sophisticated graph convolution to exploit the higher order correlations in the graph structure, and thus is more robust to the modified edges our attacks generate.

\subsection{Comparing with State-Of-The-Art Attack} 
\label{compare_attack}

%

We compare our attacks with state-of-the-art attack called Nettack~\cite{zugner2018adversarial} designed for GCN. 



\myparatight{Nettack~\cite{zugner2018adversarial}}  Nettack was designed to attack GCN~\cite{kipf2017semi}. Nettack trains a surrogate linear model of GCN. Then, Nettack defines a graph structure preserving perturbation, which constrains that the node degree distribution of the graph before and after attack should be similar. Nettack has two versions. One version only modifies the graph structure, and the other version also modifies the node attributes when they are available. We adopt the first version that modifies the graph structure. We obtained the publicly available source code of Nettack from the authors.
Note that Nettack assumes  Equal cost  for modifying connection states between nodes.

\myparatight{Our attack} It is designed based on LinLBP.  We also assume {Equal cost} to provide a fair comparison with Nettack.

\begin{table}[tbp]\renewcommand{\arraystretch}{1.4}
 \vspace{-6mm}
 \centering
 \ssmall
 \caption{Comparing FNRs of our attack and Nettack for different graph-based classification methods on Facebook.} 
 \begin{tabular}{|c|c|c|c|c|c|c|c|}
 \hline
  \textbf{Method} &   \textbf{GCN}        &    \textbf{LINE}    &  \textbf{RW}   &  \textbf{LBP}  &    \textbf{JWP} &    \textbf{LinLBP}    &   \textbf{Time}     \\ \hline
 \textbf{No attack} &      0.05   	& 	0.01  & 0.03 &  0.01 & 0 & 0 	&  0 sec \\ \hline
 \textbf{Nettack} &   {\bf 0.64}  &  0.58 & 0.33 & 0.28 & 0.13 & 0.22 & 9 hrs    \\ \hline
 \textbf{Our attack} &  0.54 & 	{\bf 0.85}	& {\bf 0.92} & {\bf 0.92} & {\bf 0.93} & \bf{0.94}   & \bf{10 secs} \\ \hline
 \end{tabular}
 \label{compare_STA}
 \vspace{-4mm}
 \end{table}

Table~\ref{compare_STA} shows the FNRs and runtime of each attack for attacking/transferring to different graph-based classification methods on Facebook, where the target nodes are selected by CC. \ALAN{Note that, as LINE outperforms DeepWalk and node2vec without attacks (see Table~\ref{transfer_FB}), we  only show the transferability results of these attacks to LINE for conciseness.} We only show results on Facebook, because Nettack is not scalable to the other three graphs. We note that the compared attacks have the same costs, e.g., they all modify $K$ edge states for each target node.  We have several observations. 

First, for attacking GCN, Nettack is the most effective, i.e., Nettack achieves the highest FNR for attacking GCN. This is because Nettack is specifically designed for GCN. Our attack can effectively transfer to GCN, though our attack is slightly less effective than Nettack at attacking GCN (0.54 vs. 0.64). 
Second, for attacking collective classification methods, i.e., RW, LBP, JWP, and LinLBP, our attack is much more effective than Nettack. Specifically, for all collective classification methods, our attack achieves FNRs above 0.90. However, Nettack only achieves FNRs $\leq 0.33$. The reason is that our attack is specifically designed for LinLBP and can well transfer to other collective classification methods. Third, for attacking LINE, our attack is more effective than Nettack. Specifically, our attack achieves a FNR of 0.85, while Nettack achieves a FNR of 0.58.  Fourth, our attack is orders of magnitude faster than Nettack.

\section{Discussion and Limitations}
\label{discussion}

\myparatight{Attacks to other graph-based classification methods} In this paper, we focus on designing attacks against LinLBP. The reason is that LinLBP assigns the same weight to all edges, making our formulated optimization problem easier to solve. In RW-based methods and JWP, the edge weights or propagation depend on the graph structure. Therefore, it is harder to optimize the adversarial matrix, because the gradient of the adversarial matrix also depends on the edge weights, which are implicit functions of the adversarial matrix.  However, it is an interesting future work to study specialized attacks for RW-based methods and JWP. 



\begin{table}[!t]
\centering
\small
\vspace{-6mm}
\caption{FPRs of our attacks on the three graphs with synthesized positive nodes for Equal cost.}
\label{FPR}
\begin{tabular}{|c|c|c|c|c|} 
 \hline
 { \textbf{Dataset}} &{\textbf{No attack}} & {\textbf{RAND}} & {\textbf{CC}}  &  { \textbf{CLOSE}} \\ \hline
 { \textbf{Facebook}} & 0.01 & 0.09 & 0.11 & 0.11 \\ \hline
 { \textbf{Enron}} & 0.01 & 0.04 & 0.03 & 0.03 \\ \hline
 { \textbf{Epinions}}& 0 & 0.10 & 0.10 & 0.10 \\ \hline
\end{tabular}
\vspace{-4mm}
\end{table}

\myparatight{Unavailability attacks} In this paper, we  focus on evasion attack that aims to increase the FNR of an attacker's target nodes. 
In practice, an attacker could also launch the \emph{unavailability attacks}. In particular, the attacker manipulates the graph structure to achieve a high \emph{False Positive Rate} (FPR).  A high FPR means a large amount of negative nodes are falsely classified as positive, which eventually makes the system unavailable and abandoned. 
\ALan{We note that although our attacks are specially designed to increase the FNR of attacker's target nodes, our attacks can also 
(slightly) increase the FPR. Table~\ref{FPR} shows the FPRs of our attacks on the three graphs with synthesized positive nodes. 
 However, it is still an interesting future work to design attacks to specifically increase the FPR. }

\myparatight{Poisoning training dataset} Our work focuses on manipulating the graph structure to attack graph-based classification methods. An attacker could also poison the training dataset to attack graph-based classification methods. Specifically, in some applications, the training dataset is dynamically updated and obtained via crowdsourcing. For instance, in 
fraudulent user 
detection on social networks, a user could report other users as 
fraudulent users. Thus, an attacker could flag normal users as 
fraudster, which poison the training dataset. It is an interesting future work to poison the training dataset and compare the robustness of different graph-based classification methods against
training dataset poisoning attacks.

\myparatight{Countermeasures} Our work focuses on attacks to graph-based classification methods. 
We discuss two possible directions for defenses, and leave detailed exploration of these defenses as future work. These defense ideas are inspired by existing defenses against adversarial examples (e.g.,~\cite{goodfellow2014explaining,Papernot16Distillation,MagNet}) and data poisoning attacks (e.g.,~\cite{barreno2010security,Jagielski18,Suciu18}). Generally speaking, we could \emph{prevent} and/or \emph{detect} our attacks. In the prevention direction, we could explore new graph-based classification methods that are more robust to graph structure manipulation by design. In the detection direction, we could design methods to detect the inserted fake edges and deleted edges. For instance, we could extract features for each node pair and train a binary classifier, which predicts whether the connection state between the node pair was modified or not. However, when designing such a detection method, we should also consider an adaptive attacker who strategically adjusts its attacks to evade both a graph-based classification method and the detector. 

Moreover, an attacker could also analyze the local structure of nodes to detect positive nodes. In particular, our attacks may change the local structure of positive nodes after inserting/deleting edges for them. For example, Table~\ref{avecc} shows the average clustering coefficients of the target nodes before and after our attacks on the three social graphs with synthesized positive nodes with Equal cost. We observe that the  clustering coefficients slightly decrease after our attacks. This is because our attacks add neighbors to a target node, but these neighbors may not be connected. It is an interesting future work to explore the feasibility of such local structures to detect positive nodes and our attacks. 

\begin{table}[!t]
\centering
\small
\caption{Average clustering coefficients of target nodes before and after our attacks on the three graphs with synthesized positive nodes for Equal cost.}
\label{avecc}
\begin{tabular}{|c|c|c|c|} 
 \hline
 { \textbf{Dataset}} & {\textbf{RAND}} & {\textbf{CC}}  &  { \textbf{CLOSE}} \\ \hline
 { \textbf{Facebook}} & {0.492/0.464} & {0.540/0.492} & {0.591/0.573} \\ \hline
 { \textbf{Enron}}& {0.567/0.549} & { 0.634/0.632} & {0.559/0.554} \\ \hline
 { \textbf{Epinions}}& {0.433/0.417} & {0.511/0.497} & {0.473/0.472} \\ \hline
\end{tabular}
\vspace{-4mm}
\end{table}

\ALan{
\myparatight{Attacks to graph-level classification methods} In this paper, our attacks focus on evading node-level classification methods. In practice, graph-level classification methods have also been used for security analytics. For example, 
several approaches~\cite{cesare2010classification,gascon2013structural} proposed to analyze and classify control-flow graphs for malware detection. 
These approaches typically make predictions on the level of graphs and hence our attacks are not applicable. We acknowledge that it is an interesting future work to design attacks against graph-level classification methods. 
}


\section{Conclusion and Future Work}

We perform the first systematic study on attacks to collective classification methods via manipulating the graph structure. We propose a threat model to characterize the attack surface of collective classification methods, e.g., an attacker's background knowledge can be characterized along the Parameter, Training dataset, and Graph dimensions. We show that our attack can be formulated as an optimization problem, which can be solved approximately by the techniques we propose. 
Our 
evaluation results on a large-scale Twitter graph with real positive nodes and three graphs with 
synthesized positive nodes show that our attacks can make collective classification methods misclassify an attacker's target nodes; our attacks can transfer to graph neural network methods; our attacks do not require access to the true parameters of LinLBP, the true training dataset, and/or the complete graph; and our attacks outperform existing attacks for attacking collective classification methods and certain graph neural network methods. 

Interesting future work includes 1) designing specialized attacks to other graph-based classification methods, 2) adjusting our attacks as unavailability attacks, 3) designing training data poisoning attacks to graph-based classification methods, 4) designing attacks against graph-level classification methods, and 5) enhancing robustness of graph-based classification methods. 

\section*{Acknowledgements}
We would like to thank the anonymous reviewers for their constructive feedback.  This work was supported by the National Science
Foundation under grants No. 1937787 and 1937786. Any opinions, findings and conclusions or
recommendations expressed in this material are those of the
author(s) and do not necessarily reflect the views of the funding
agencies. 


{
{
\bibliographystyle{ACM-Reference-Format}
\bibliography{refs,refs_NDSS}
}}


\appendix




\begin{figure}[!htbp]
\center
\subfloat[Uniform cost]{\includegraphics[width=0.24\textwidth]{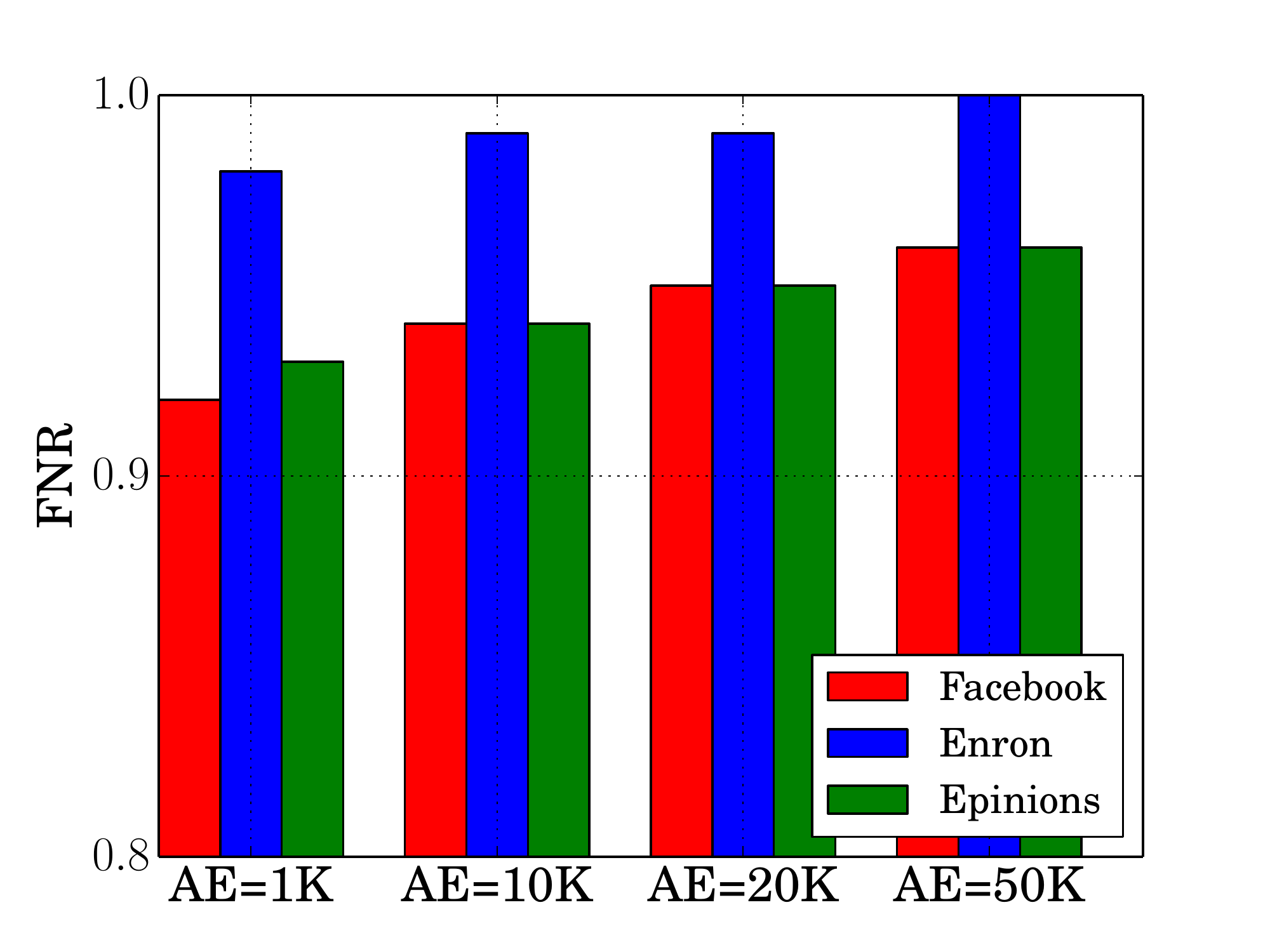}\label{}} 
\subfloat[Categorical cost]{\includegraphics[width=0.24\textwidth]{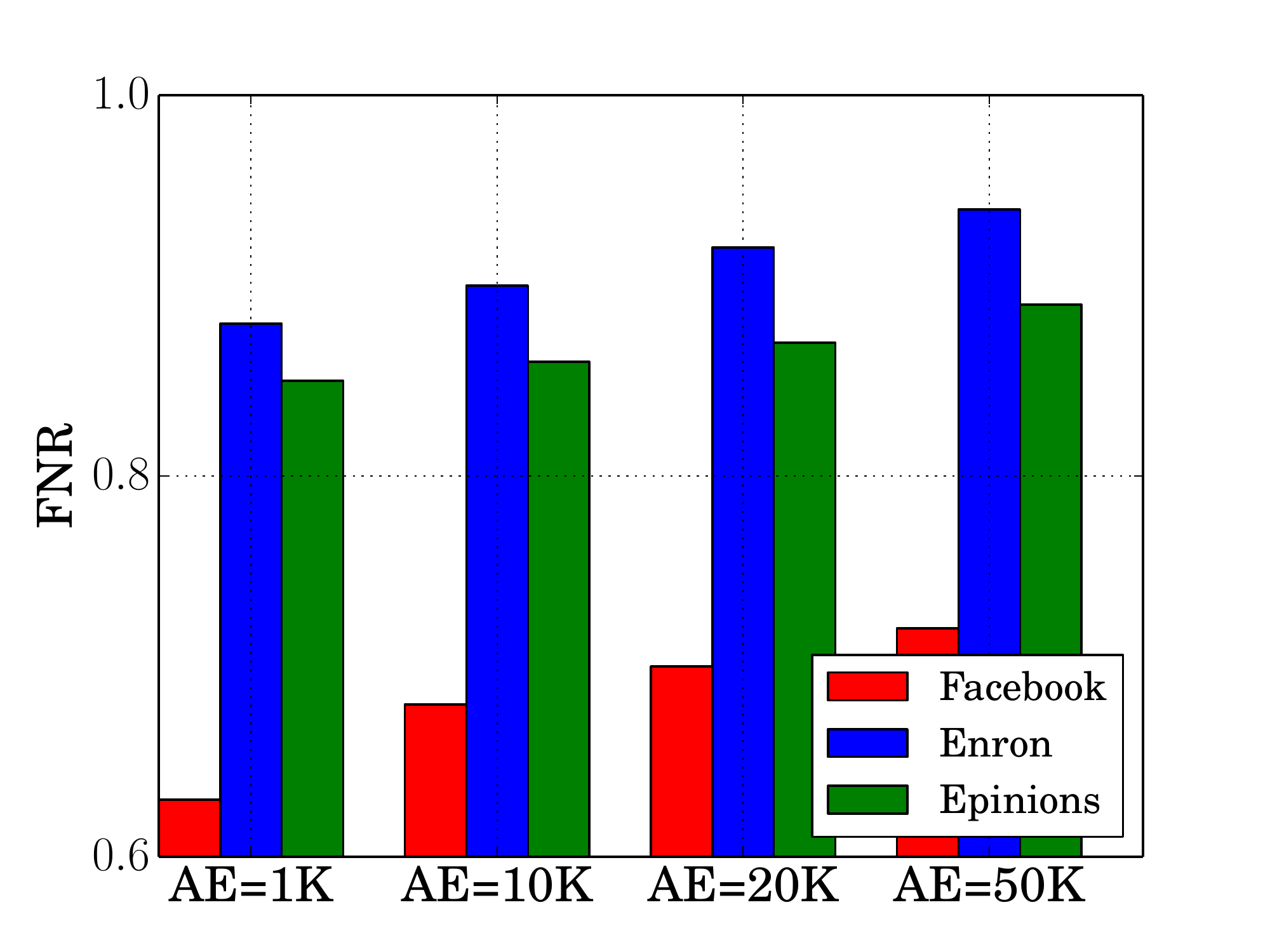}\label{}} 
\caption{Impact of $AE$ on the three social graphs with synthesized positive nodes with (a) Uniform cost and (b) Categorical cost.} 
\label{impact-AE-Uni-Cat}
\vspace{-2mm}
\end{figure}


\begin{figure}[!htbp]
\center
\subfloat[{Uniform cost}]{\includegraphics[width=0.24\textwidth]{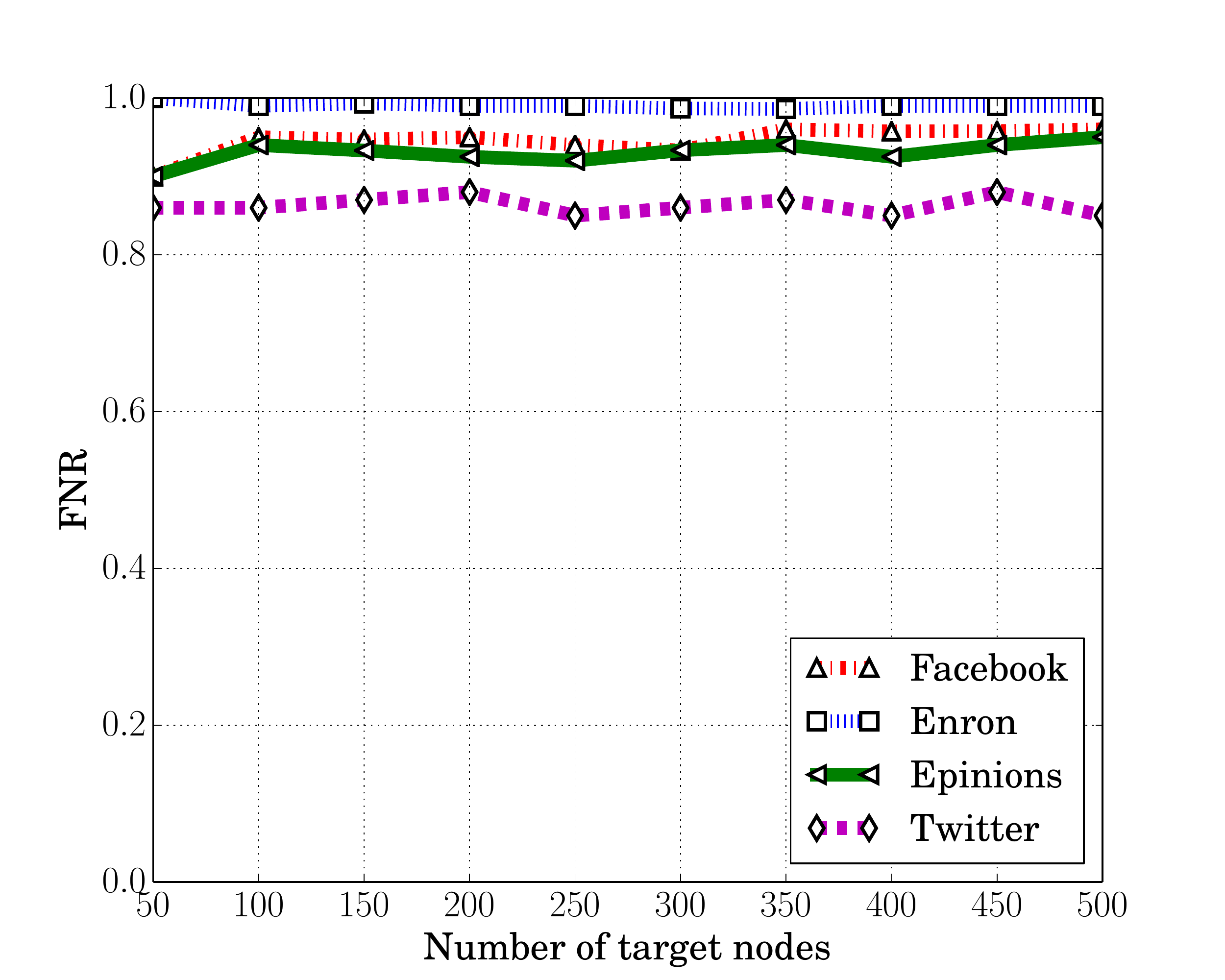} \label{target-uni}}
\subfloat[{Categorical cost}]{\includegraphics[width=0.24\textwidth]{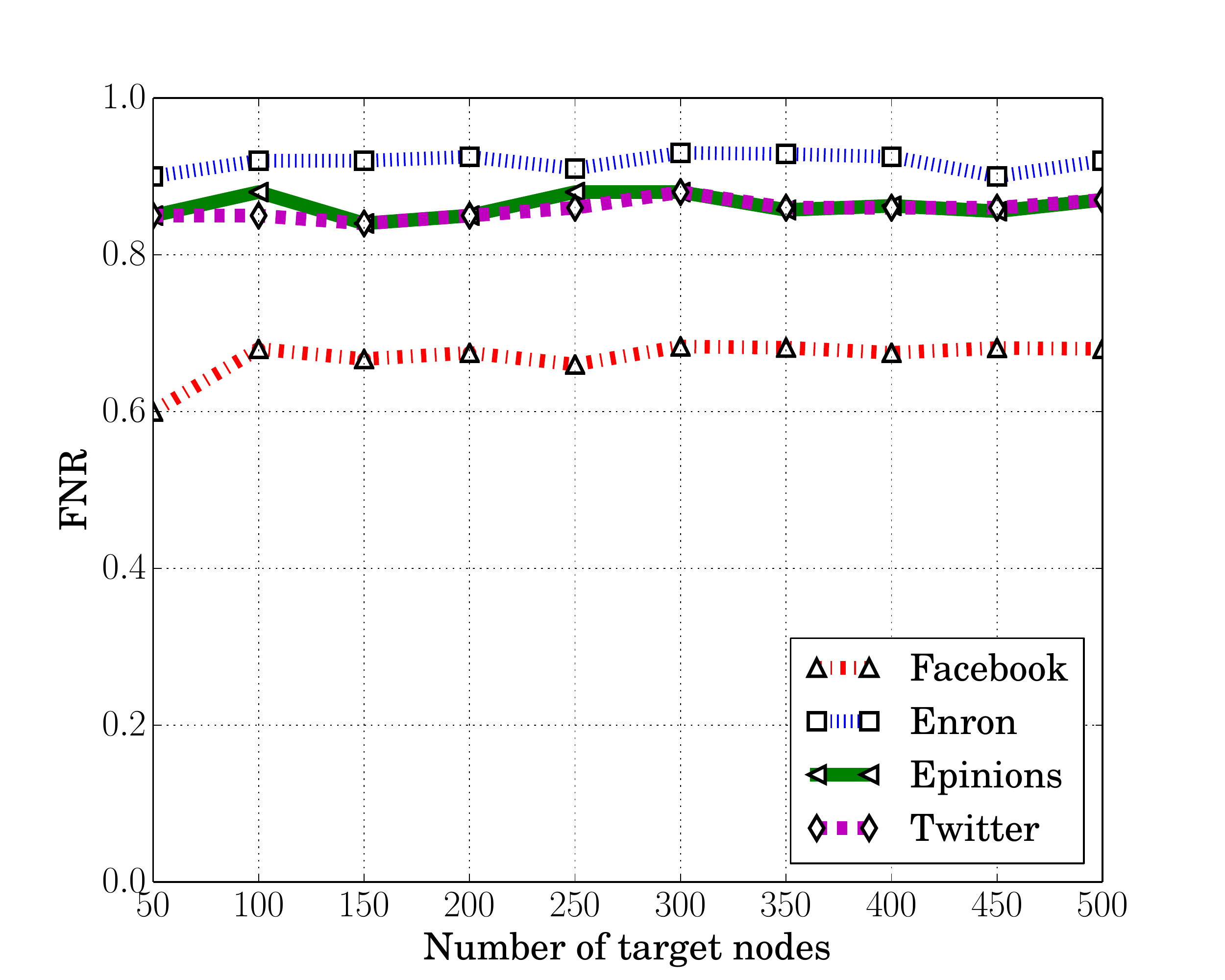} \label{target-cat}}
\caption{Impact of the number of target nodes on the three graphs with CC as  the method of selecting target nodes. (a) Uniform cost and (b) Categorical cost. 
}
\label{impact-attacker-nodes-other-cc}
\end{figure}

\begin{figure*}[!htbp]
\center
\subfloat[Facebook]{\includegraphics[width=0.24\textwidth]{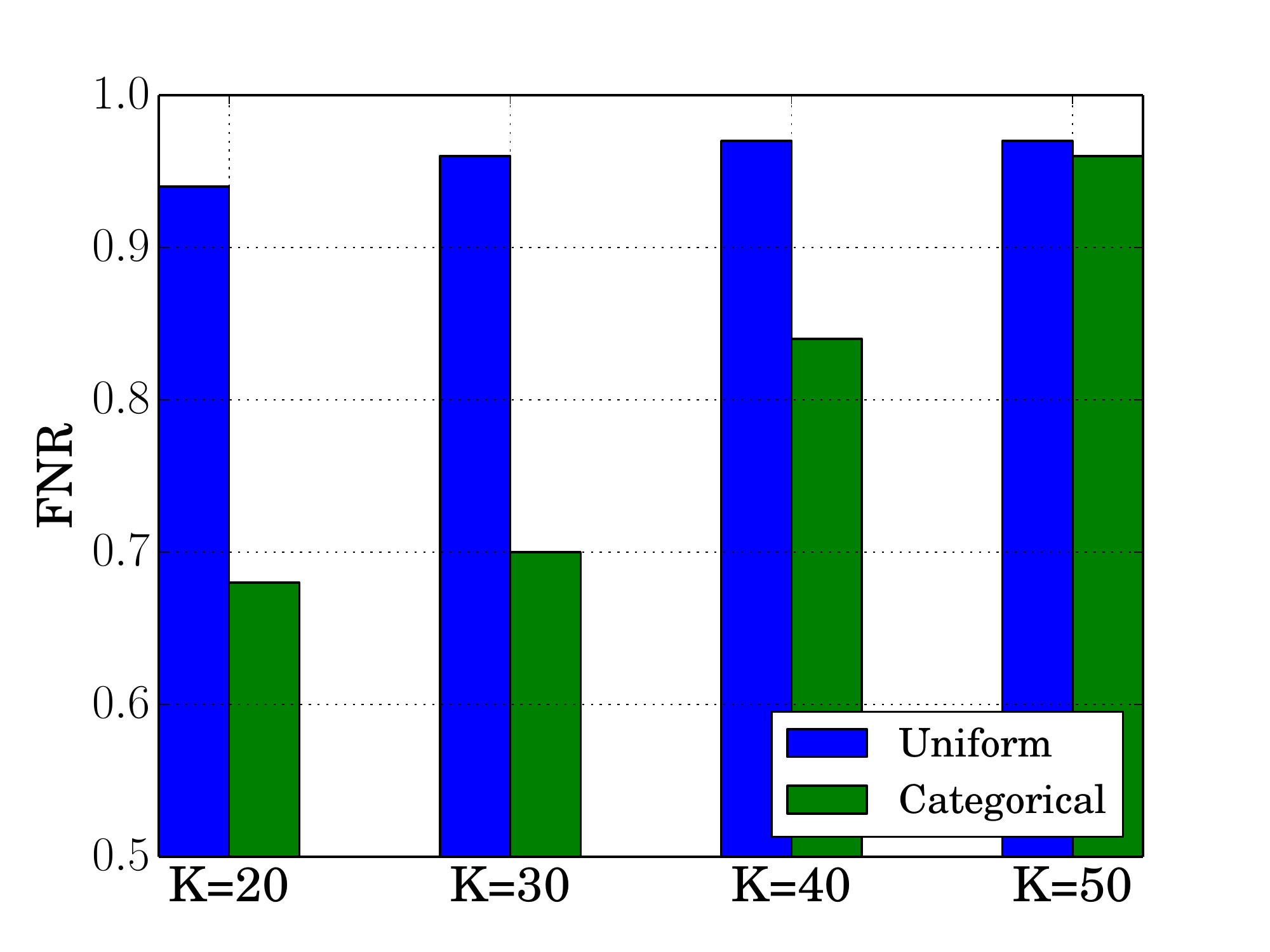}\label{}} 
\subfloat[Enron]{\includegraphics[width=0.24\textwidth]{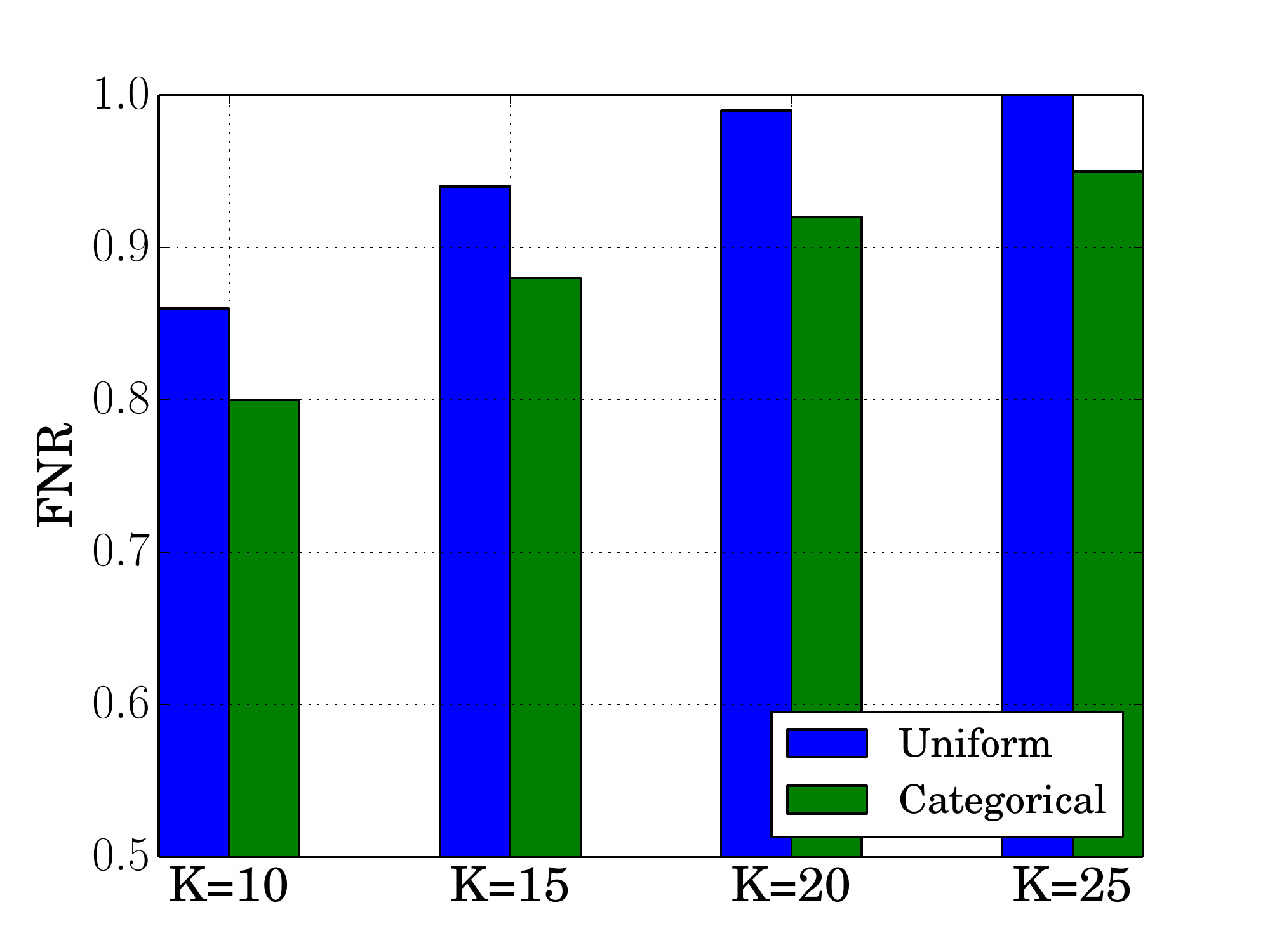}\label{}} 
\subfloat[Epinions]{\includegraphics[width=0.24\textwidth]{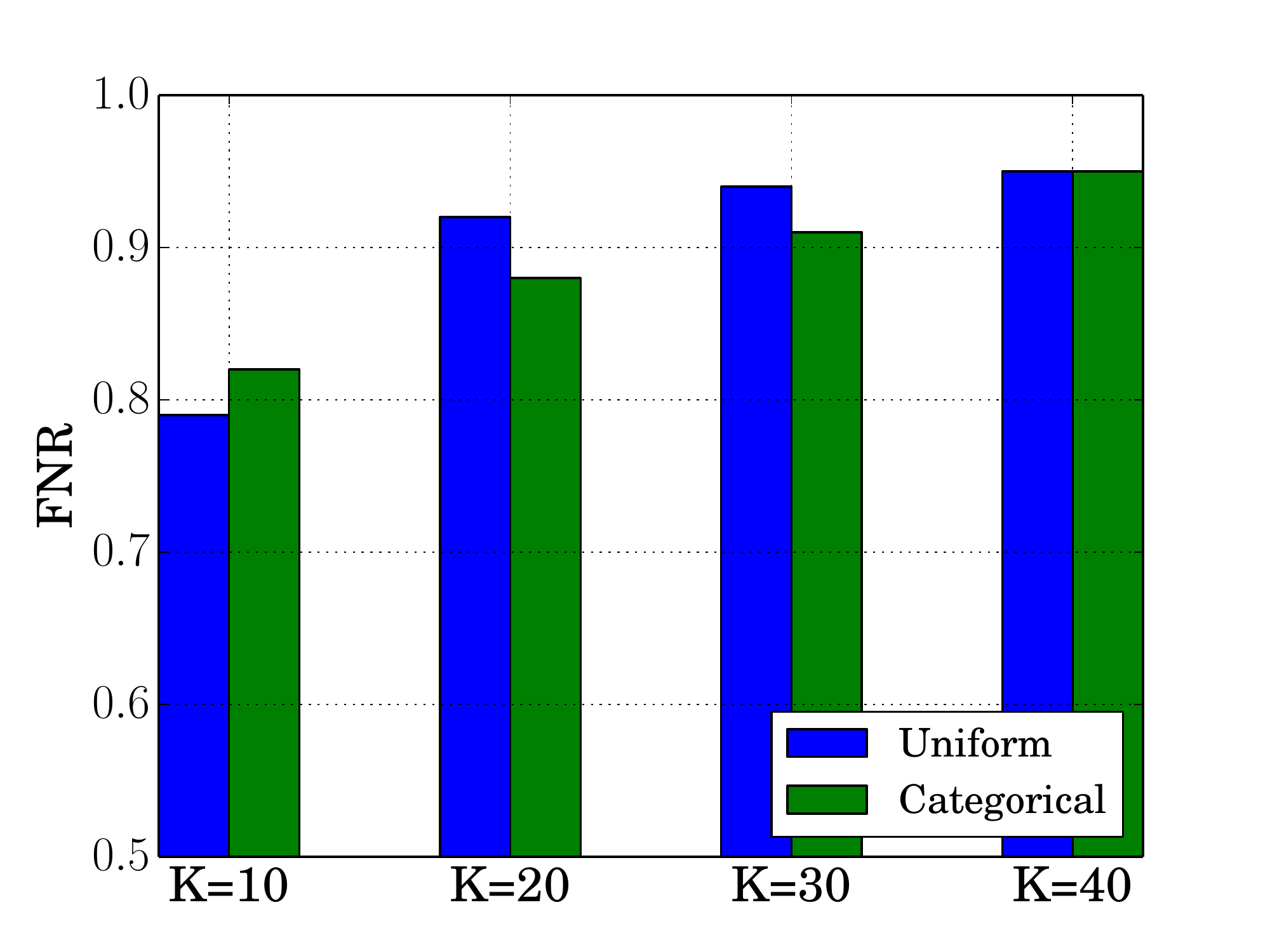}\label{}} 
\subfloat[Twitter]{\includegraphics[width=0.24\textwidth]{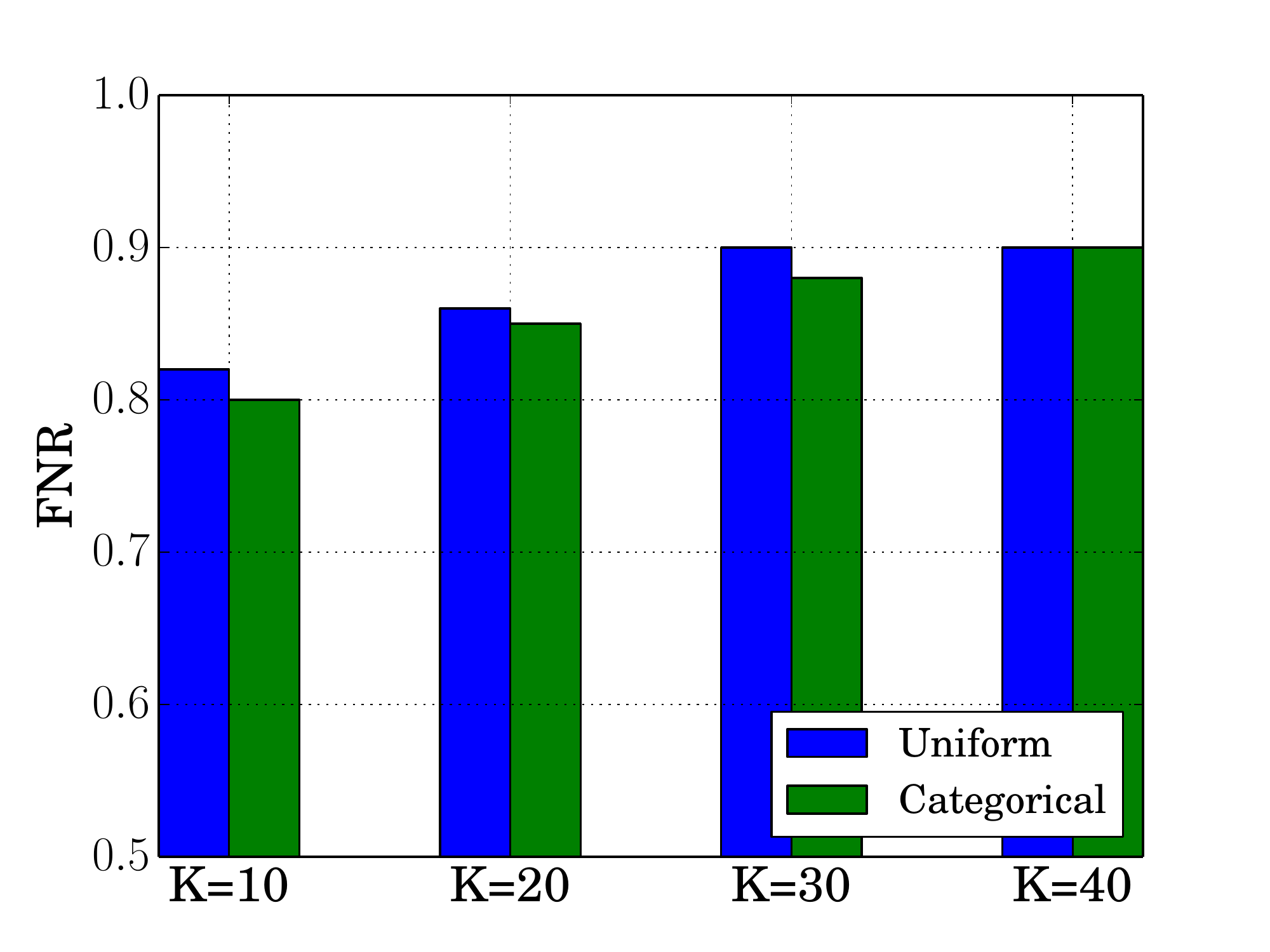}\label{}} 
\caption{Impact of $K$ on the four graphs with Uniform cost and Categorical cost and with CC as the method of selecting target nodes.} 
\label{impact-K-CC-Other}
\vspace{-2mm}
\end{figure*}

\begin{figure}[htbp]
\center
\subfloat[$\lambda$]{\includegraphics[width=0.24\textwidth]{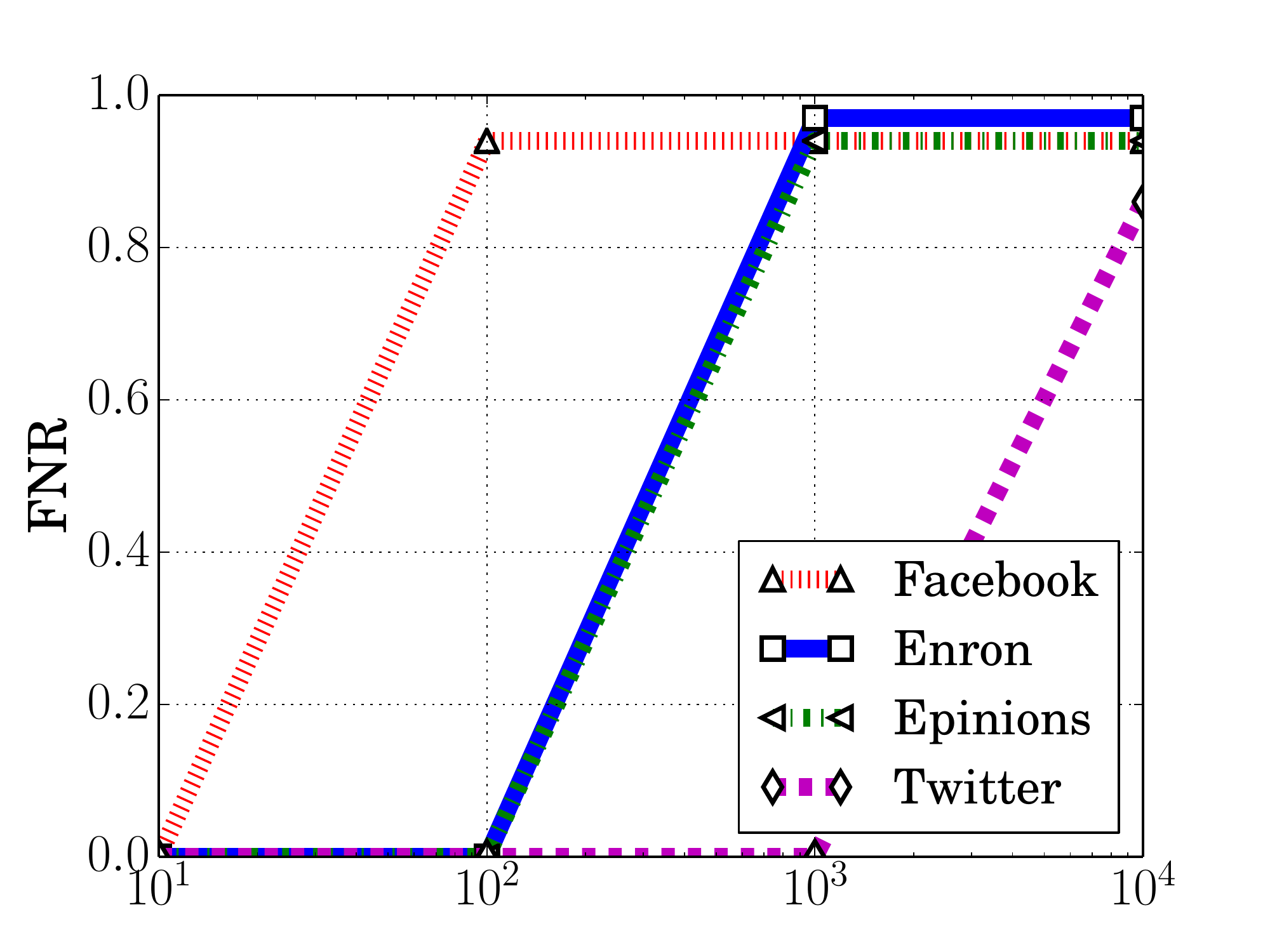}}
\subfloat[$\eta$]{\includegraphics[width=0.24\textwidth]{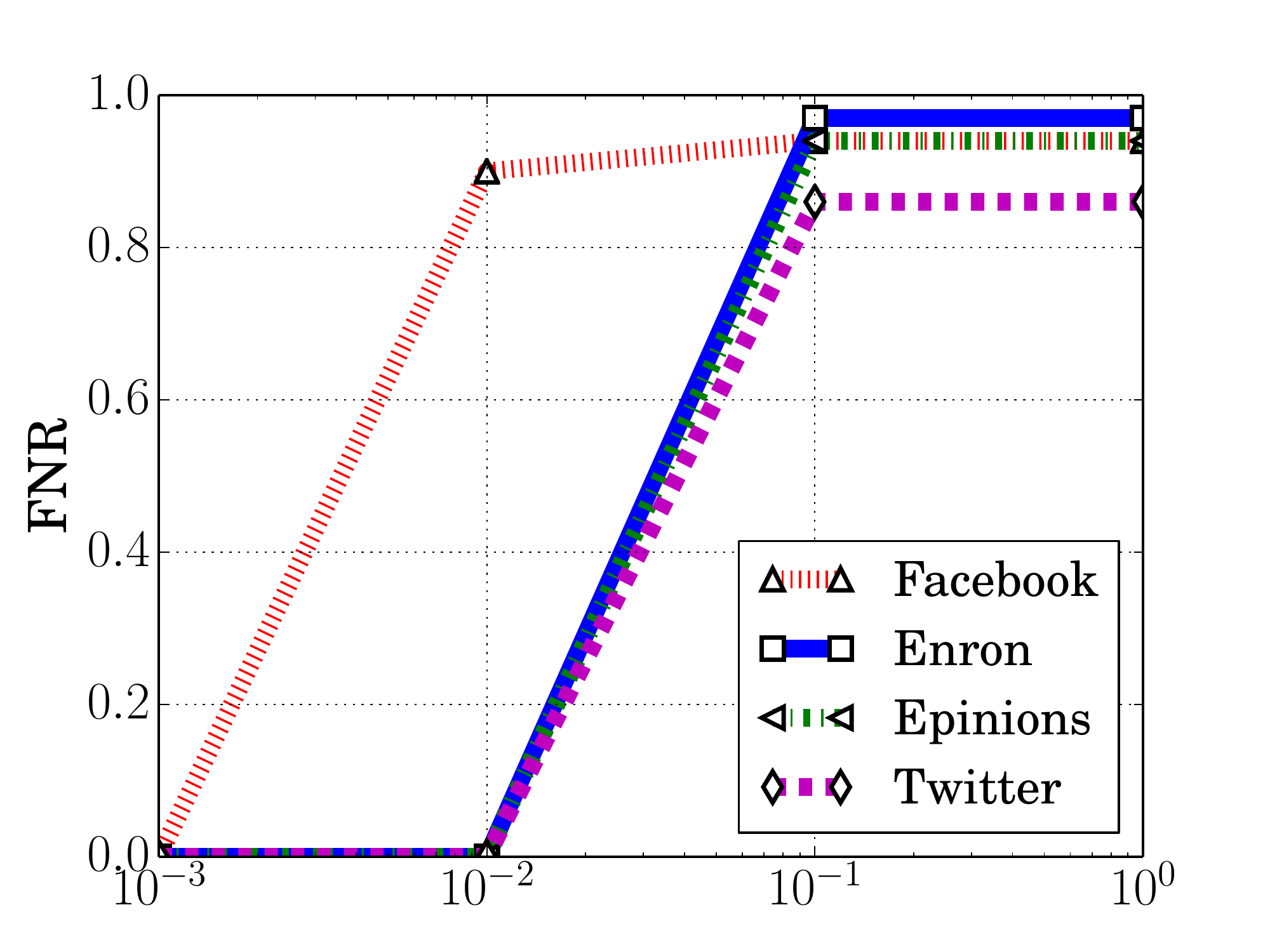}}
\caption{Impact of $\lambda$ and $\eta$ on our attacks with Uniform cost.}
\label{impact-lambda-eta-Uni}
\end{figure}

\begin{figure}[htbp]
\center
\subfloat[$\lambda$]{\includegraphics[width=0.24\textwidth]{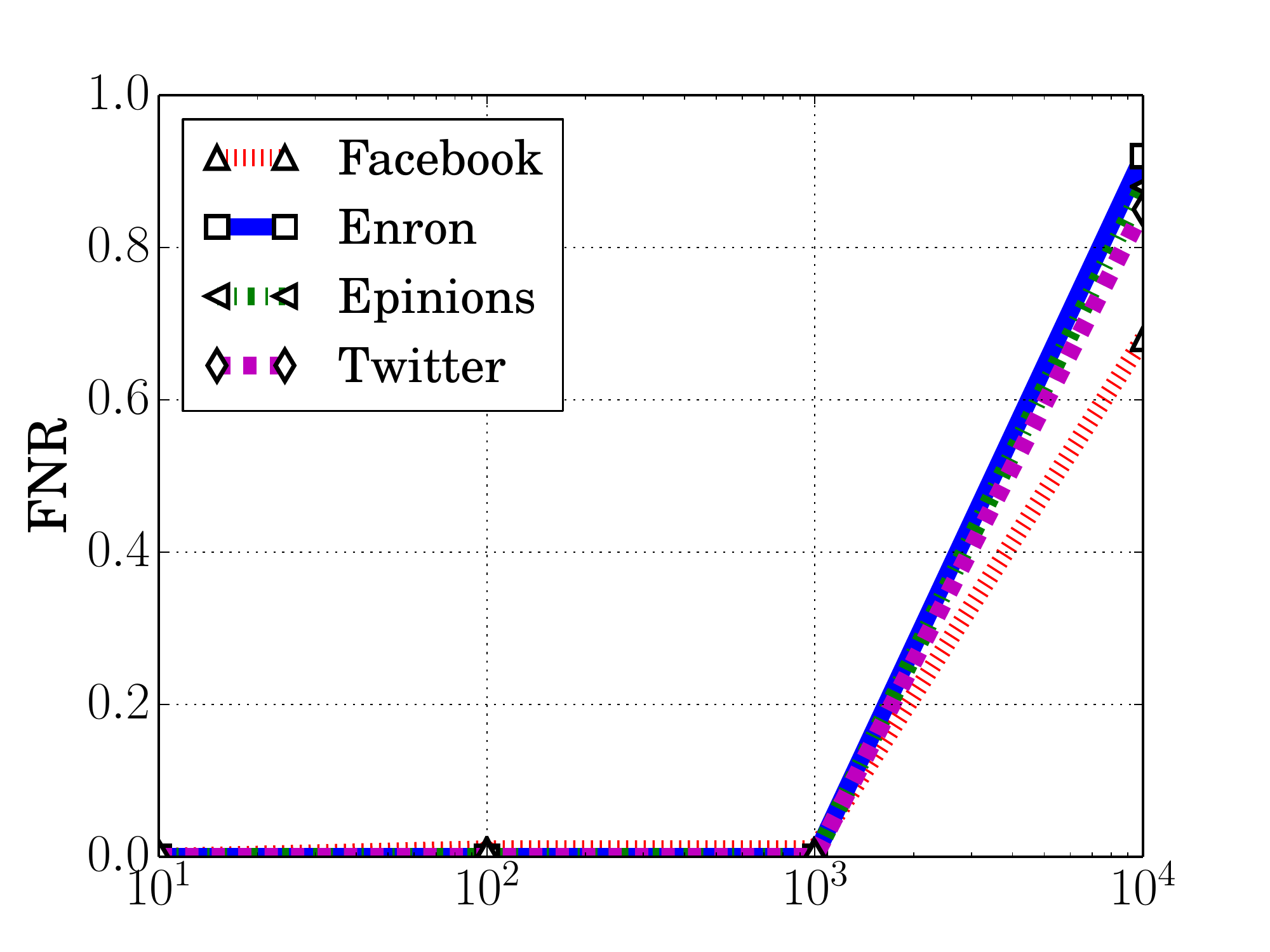}}
\subfloat[$\eta$]{\includegraphics[width=0.24\textwidth]{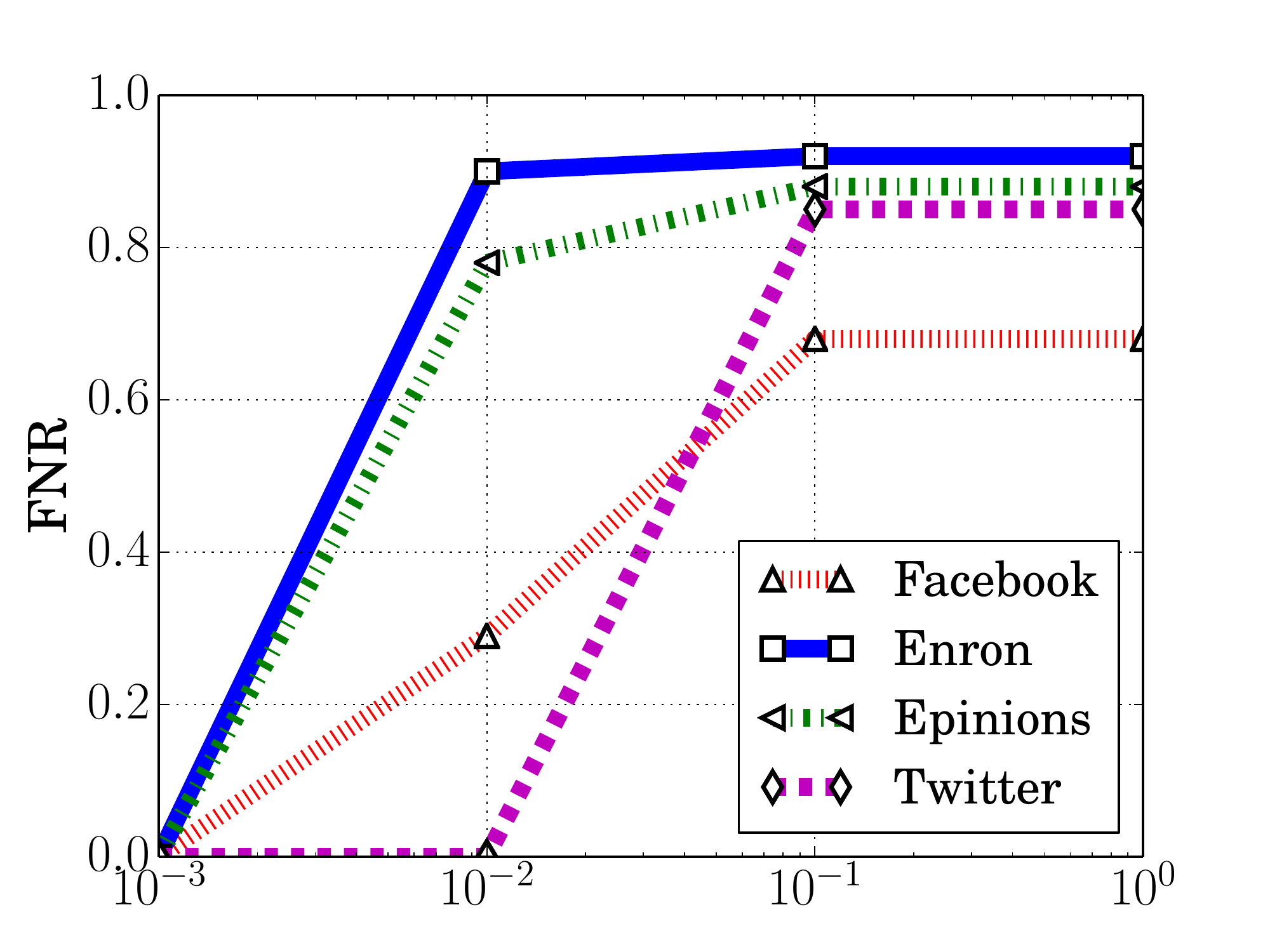}}
\caption{Impact of $\lambda$ and $\eta$ on our attacks with Categorical cost.}
\label{impact-lambda-eta-Cat}
\end{figure}

\begin{figure}[htbp]
\center
\subfloat[$\theta$]{\includegraphics[width=0.24\textwidth]{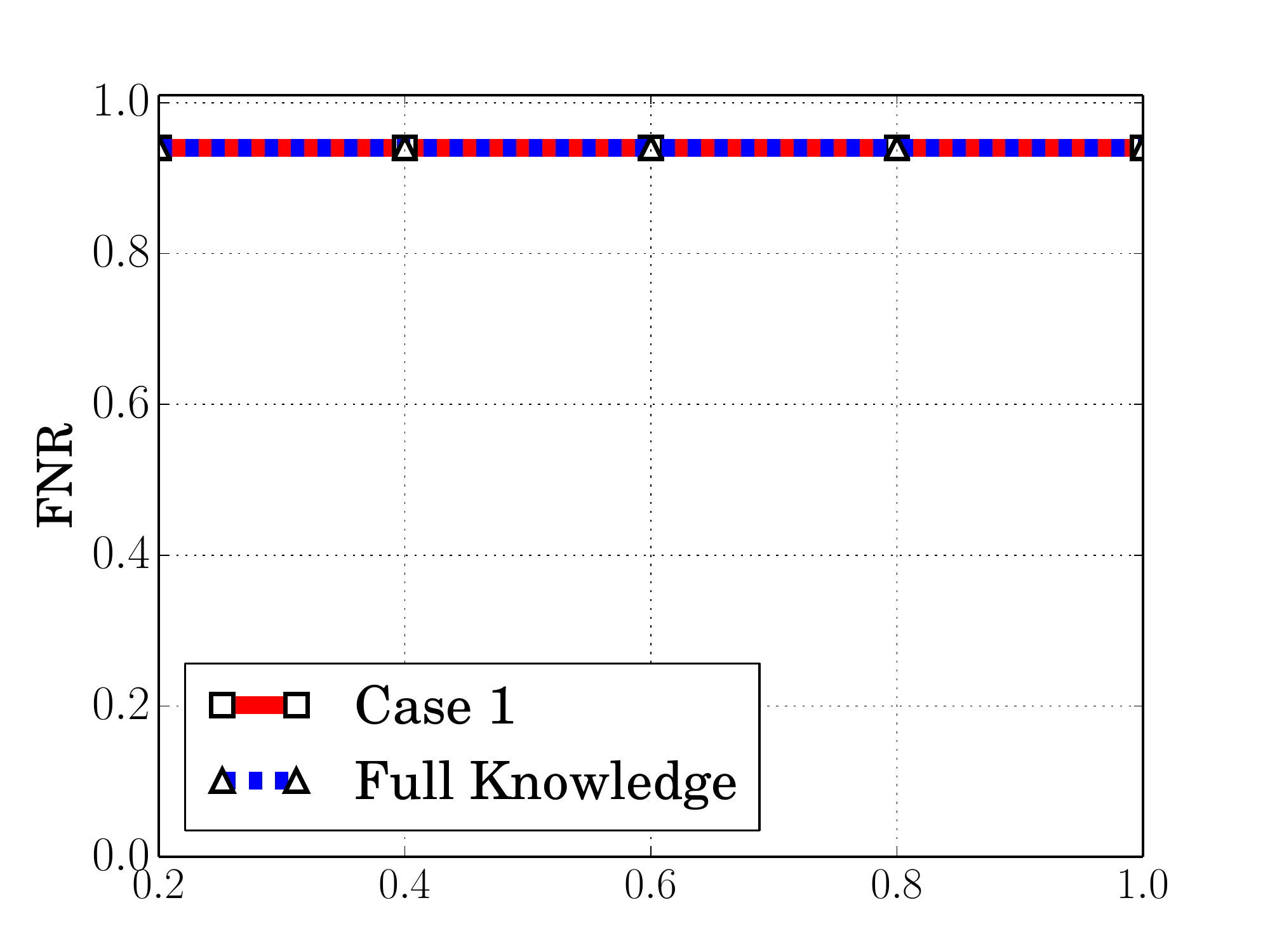}\label{impact-substitute-prior-FB}}
\subfloat[$w$]{\includegraphics[width=0.24\textwidth]{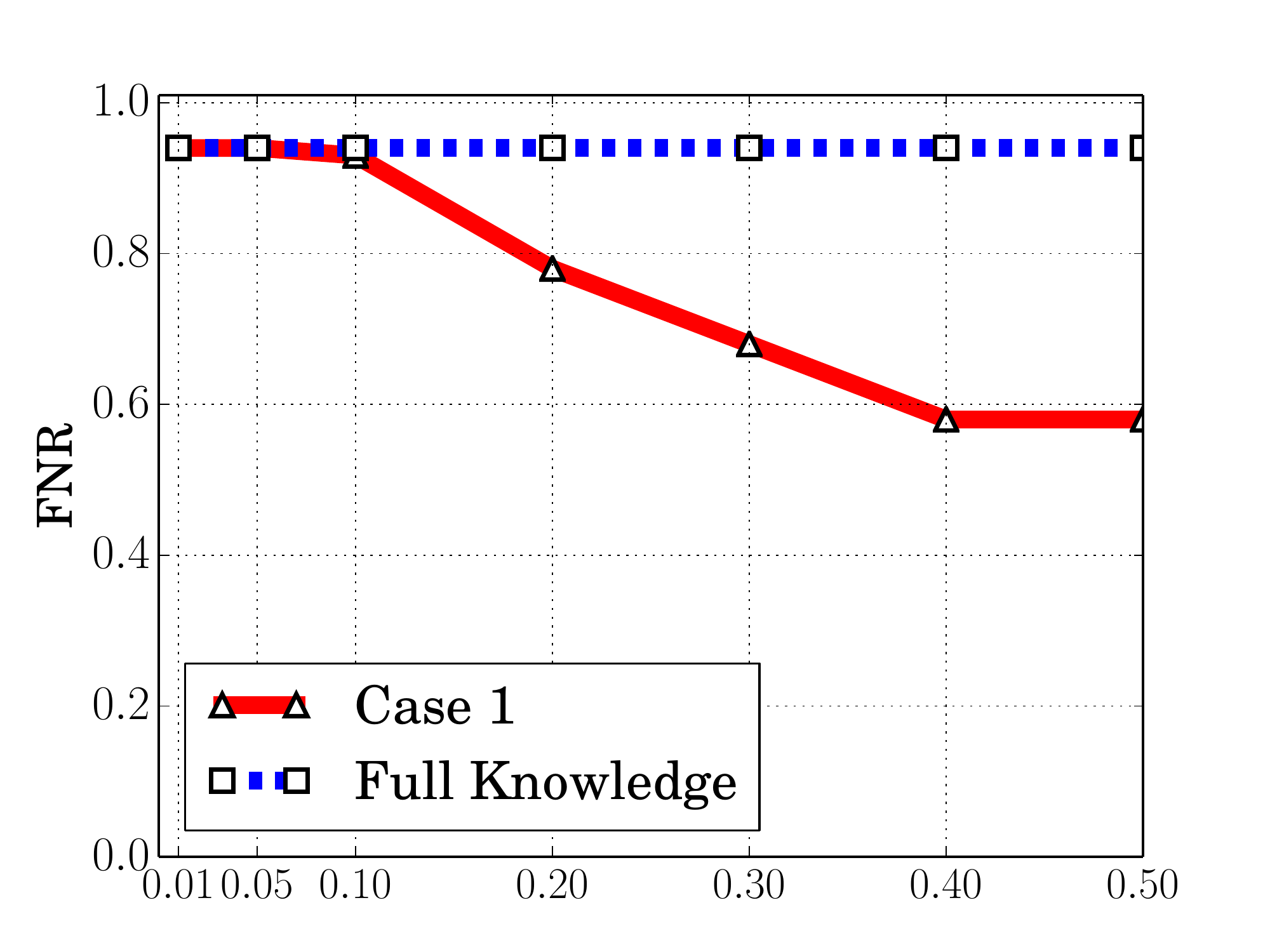}\label{impact-substitute-weight-FB}}
\caption{FNRs of our attacks with different substitute parameters of LinLBP on Facebook with Equal cost. 
}
\label{impact-substitute-parameter-FB}
\end{figure}

\begin{figure}[htbp]
\center
\subfloat[$\theta$]{\includegraphics[width=0.24\textwidth]{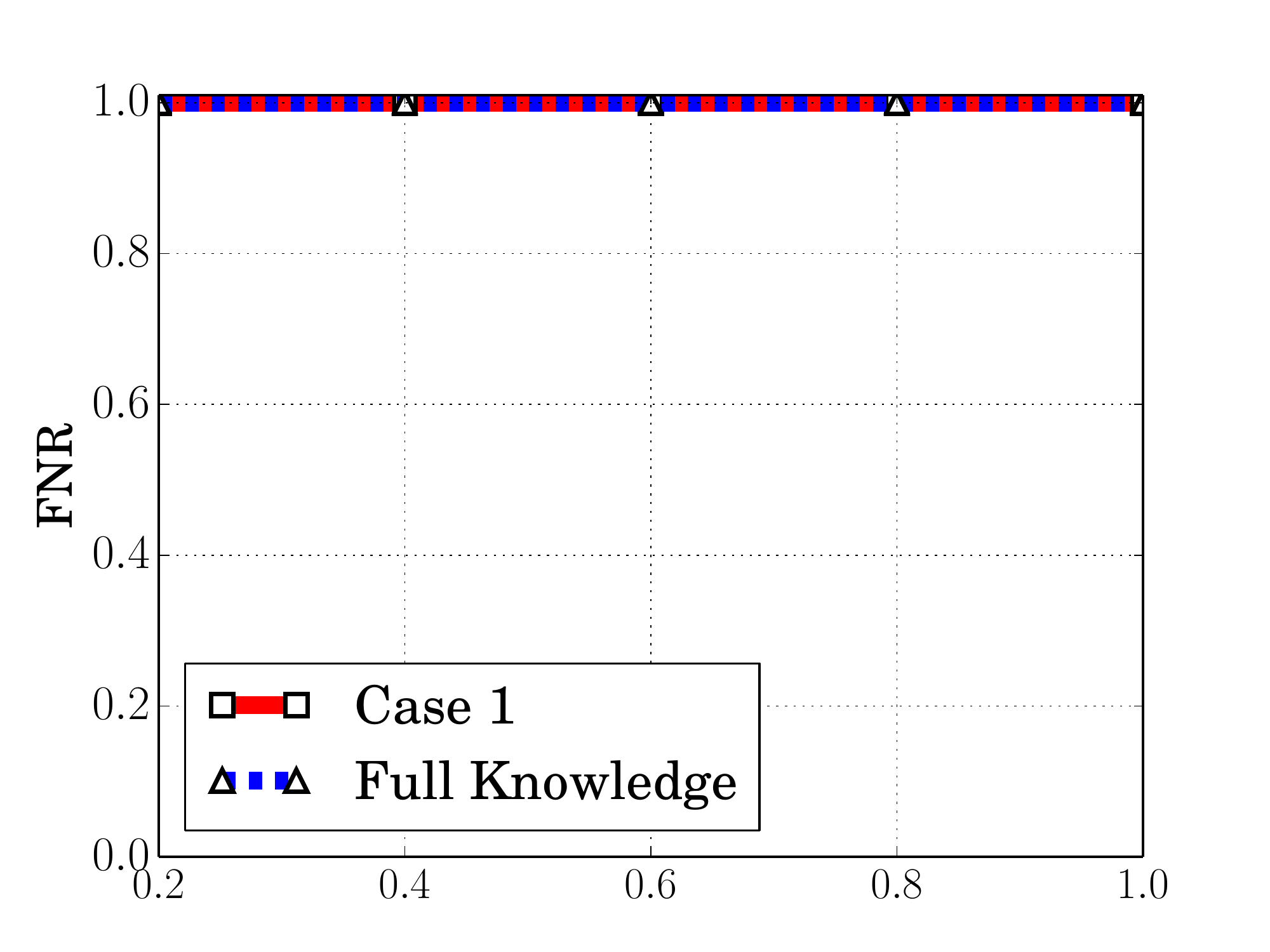}\label{impact-substitute-prior-Enron}}
\subfloat[$w$]{\includegraphics[width=0.24\textwidth]{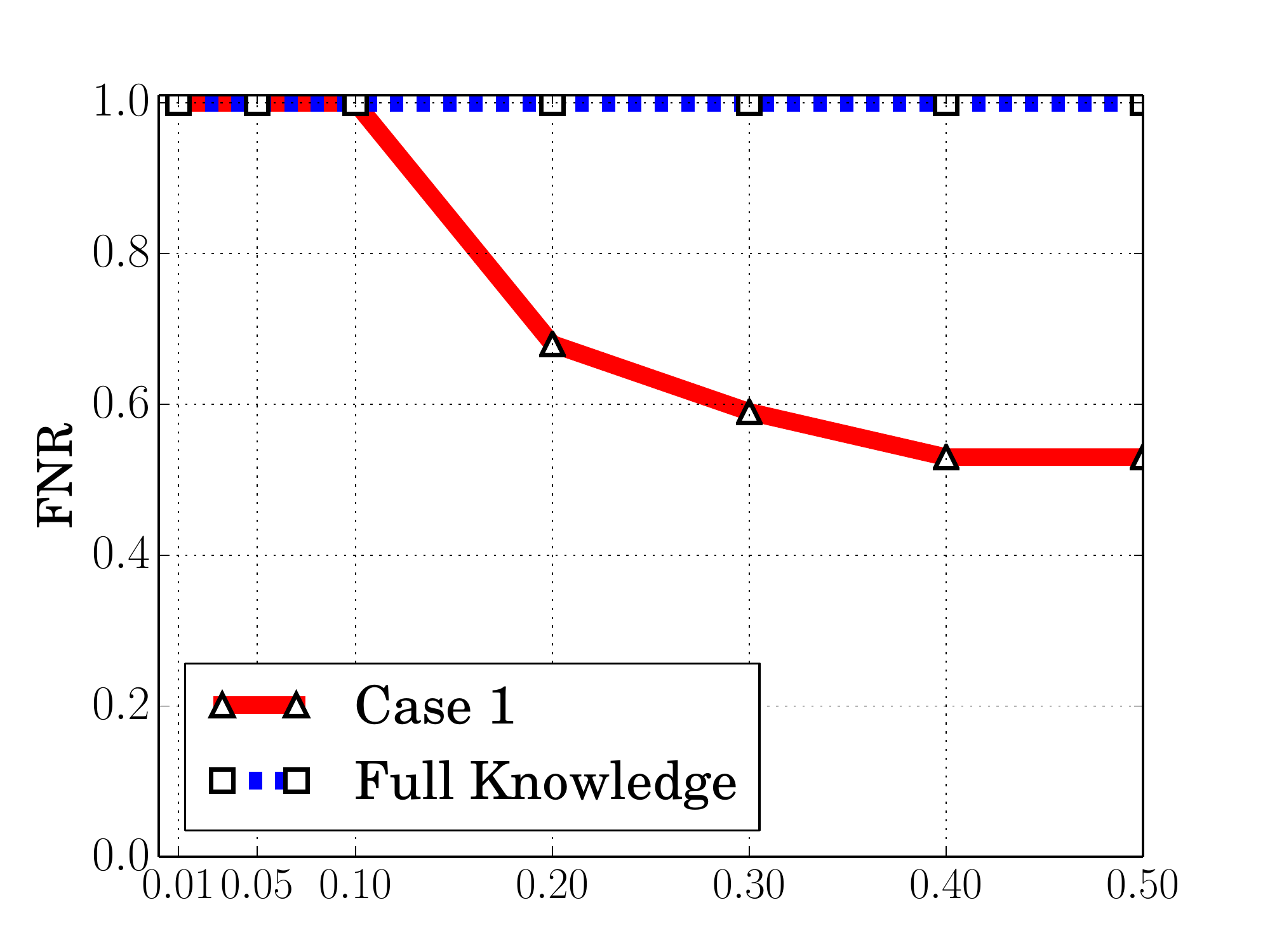}\label{impact-substitute-weight-Enron}}
\caption{FNRs of our attacks with different substitute parameters of LinLBP on Enron with Equal cost. 
}
\label{impact-substitute-parameter-Enron}
\end{figure}

\begin{figure}[htbp]
\center
\subfloat[$\theta$]{\includegraphics[width=0.24\textwidth]{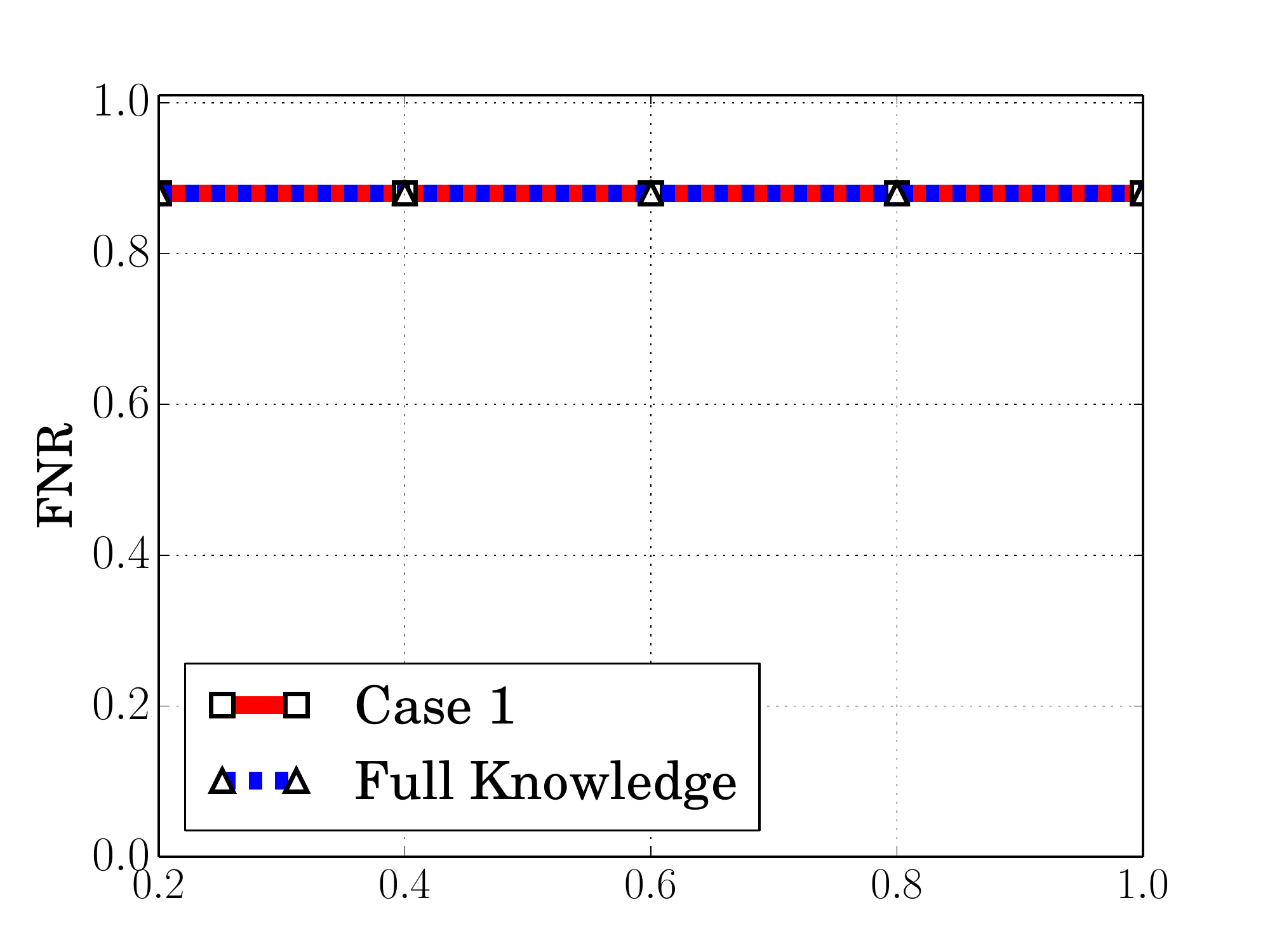}\label{impact-substitute-prior-Twitter}}
\subfloat[$w$]{\includegraphics[width=0.24\textwidth]{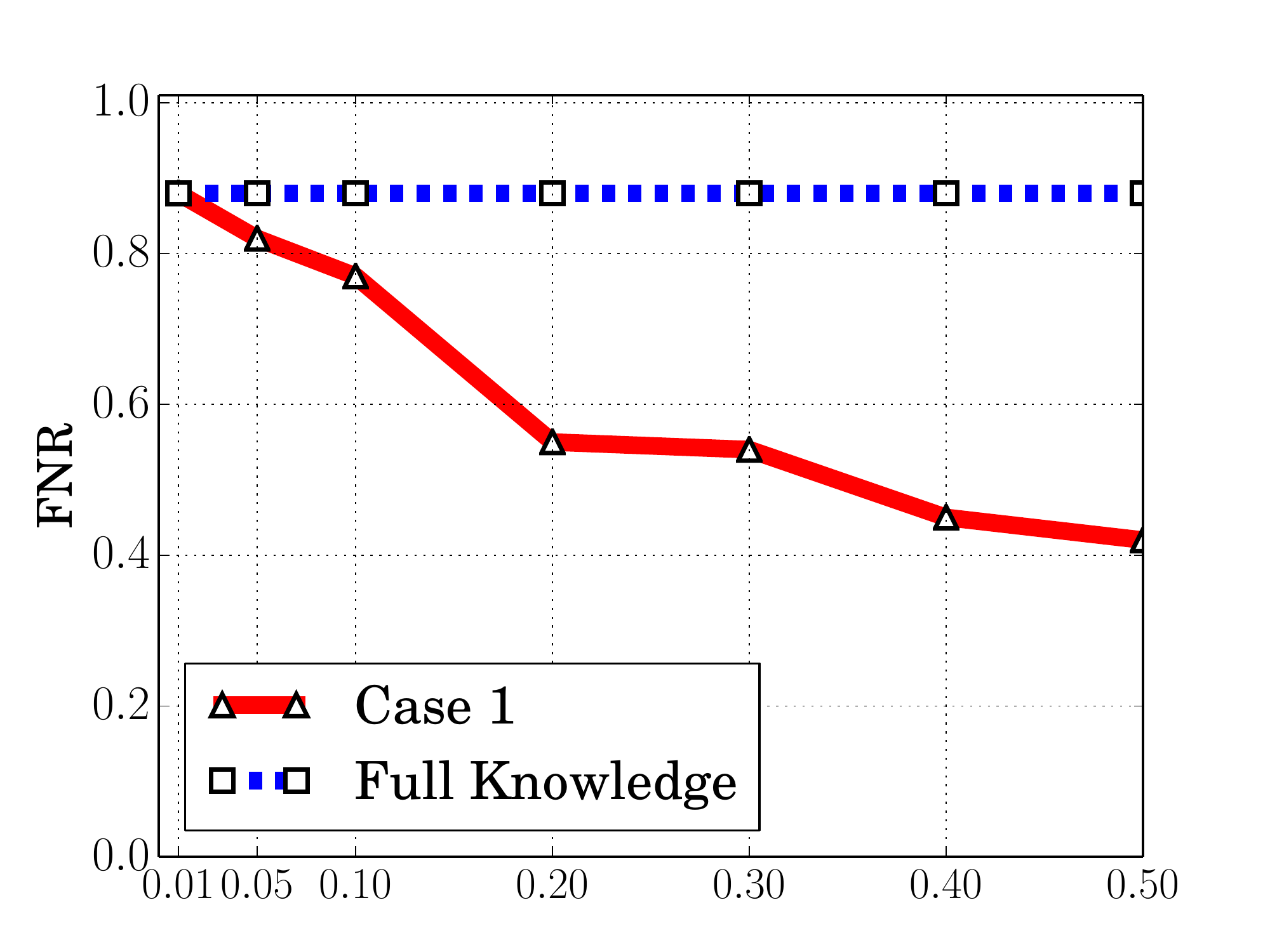}\label{impact-substitute-weight-Twitter}}
\caption{FNRs of our attacks with different substitute parameters of LinLBP on Twitter with Equal cost. 
}
\label{impact-substitute-parameter-Twitter}
\end{figure}

\begin{figure*}[!tbp]
\vspace{-6mm}
\center
\subfloat[Facebook]{\includegraphics[width=0.32\textwidth]{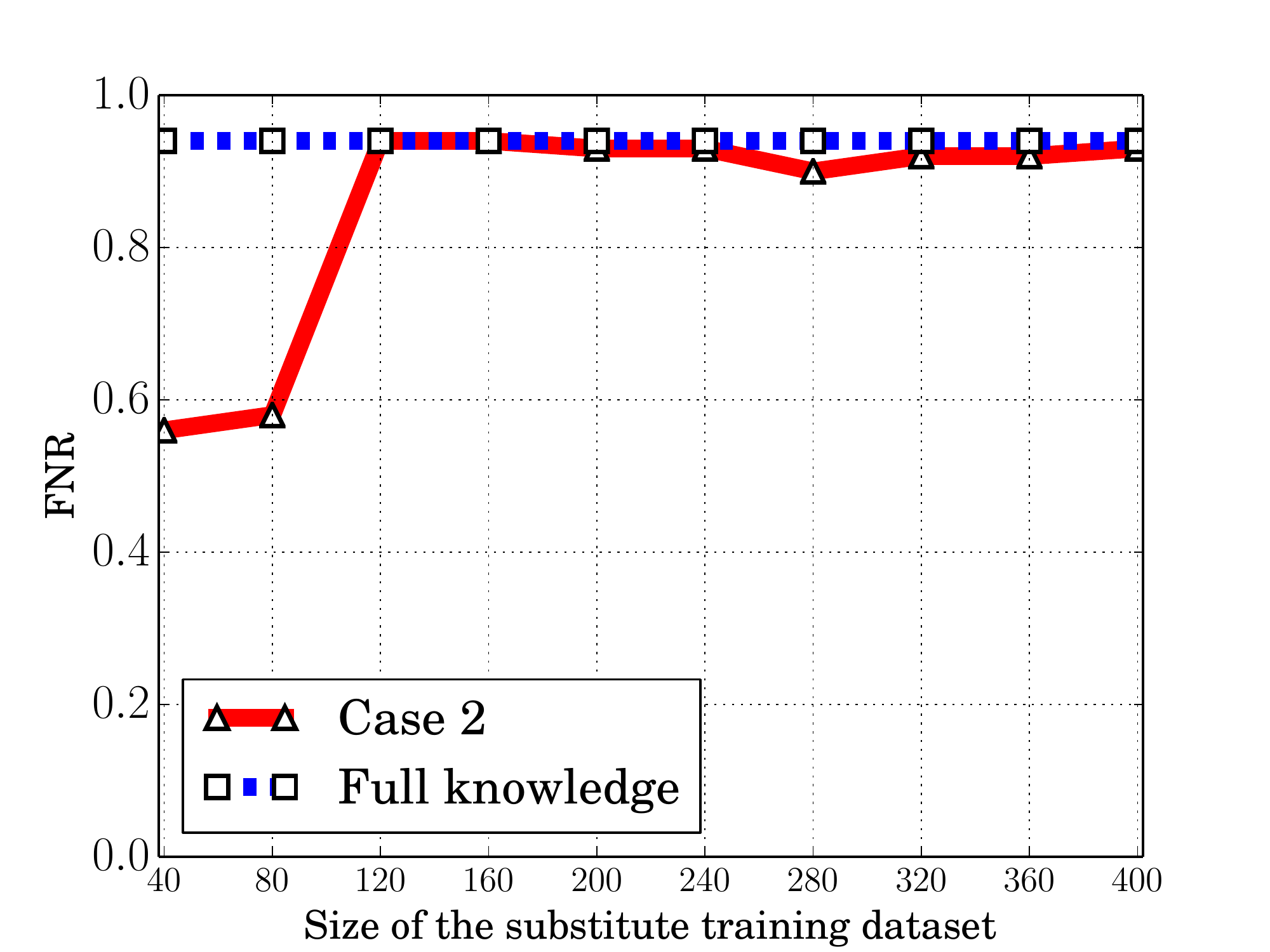} \label{substitute-FB}}
\subfloat[Enron]{\includegraphics[width=0.32\textwidth]{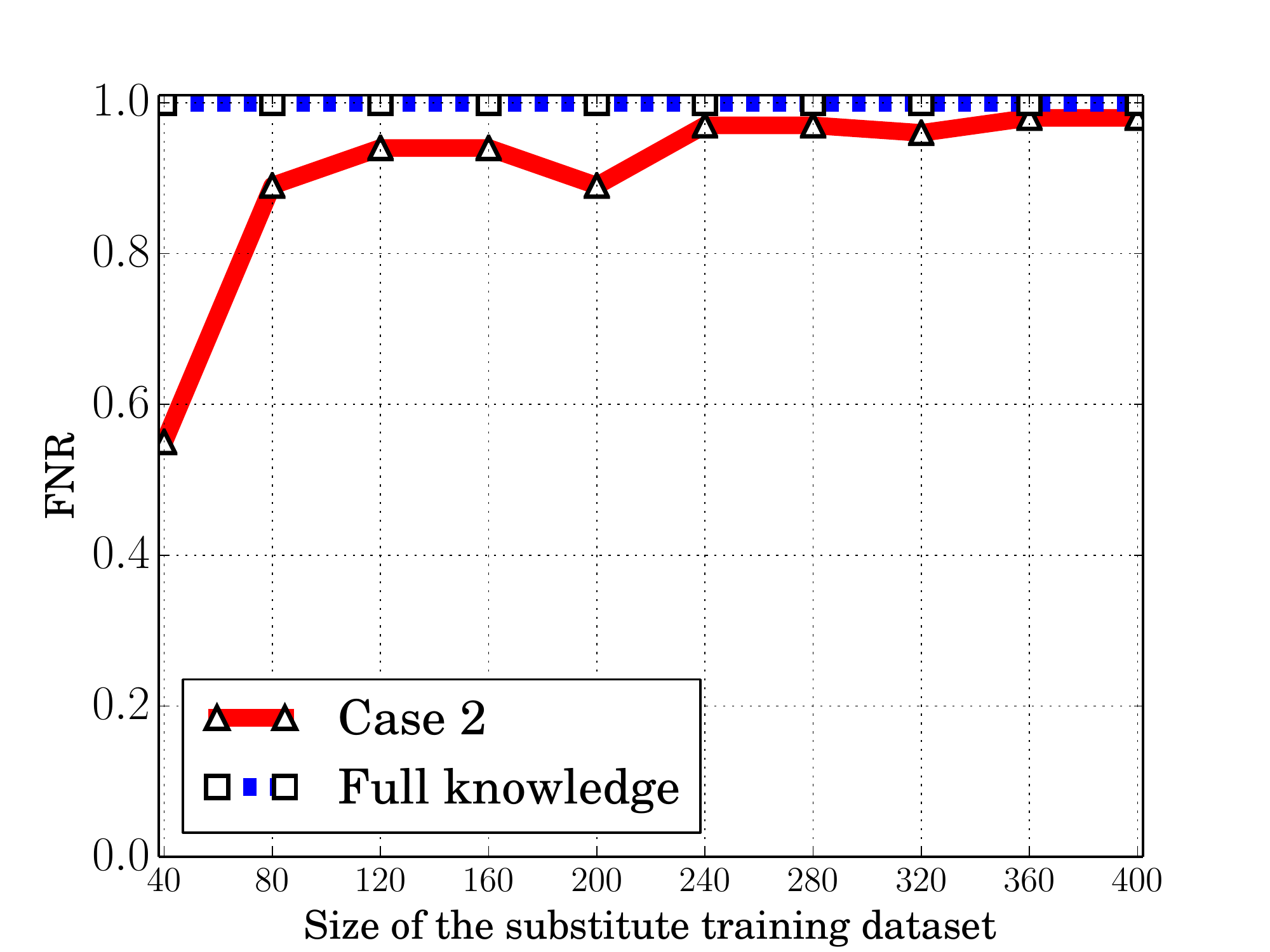}\label{substitute-Enron}}
\subfloat[Twitter]{\includegraphics[width=0.32\textwidth]{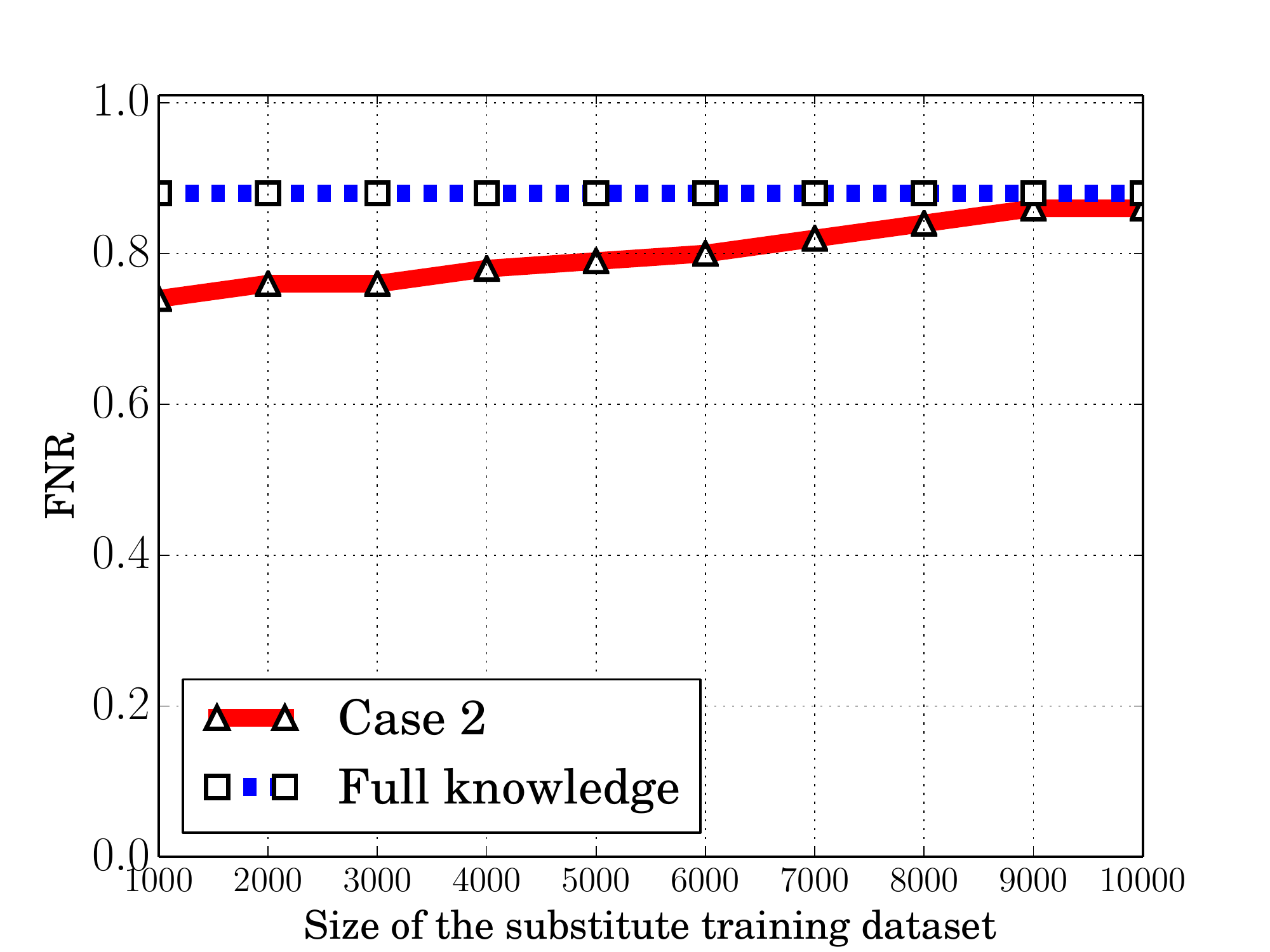}\label{substitute-Twitter}}
\caption{FNRs of our attacks vs. size of the substitute training dataset on (a) Facebook,  (b) Enron, and (c) Twitter with Equal cost. 
}
\label{impact-substitute-nodes-other}
\end{figure*}

\begin{figure*}[!tbp]
\center
\subfloat[Facebook]{\includegraphics[width=0.32\textwidth]{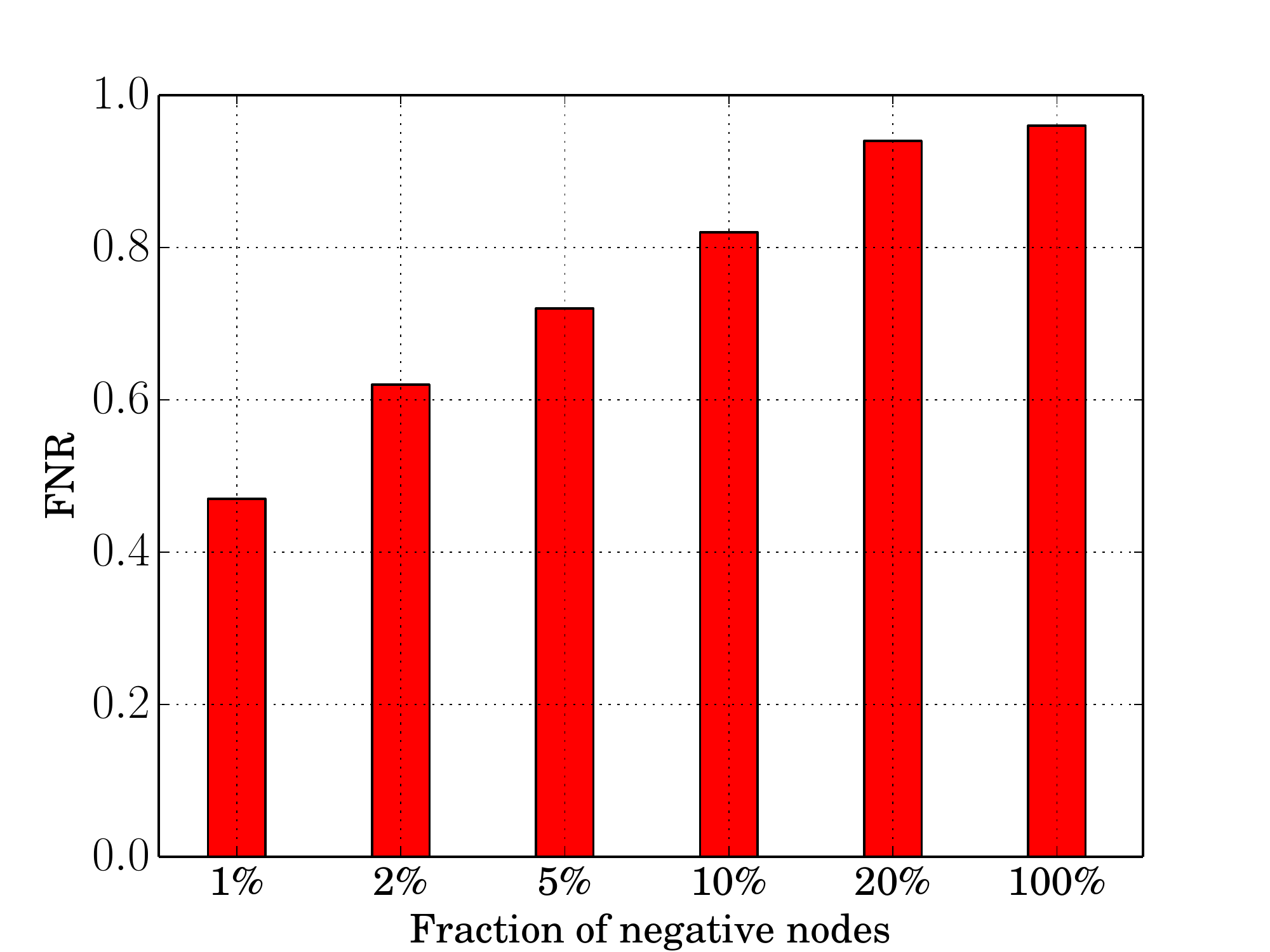}\label{partial-FB}} 
\subfloat[Enron]{\includegraphics[width=0.32\textwidth]{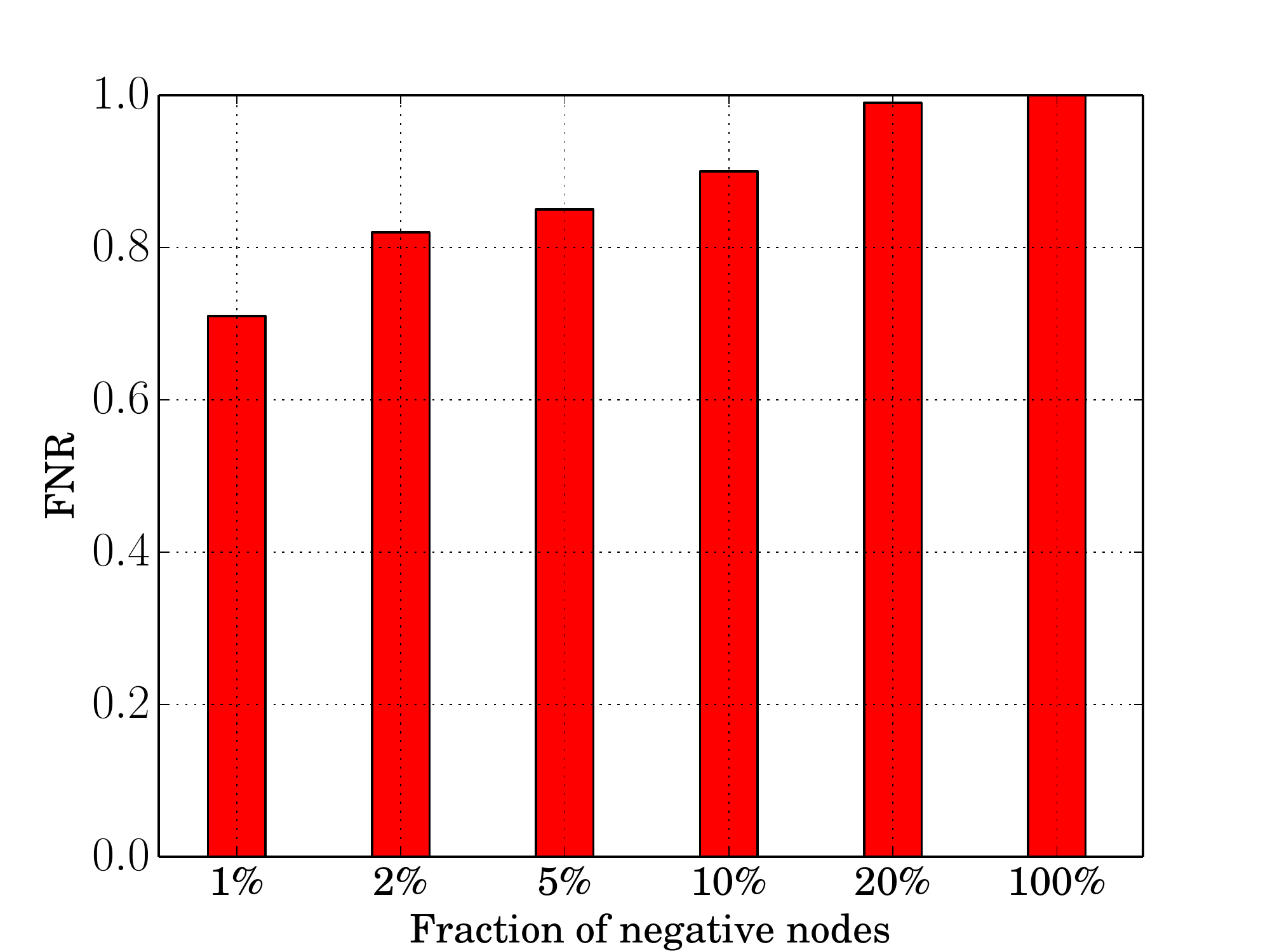}\label{partial-Enron}} 
\subfloat[Twitter]{\includegraphics[width=0.32\textwidth]{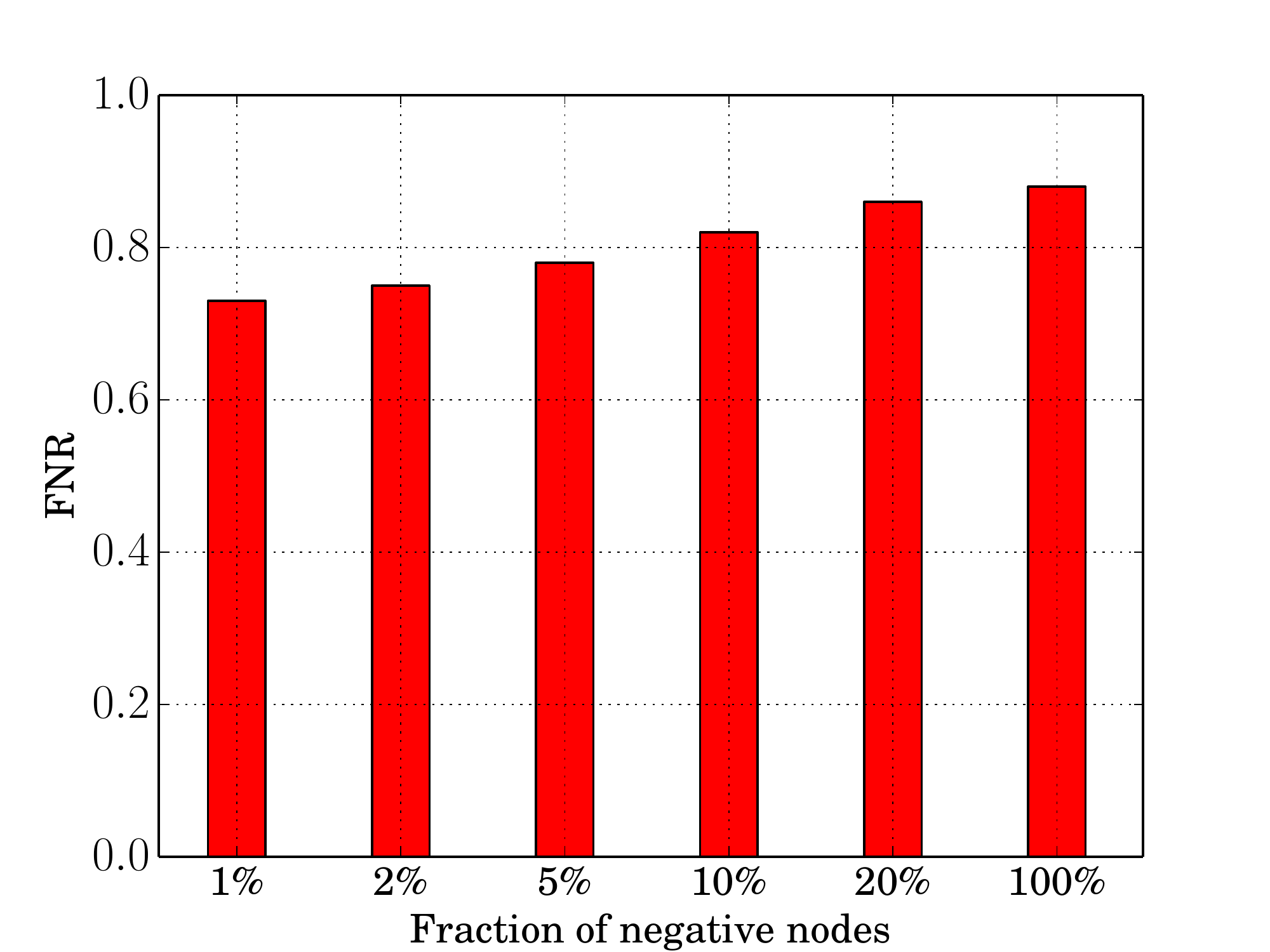}\label{partial-Twitter}} 
\caption{FNRs of our attacks on (a) Facebook,  (b) Enron, and (c) Twitter with Equal cost when the attacker knows $\tau$\% of the negative nodes and edges between them.}
\label{impact-partial-Equal-Other}
\end{figure*}

\begin{figure*}[!tbp]
\center
\subfloat[]{\includegraphics[width=0.32\textwidth]{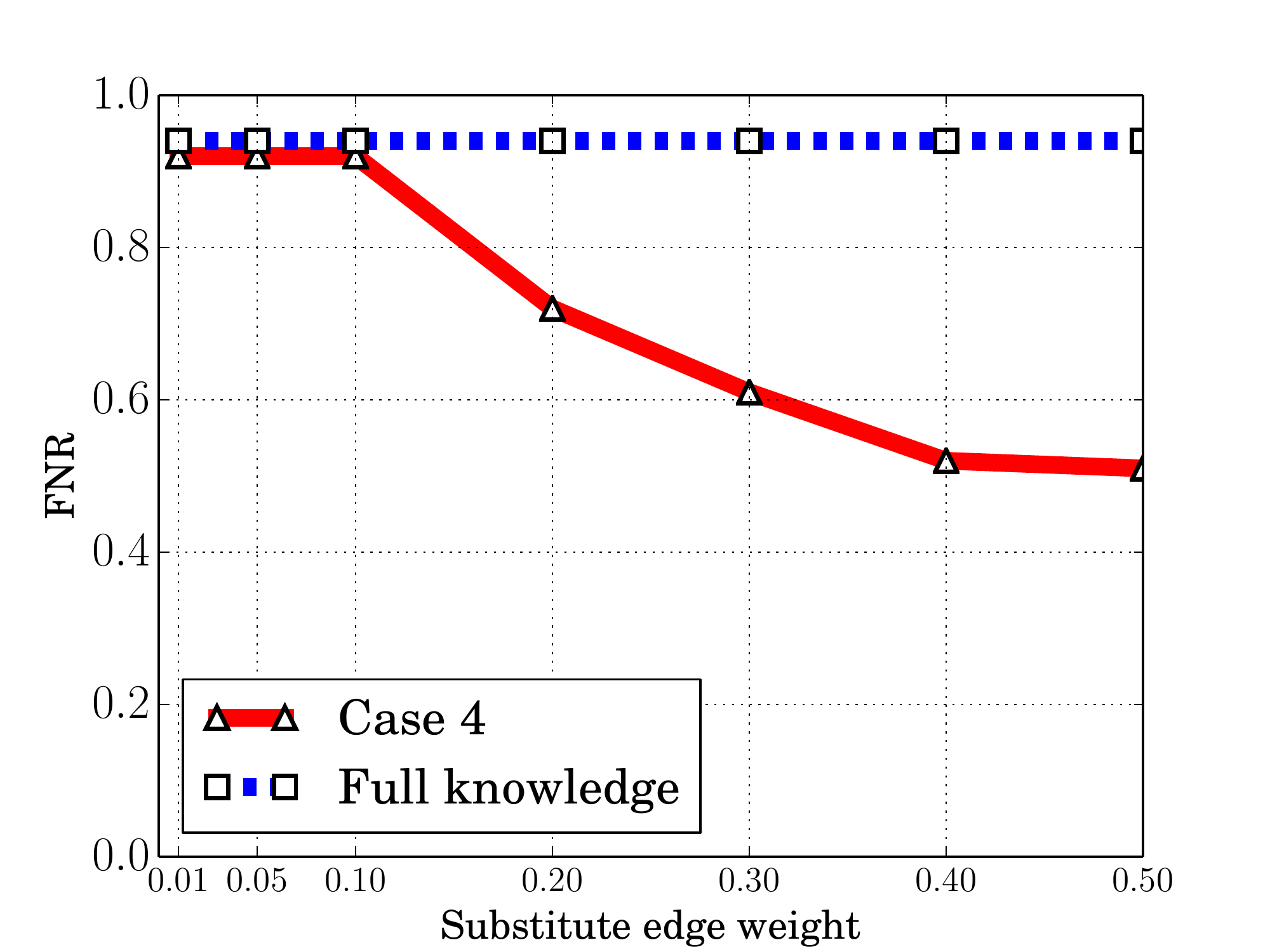} \label{least-FB-weight}} 
\subfloat[]{\includegraphics[width=0.32\textwidth]{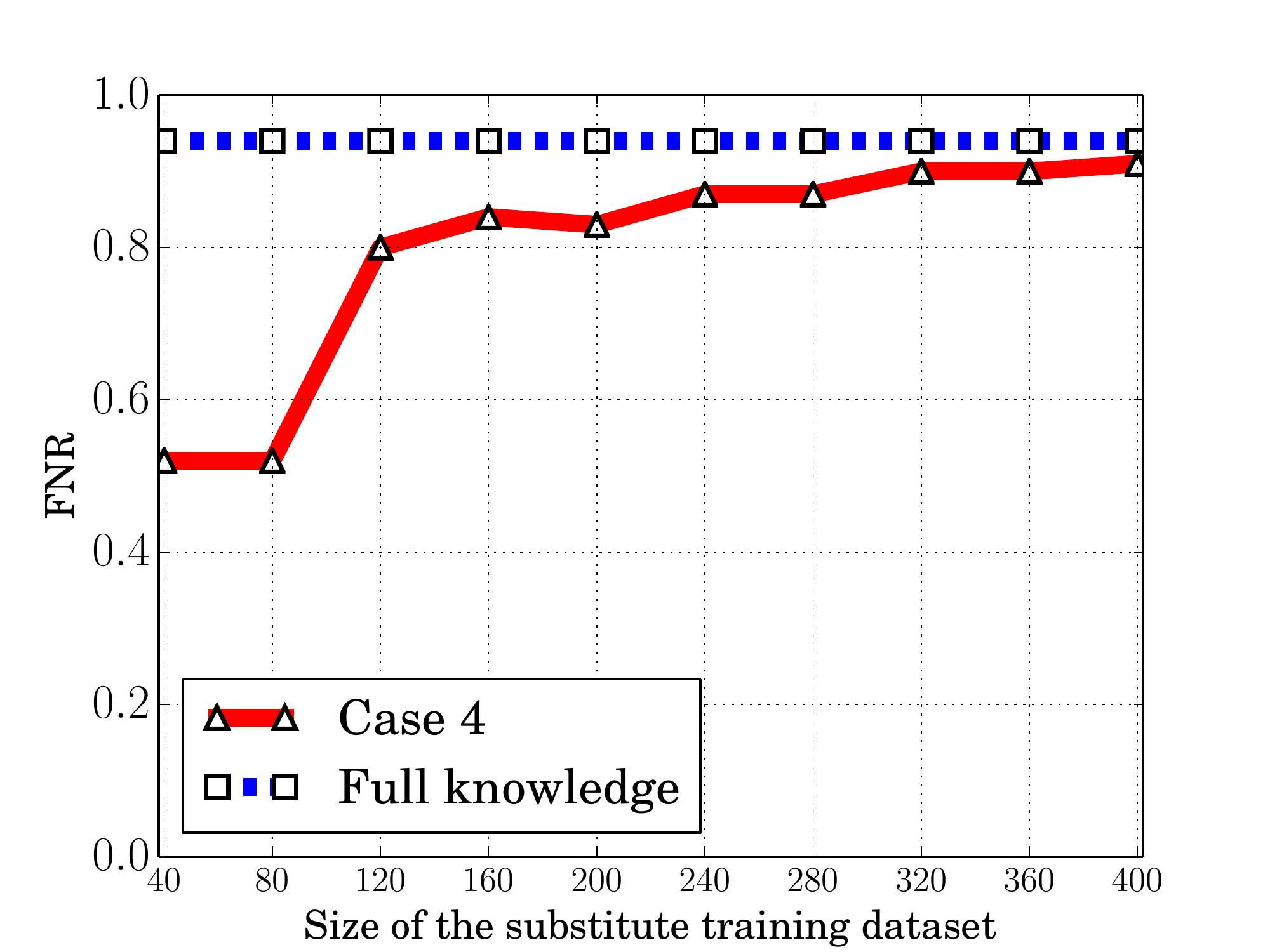} \label{least-FB-training}}
\subfloat[]{\includegraphics[width=0.32\textwidth]{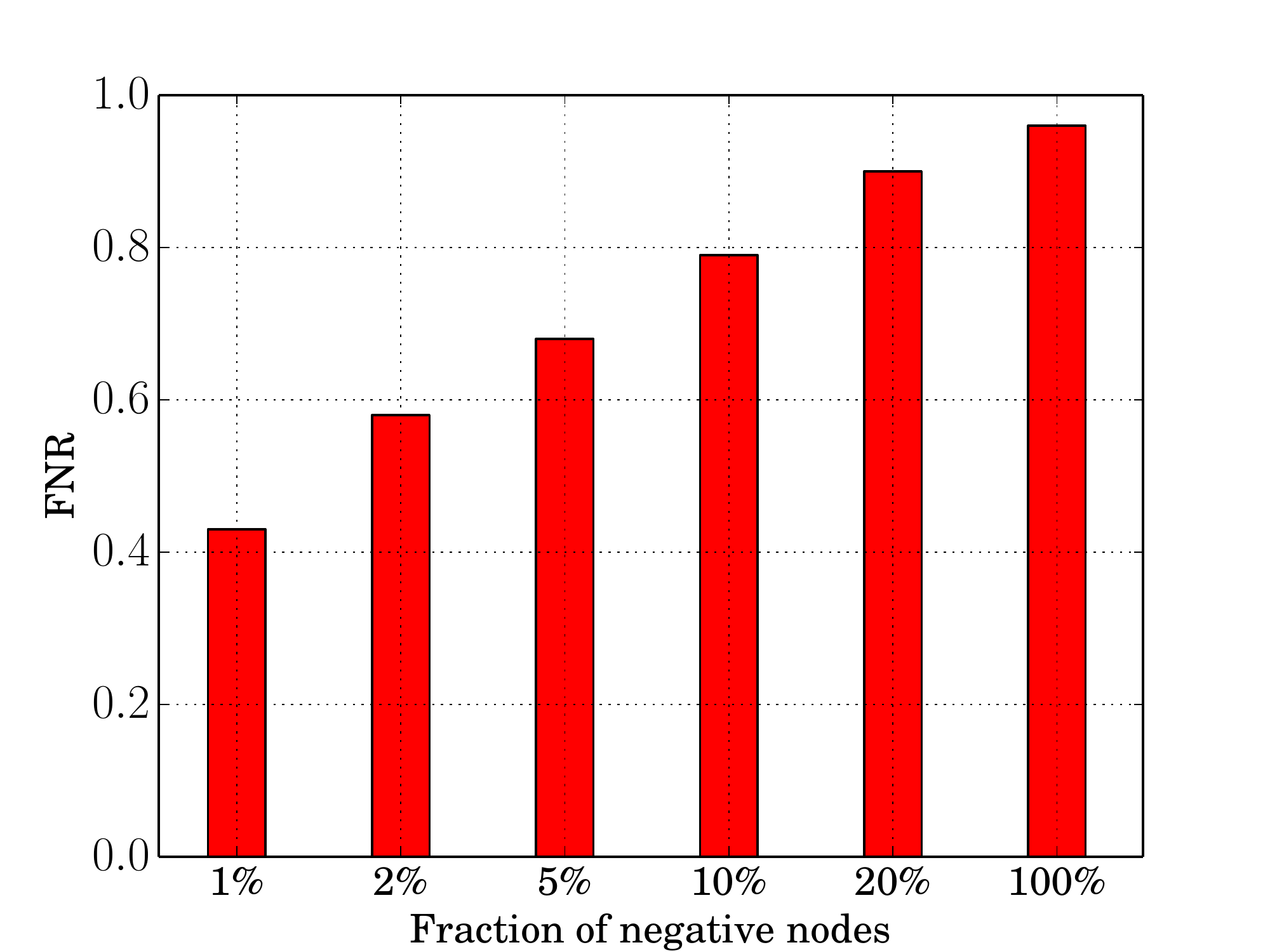} \label{least-FB-partial}} 
\caption{FNRs of our attacks vs. (a) the substitute edge weight; (b) size of the substitute training dataset; (c) fraction of negative nodes on Facebook when the attacker uses substitute parameters, substitute training dataset, and a partial graph.}
\label{least-FB}
\end{figure*}

\begin{figure*}[!tbp]
\vspace{-6mm}
\center
\subfloat[]{\includegraphics[width=0.32\textwidth]{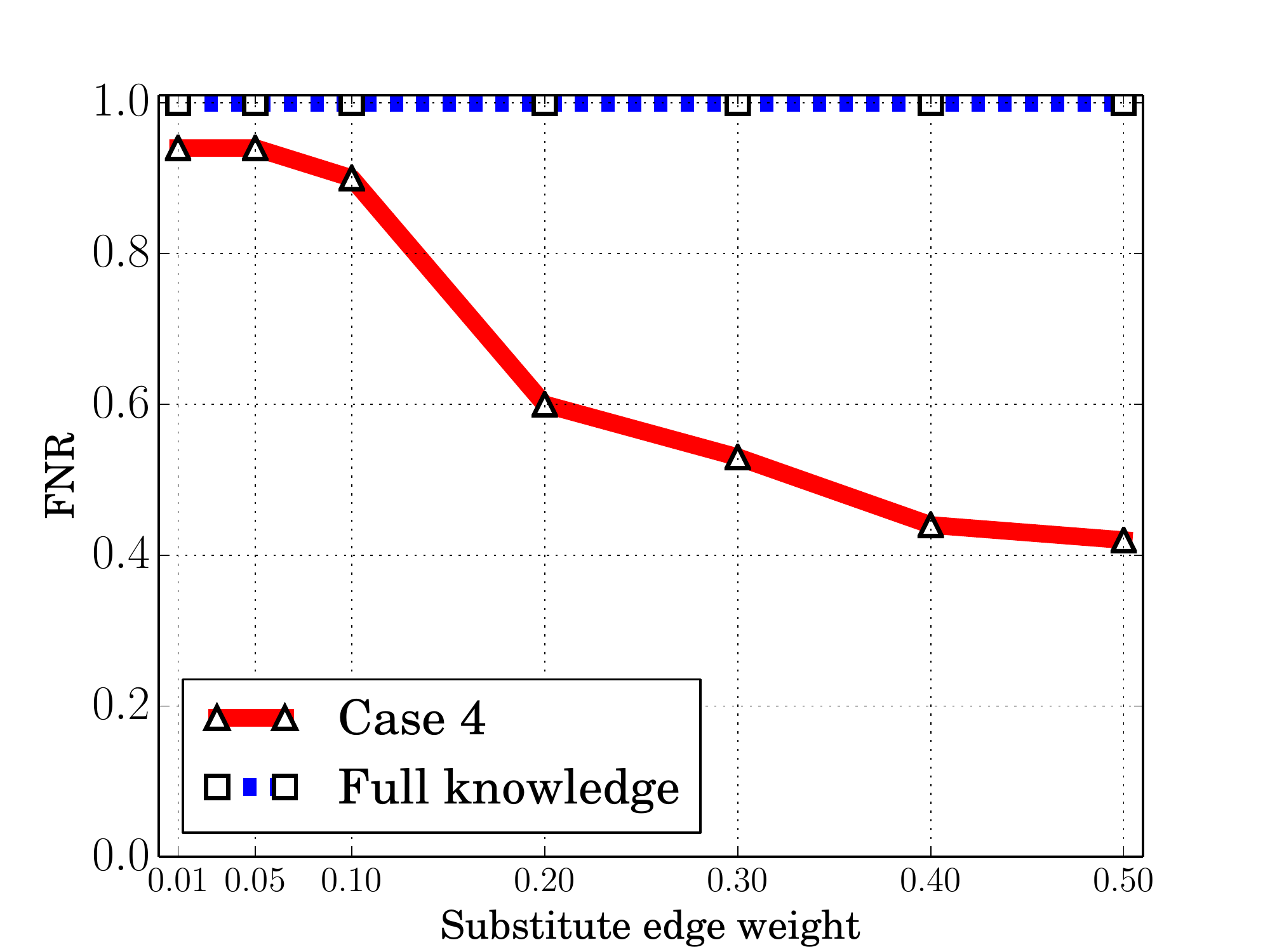} \label{least-Enron-weight}} 
\subfloat[]{\includegraphics[width=0.32\textwidth]{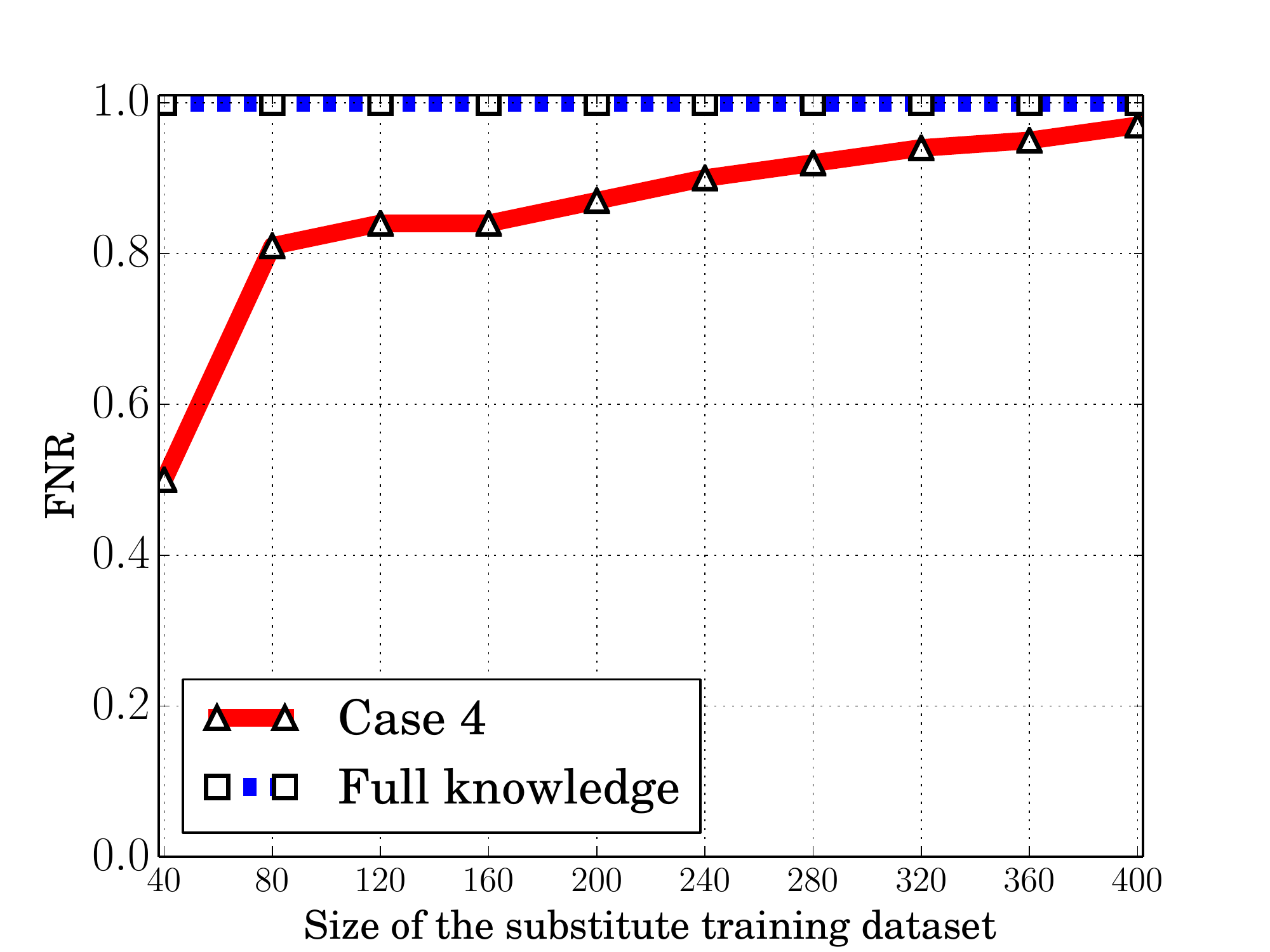} \label{least-Enron-training}}
\subfloat[]{\includegraphics[width=0.32\textwidth]{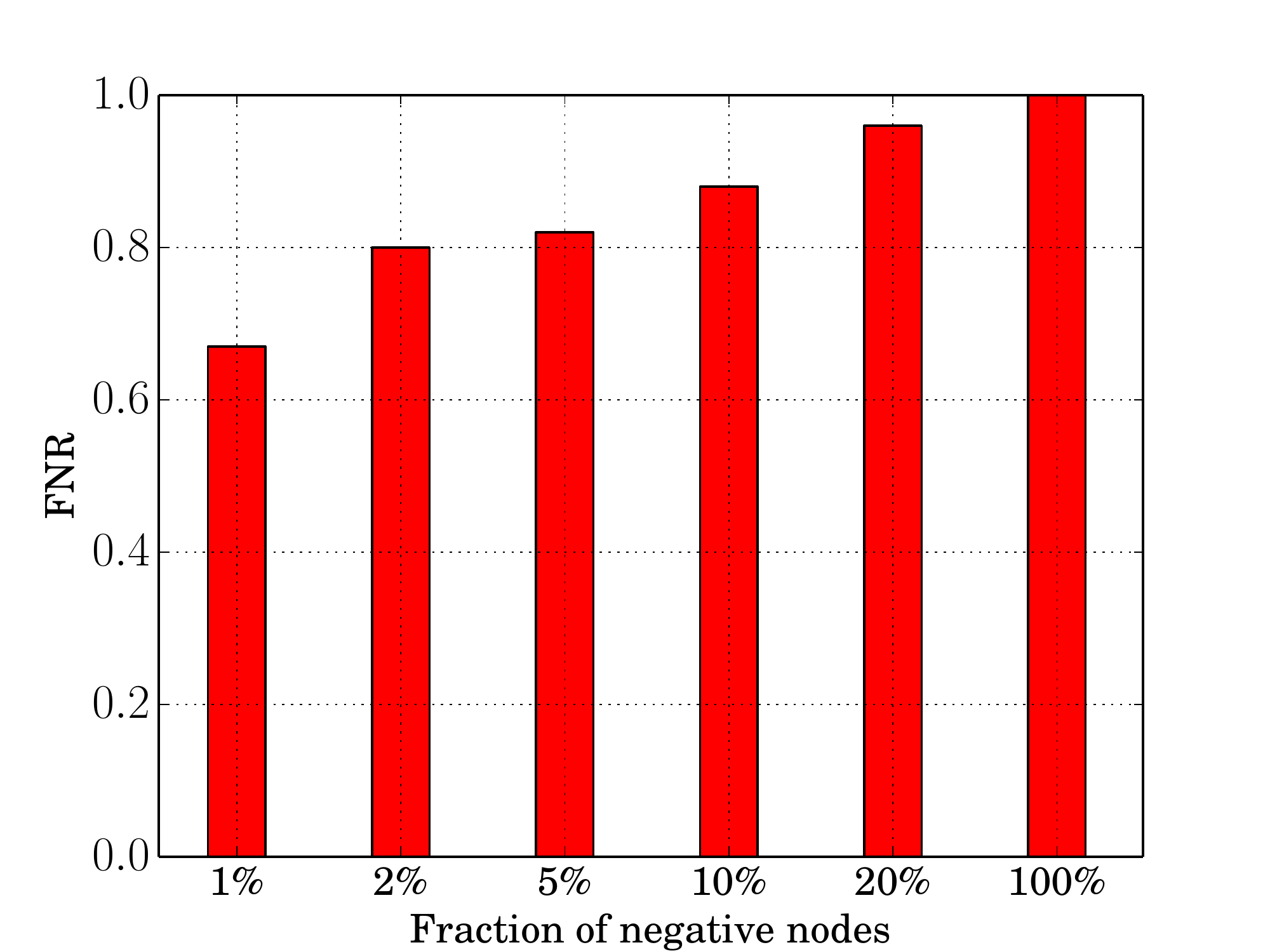} \label{least-Enron-partial}} 
\caption{FNRs of our attacks vs. (a) the substitute edge weight; (b) size of the substitute training dataset; (c) fraction of negative nodes on Enron when the attacker uses substitute parameters, substitute training dataset, and a partial graph.}
\label{least-Enron}
\end{figure*}

\begin{figure*}[!tbp]
\center
\subfloat[]{\includegraphics[width=0.32\textwidth]{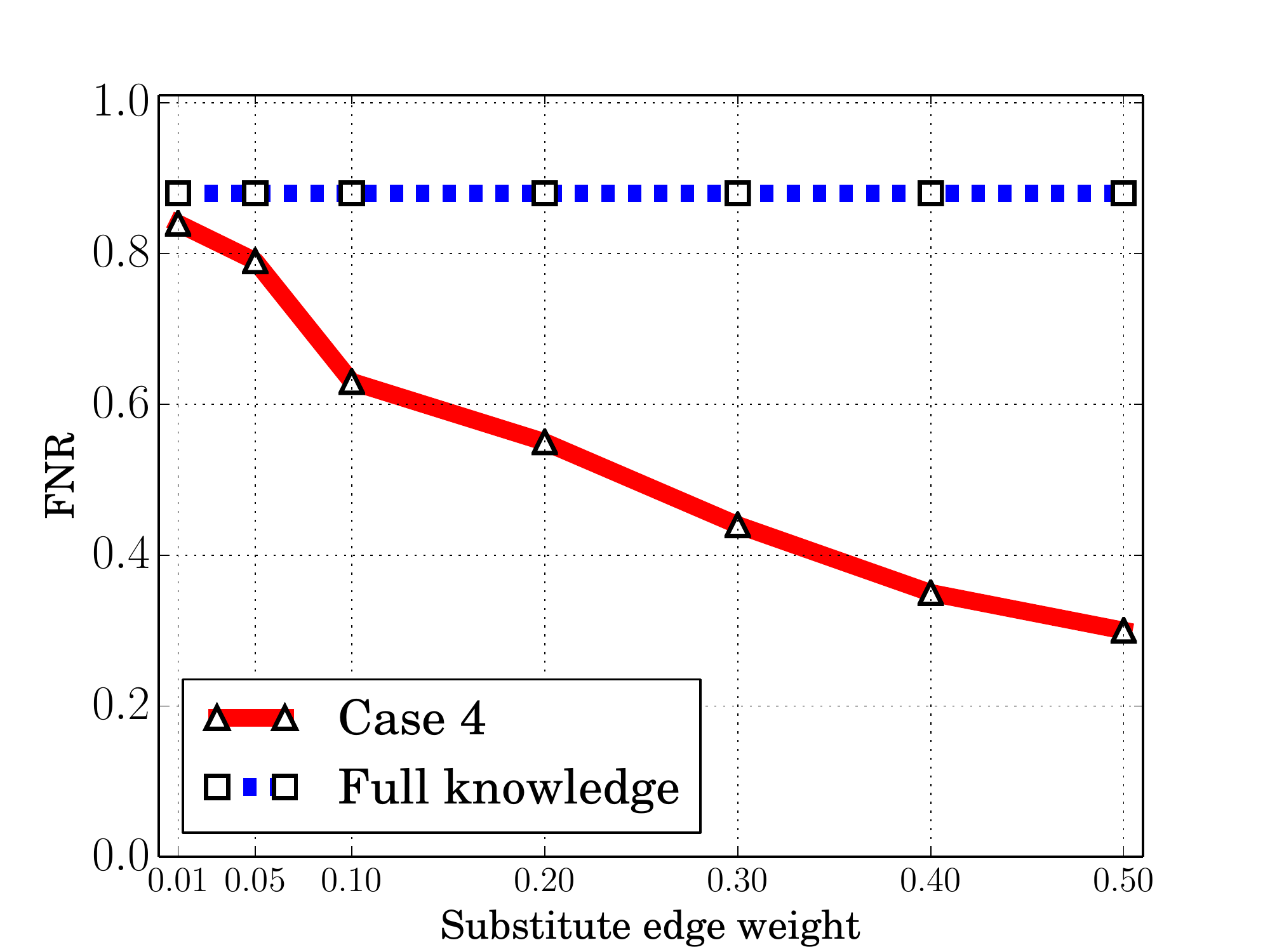} \label{least-Twitter-weight}} 
\subfloat[]{\includegraphics[width=0.32\textwidth]{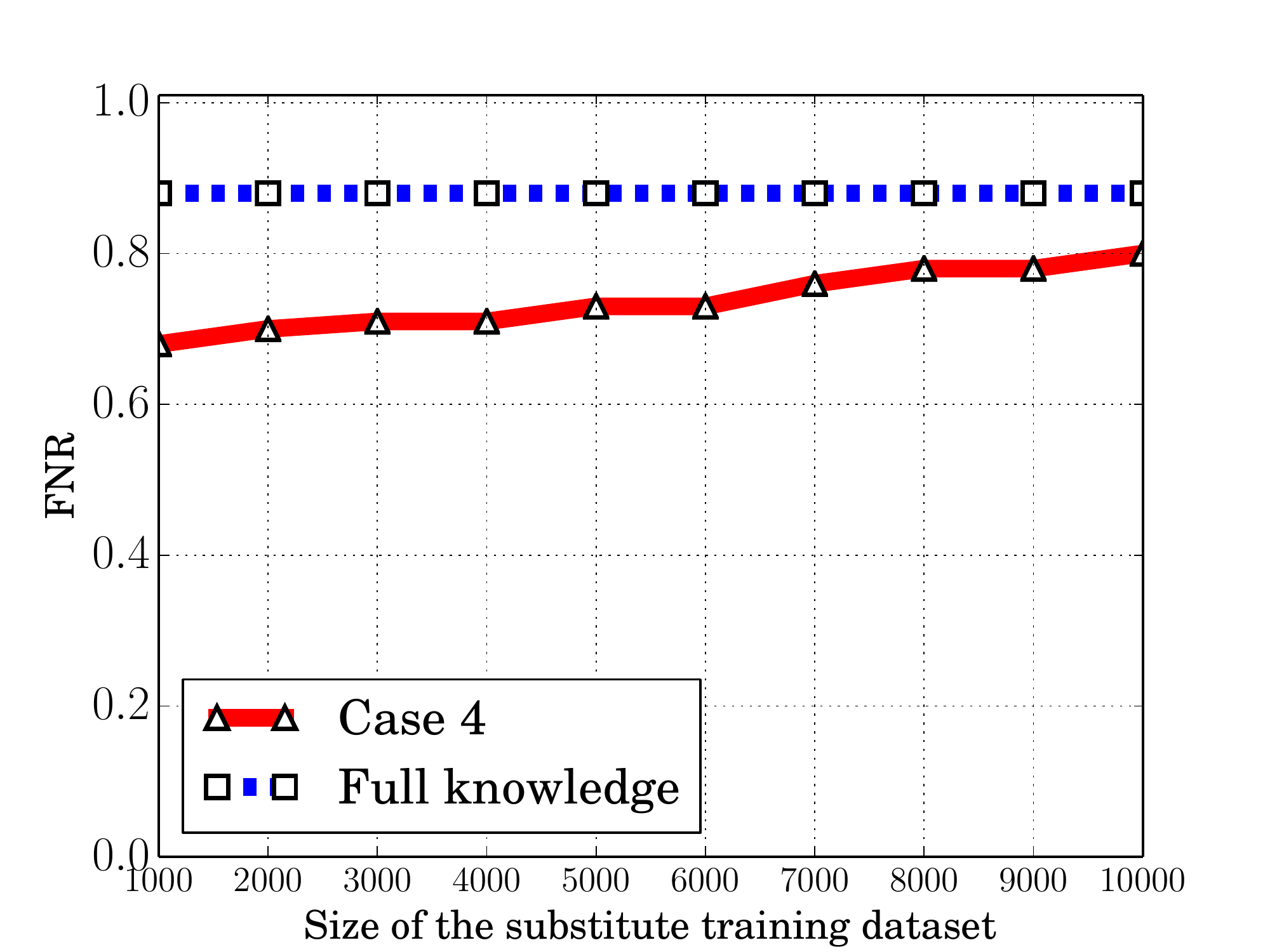} \label{least-Twitter-training}}
\subfloat[]{\includegraphics[width=0.32\textwidth]{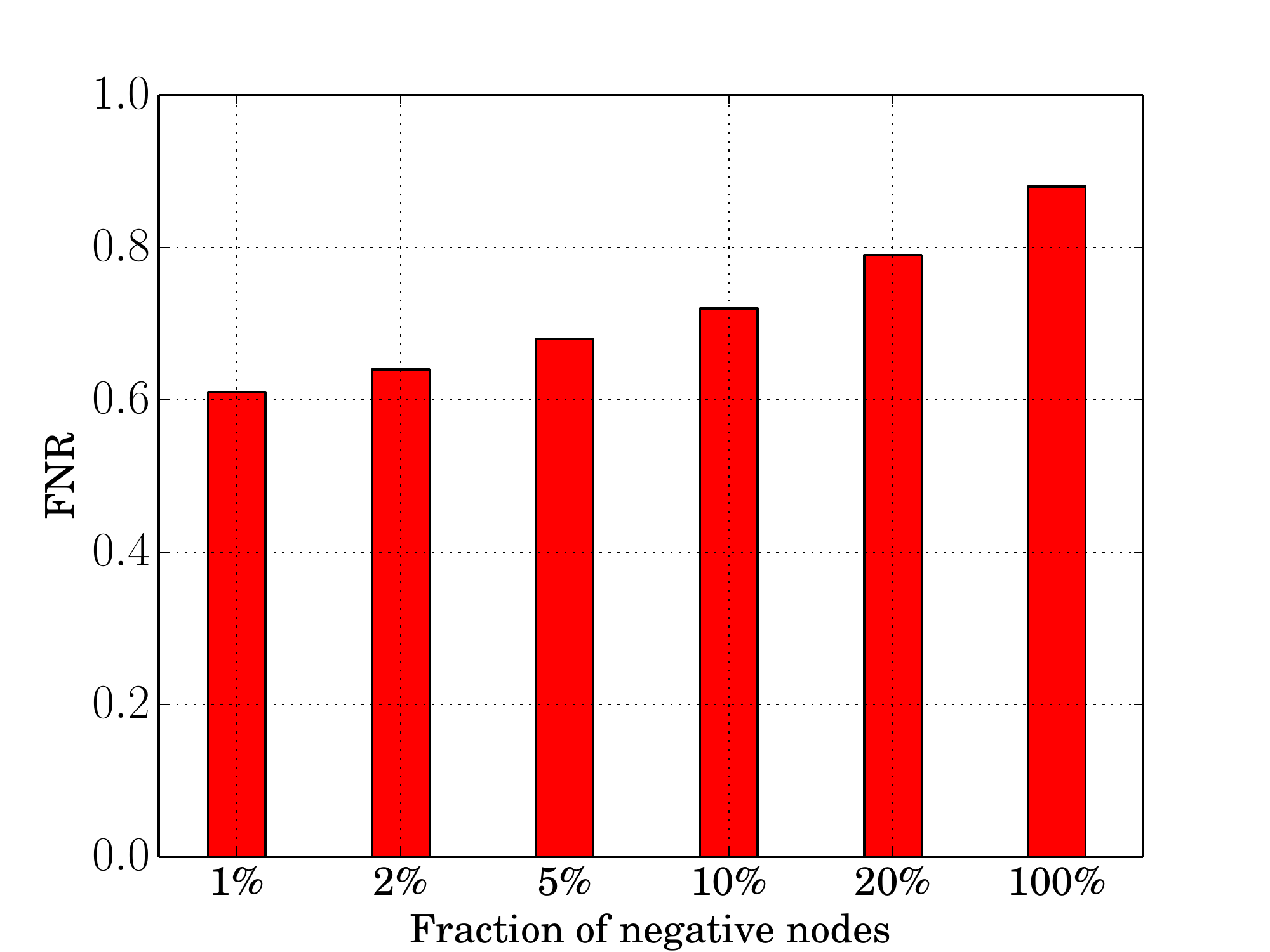} \label{least-Twitter-partial}} 
\caption{FNRs of our attacks vs. (a) the substitute edge weight; (b) size of the substitute training dataset; (c) fraction of negative nodes on Twitter when the attacker uses substitute parameters, substitute training dataset, and a partial graph.}
\label{least-Twitter}
\end{figure*}

\begin{table}[!tbp]
\centering
\ssmall
\caption{Transferability of our attacks to other graph-based classification methods on Enron. 
}
\addtolength{\tabcolsep}{-2pt}
\begin{tabular}{|c|c|c|c|c|c|}
\hline
\multicolumn{6}{|c|}{\textbf{RAND}}             \\ \hline
\multicolumn{2}{|c|}{\textbf{Method}} &  \textbf{No attack}  &  \textbf{Equal} & \textbf{Uniform}   & \textbf{Categorical}  \\ \hline
\multirow{4}{*}{{\bf \makecell{Collective \\ Classification}}}  & {\bf LinLBP}  & 0 & 1.00 & 0.96 & 0.89 \\ \cline{2-6} 
                   &  {\bf JWP} & 0 & 0.97 & 0.96 & 0.86 \\ \cline{2-6} 
                   &  {\bf LBP} & 0.02 & 0.96 & 0.96 & 0.89 \\ \cline{2-6} 
                   & {\bf RW}  & 0.05 & 0.95 & 0.95 & 0.89 \\ \hline
\multirow{4}{*}{{\bf \makecell{Graph  \\ Neural Network}}}  &  {\bf LINE} & 0.06 & 0.94 & 0.92 & 0.85 \\ \cline{2-6}  
				& {\bf DeepWalk}  & 0.25 & 0.6 & 0.58 & 0.5 \\ \cline{2-6} 
				& {\bf node2vec}  & 0.23 & 0.6 & 0.59 & 0.5 \\ \cline{2-6} 
				& {\bf GCN}  & 0.22 & 0.49 & 0.49 & 0.45 \\ \hline
\multicolumn{6}{|c|}{\textbf{CC}}             \\ \hline
\multicolumn{2}{|c|}{\textbf{Method}} &  \textbf{No attack}  &  \textbf{Equal} & \textbf{Uniform}   & \textbf{Categorical}  \\ \hline
\multirow{4}{*}{{\bf \makecell{Collective \\ Classification}}}  & {\bf LinLBP}  & 0 & 1.00 & 0.99 & 0.92 \\ \cline{2-6} 
                   &  {\bf JWP} & 0 & 1.00 & 0.98 & 0.89 \\ \cline{2-6} 
                   &  {\bf LBP} & 0 & 1.00 & 0.91 & 0.85 \\ \cline{2-6} 
                   & {\bf RW}  & 0.04 & 1.00 & 0.97 & 0.81 \\ \hline
\multirow{4}{*}{{\bf \makecell{Graph  \\ Neural Network}}}  &  {\bf LINE} & 0.04 & 0.98 & 0.95 & 0.63 \\ \cline{2-6}  
				& {\bf DeepWalk}  & 0.23 & 0.7 & 0.65 & 0.45 \\ \cline{2-6} 
				& {\bf node2vec}  & 0.25 & 0.68 & 0.6 & 0.5 \\ \cline{2-6} 
                  & {\bf GCN}  & 0.28 & 0.58 & 0.58 & 0.52 \\ \hline
\multicolumn{6}{|c|}{\textbf{CLOSE}}             \\ \hline
\multicolumn{2}{|c|}{\textbf{Method}} &  \textbf{No attack}  &  \textbf{Equal} & \textbf{Uniform}   & \textbf{Categorical}  \\ \hline
\multirow{4}{*}{{\bf \makecell{Collective \\ Classification}}}  & {\bf LinLBP}  & 0 & 1.00 & 1.00 & 0.90 \\ \cline{2-6} 
                   &  {\bf JWP} & 0 & 1.00 & 0.99 & 0.90 \\ \cline{2-6} 
                   &  {\bf LBP} & 0 & 1.00 & 0.99 & 0.84 \\ \cline{2-6} 
                   & {\bf RW}  & 0 & 0.98 & 0.97 & 0.81 \\ \hline
\multirow{4}{*}{{\bf \makecell{Graph  \\ Neural Network}}}  &  {\bf LINE} & 0.13 & 1.00 & 0.95 & 0.6 \\ \cline{2-6}  
				& {\bf DeepWalk}  & 0.3 & 1.00 & 0.93 & 0.48 \\ \cline{2-6} 
				& {\bf node2vec}  & 0.30 & 0.95 & 0.89 & 0.50 \\ \cline{2-6} 
                  & {\bf GCN}  & 0.3 & 0.65 & 0.65 & 0.45 \\ \hline
\end{tabular}
\label{transfer_Enron}
\end{table}

\begin{table}[!tbp]
\centering
\ssmall
\caption{Transferability of our attacks to other graph-based classification methods on Epinions. 
"--" means that the method cannot be executed on our machine due to the insufficient memory. 
}
\addtolength{\tabcolsep}{-2pt}
\begin{tabular}{|c|c|c|c|c|c|}
\hline
\multicolumn{6}{|c|}{\textbf{RAND}}             \\ \hline
\multicolumn{2}{|c|}{\textbf{Method}} &  \textbf{No attack}  &  \textbf{Equal} & \textbf{Uniform}   & \textbf{Categorical}  \\ \hline
\multirow{4}{*}{{\bf \makecell{Collective \\ Classification}}}  & {\bf LinLBP}  & 0 & 0.98 & 0.94 & 0.92 \\ \cline{2-6} 
                   &  {\bf JWP} & 0 & 0.96 & 0.94 & 0.83 \\ \cline{2-6} 
                   &  {\bf LBP} & 0.07 & 0.96 & 0.94 & 0.92 \\ \cline{2-6} 
                   & {\bf RW}  & 0.08 & 0.95 & 0.87 & 0.9 \\ \hline
\multirow{4}{*}{{\bf \makecell{Graph  \\ Neural Network}}}   &  {\bf LINE} & 0.33 & 0.96 & 0.95 & 0.92 \\ \cline{2-6} 
				& {\bf DeepWalk}  & 0.5 & 0.65 & 0.85 & 0.68 \\ \cline{2-6} 
				& {\bf node2vec}  & 0.48 & 0.65 & 0.82 & 0.65 \\ \cline{2-6} 
                   & {\bf GCN}  &  -- & -- &  -- & --  \\ \hline
\multicolumn{6}{|c|}{\textbf{CC}}             \\ \hline
\multicolumn{2}{|c|}{\textbf{Method}} &  \textbf{No attack}  &  \textbf{Equal} & \textbf{Uniform}   & \textbf{Categorical}  \\ \hline
\multirow{4}{*}{{\bf \makecell{Collective \\ Classification}}}  & {\bf LinLBP}  & 0 &  0.99 &  0.94	 &  0.88 \\ \cline{2-6} 
                   &  {\bf JWP} & 0 & 0.96 & 0.94 & 0.78 \\ \cline{2-6} 
                   &  {\bf LBP} & 0 & 0.95 & 0.91 & 0.86 \\ \cline{2-6} 
                   & {\bf RW}  & 0 & 0.94 & 0.87 & 0.76 \\ \hline
\multirow{4}{*}{{\bf \makecell{Graph  \\ Neural Network}}}  &  {\bf LINE} & 0.31 & 0.96 & 0.9 & 0.86 \\ \cline{2-6}  
				& {\bf DeepWalk}  & 0.45 & 0.75 & 0.78 & 0.68 \\ \cline{2-6} 
				& {\bf node2vec}  & 0.42 & 0.75 & 0.75 & 0.7 \\ \cline{2-6} 
                   & {\bf GCN}  & -- & -- & -- & -- \\ \hline
\multicolumn{6}{|c|}{\textbf{CLOSE}}             \\ \hline
\multicolumn{2}{|c|}{\textbf{Method}} &  \textbf{No attack}  &  \textbf{Equal} & \textbf{Uniform}   & \textbf{Categorical}  \\ \hline
\multirow{4}{*}{{\bf \makecell{Collective \\ Classification}}}  & {\bf LinLBP}  & 0 &  1.00 &  0.96	 &  0.86 \\ \cline{2-6} 
                   &  {\bf JWP} & 0 & 1.00 & 0.96 & 0.86 \\ \cline{2-6} 
                   &  {\bf LBP} & 0 & 1.00 & 0.94 & 0.85 \\ \cline{2-6} 
                   & {\bf RW}  & 0 & 1.00 & 0.95 & 0.8 \\ \hline
\multirow{4}{*}{{\bf \makecell{Graph  \\ Neural Network}}}  &  {\bf LINE} & 0.39 & 0.96 & 0.69 & 0.76 \\ \cline{2-6}  
				& {\bf DeepWalk}  & 0.45 & 0.58 & 0.55 & 0.46 \\ \cline{2-6} 
				& {\bf node2vec}  & 0.42 & 0.59 & 0.59 & 0.49 \\ \cline{2-6} 
                & {\bf GCN}  & -- & -- & -- & -- \\ \hline
\end{tabular}
\label{transfer_Epinions}
\end{table}

\begin{table}[!tbp]
\centering
\ssmall
\caption{Transferability of our attacks to other graph-based classification methods on Twitter.
}
\addtolength{\tabcolsep}{-2pt}
\begin{tabular}{|c|c|c|c|c|c|}
\hline
\multicolumn{6}{|c|}{\textbf{RAND}}             \\ \hline
\multicolumn{2}{|c|}{\textbf{Method}} &  \textbf{No attack}  &  \textbf{Equal} & \textbf{Uniform}   & \textbf{Categorical}  \\ \hline
\multirow{4}{*}{{\bf \makecell{Collective \\ Classification}}}  & {\bf LinLBP}  & 0 & 0.85 & 0.85 & 0.82 \\ \cline{2-6} 
                   &  {\bf JWP} & 0 & 0.82 & 0.81 & 0.76 \\ \cline{2-6} 
                   &  {\bf LBP} & 0.01 & 0.82 & 0.8 & 0.72 \\ \cline{2-6} 
                   & {\bf RW}  & 0.03 & 0.78 & 0.77 & 0.69 \\ \hline
\multirow{4}{*}{{\bf \makecell{Graph  \\ Neural Network}}}  &  {\bf LINE} & -- & -- & -- & -- \\ \cline{2-6}  
				& {\bf DeepWalk}  & -- & -- & -- & -- \\ \cline{2-6} 
				& {\bf node2vec}  & -- & -- & -- & -- \\ \cline{2-6} 
                & {\bf GCN}  & -- & -- & -- & -- \\ \hline

\multicolumn{6}{|c|}{\textbf{CC}}             \\ \hline
\multicolumn{2}{|c|}{\textbf{Method}} &  \textbf{No attack}  &  \textbf{Equal} & \textbf{Uniform}   & \textbf{Categorical}  \\ \hline
\multirow{4}{*}{{\bf \makecell{Collective \\ Classification}}}  & {\bf LinLBP}  & 0 &  0.88 &  0.86	 &  0.85 \\ \cline{2-6} 
                   &  {\bf JWP} & 0 & 0.82 & 0.82 & 0.80 \\ \cline{2-6} 
                   &  {\bf LBP} & 0 & 0.81 & 0.81 & 0.78 \\ \cline{2-6} 
                   & {\bf RW}  & 0.03 & 0.80 & 0.80 & 0.76 \\ \hline
\multirow{4}{*}{{\bf \makecell{Graph  \\ Neural Network}}}  &  {\bf LINE} & -- & -- & -- & -- \\ \cline{2-6}  
				& {\bf DeepWalk}  & -- & -- & -- & -- \\ \cline{2-6} 
				& {\bf node2vec}  & -- & -- & -- & -- \\ \cline{2-6} 
                & {\bf GCN}  & -- & -- & -- & -- \\ \hline

\multicolumn{6}{|c|}{\textbf{CLOSE}}             \\ \hline
\multicolumn{2}{|c|}{\textbf{Method}} &  \textbf{No attack}  &  \textbf{Equal} & \textbf{Uniform}   & \textbf{Categorical}  \\ \hline
\multirow{4}{*}{{\bf \makecell{Collective \\ Classification}}}  & {\bf LinLBP}  & 0 &  0.87 &  0.85	 &  0.83 \\ \cline{2-6} 
                   &  {\bf JWP} & 0 & 0.87 & 0.85 & 0.83 \\ \cline{2-6} 
                   &  {\bf LBP} & 0 & 0.82 & 0.82 & 0.8 \\ \cline{2-6} 
                   & {\bf RW}  & 0.01 & 0.82 & 0.81 & 0.78 \\ \hline
\multirow{4}{*}{{\bf \makecell{Graph  \\ Neural Network}}}  &  {\bf LINE} & -- & -- & -- & -- \\ \cline{2-6}  
				& {\bf DeepWalk}  & -- & -- & -- & -- \\ \cline{2-6} 
				& {\bf node2vec}  & -- & -- & -- & -- \\ \cline{2-6} 
                & {\bf GCN}  & -- & -- & -- & -- \\ \hline
\end{tabular}
\label{transfer_Twitter}
\end{table}

\end{document}